\documentclass[11pt]{article}
\pdfoutput=1
\usepackage{ifpdf}
\usepackage{amssymb}
\usepackage{amsbsy}
\usepackage{graphicx}
\usepackage[numbers]{natbib}
\usepackage{color}
\usepackage{amsmath}

\title{Exact regularized point particle method for multi-phase flows in the 
two-way coupling regime}

\author{P.~Gualtieri$^1$
\thanks{Email address for correspondence: paolo.gualtieri@uniroma1.it},
F.~Picano$^{2,1}$, G.~Sardina$^{3,1}$ \& C.M.~Casciola$^1$ \\
$^1$Dipartimento di Ingegneria Meccanica e Aerospaziale, \\
Universit\`a di Roma {\em La Sapienza} Via Eudossiana 18, 00184 Roma {\it Italy}. \\
$^2$Linn{\'e} Flow Center, KTH Mechanics, \\Osquars Backe 18, SE-100 44 Stockholm, {\it Sweden}. \\
$^3$UKE - Universit\`a Kore di ENNA \\ Facolt\`a di Ingegneria, Architettura e
Scienze Motorie, \\ Via delle Olimpiadi,  94100 Enna, {\it Italy}.}

\newcommand {\fg}     {\hat g}

\newcommand {\vxi}    {\mbox{\boldmath $\xi$}}
\newcommand {\vzeta}  {\mbox{\boldmath $\zeta$}}

\newcommand {\vomega} {\mbox{\boldmath $\omega$}}

\newcommand {\vA}     {{\bf A}}

\newcommand {\vd}     {{\bf d}}
\newcommand {\vD}     {{\bf D}}
\newcommand {\ve}     {{\bf e}}

\newcommand {\vF}     {{\bf F}}
\newcommand {\vg}     {{\bf g}}

\newcommand {\vh}     {{\bf h}}
\newcommand {\vk}     {{\bf k}}

\newcommand {\vI}     {{\bf I}}

\newcommand {\vn}     {{\bf n}}

\newcommand {\vr}     {{\bf r}}
\newcommand {\vR}     {{\bf R}}

\newcommand {\vu}     {{\bf u}}

\newcommand {\vv}     {{\bf v}}
\newcommand {\vV}     {{\bf V}}
\newcommand {\vx}     {{\bf x}}

\newcommand {\vy}     {{\bf y}}

\newcommand {\vw}     {{\bf w}}

\def\Rset {{\rm I \kern-.2em R}} 

\begin{document}
\maketitle
\begin{abstract} 
Particulate flows have been largely studied under the simplifying assumptions of 
one-way coupling regime where the disperse phase do not react-back on the 
carrier fluid. For instance, in the context of turbulent flows, many
non trivial phenomena such as small scales particles clustering or preferential 
spatial accumulation have been explained and understood.
A more complete view of multiphase flows can be gained calling into play
two-way coupling effects, i.e. by accounting for the inter-phase momentum 
exchange between the carrier and the suspended phase, certainly relevant at 
increasing mass loading. In such regime, partially investigated in the past by 
the so-called Particle In Cell (PIC) method, much is still to be learned 
about the  dynamics of the disperse phase and the ensuing alteration of 
the carrier flow. 

In this paper we present a new methodology rigorously designed to capture the
inter-phase momentum exchange for particles smaller than the smallest 
hydrodynamical scale, e.g. the Kolmogorov scale in a turbulent flow.
In fact, the momentum coupling mechanism exploits the unsteady Stokes 
flow around a small 
rigid sphere where the transient disturbance produced by each particle is 
evaluated in a closed  form. The particles are described as lumped, 
point masses which would lead to the appearance of singularities. A rigorous 
regularization procedure is conceived to extract the physically relevant 
interactions between particles and fluid which avoids any ``ah hoc'' assumption. 
The approach is suited for high efficiency implementation on massively 
parallel machines since the transient disturbance produced by the particles 
is strongly  localized in space  around the actual particle position. 
As will be shown, hundred thousands particles 
can therefore be handled at an affordable computational cost as demonstrated by 
a preliminary application to a particle laden turbulent shear flow.
\end{abstract} 
\section{Introduction} \label{sec:intro}
Multiphase flows represent the cornerstone of many fields
of science and technology ranging from micro-scale devices to the large scale 
cyclonic separators of industrial plants. In the context of micro/nano 
technologies, the transport of small particles or bubbles by a carrier fluid 
is fundamental in designing micro-devices where particles must be separated, 
mixed or advected towards the sensible regions of the apparatus for detection 
purposes, see e.g \cite{stone2004}. Concerning larger scale devices, the turbulent 
transport of a disperse phase is relevant for the dynamics of small fuel 
droplets in combustion chambers, \cite{post}, or in the spatial evolution of 
sprays employed for surface coating, \cite{pawlowski2008}.

Important aspects of multiphase flows are related to the intrinsic coupling 
between the motion of the disperse phase and the carrier fluid which involves 
mass, momentum and energy exchange between the two phases. Hydrodynamic 
interactions among the particles or inter-particles collisions might also
occur. The regime where all these interactions take place is known as four-way 
coupling regime, see e.g. \cite{balachandar_rev,elgo_map}. 
The straightforward method to capture
such complex physics is represented by numerical simulations where the
fluid flow around each particle is fully resolved. This means that the actual
particle boundary has to be resolved on the computational grid and the 
coupling with the fluid occurs via the non slip boundary conditions imposed on
the particle surface. The hydrodynamic force on each particle can be directly
computed by integrating the pressure and shear stress distribution on the 
boundary. Even though this approach captures entirely the physics, it is 
computationally demanding and limited to the simulation of a relatively
small number of ``large'' particles. The adjective large means that the particle
typical size, the diameter $d_p$, is larger than the smallest physically
active hydrodynamical scale $\eta$. For instance $\eta$ could be either
the Kolmogorov dissipative scale in a turbulent flow or the smallest spatial 
scale in a micro-fluidic apparatus. In the context of the so called
resolved particles simulations many approaches are available ranging from
finite volume techniques, \cite{burton}, immersed boundary methods, 
\cite{lucci_jfm},
or approaches based on the Lattice-Boltzman equations, \cite{ten_cate,gao2011}.
Alternative approaches are however available. For instance, the PHYSALIS technique,
see e.g. \cite{zhang} and references therein, has been recently adopted to address
the interaction of solid particle and a turbulent flow, \cite{naso}. 
\cite{homann_bec} adapted the pseudo-penalization spectral method proposed by
\cite{pasquetti} to account for the coupled dynamics of neutrally buoyant
particles in a turbulent flow. The Force Coupling Method (FCM) proposed by Maxey 
and coworkers, see e.g. among many others the papers by \cite{maxeypat,lomholt}, 
is certainly worth mentioning in detail. In the FCM the effect that each particle 
exerts on the fluid is approximated by a multipole expansion of a regularized 
steady Stokes solution where the concentrated delta-function forces are mollified 
to a Gaussian.
The basic method has been continuously improved by 
including several physical effects such as lubrication forces for closely packed 
particles \cite{dance} or the effects of elongated particles \cite{liu}. 
Recently a numerical simulation of homogeneous isotropic turbulence laden with 
thousands of relatively large particles $(d_p/\eta=6\div12)$ has been reported 
by \cite{yeo}.

The opposite limit of particles much smaller than the smallest 
hydrodynamical scale is also relevant in many applications. For instance 
the mixing and combustion of a turbulent spray after that the primary
atomization phase has occurred, takes place in presence of significant
momentum coupling among the carrier fluid and the fuel droplets, see
e.g. the recent review by \cite{jenny}. In fact, 
in dilute suspensions the volume fraction of the particles is small enough
to neglect hydrodynamic interactions and collisions among particles.
However, for large values of the particle-to-fluid density ratio,
significant mass loads (ratio between the mass of the disperse phase and 
the fluid) may occur. In such regime, the so called the two-way coupling regime,
the momentum exchange between the two phases is significant and must be 
accounted for. The Particle In Cell (PIC) method, \cite{crowe}, is still a 
valuable tool to model the momentum coupling. Such approach needs substantial
care, however,  due to technical issues associated with the injection of the point-wise 
forcing of the particles on the computational grid where the continuous fluid 
phase is resolved. Indeed, the force that the particles exert on the fluid is 
regularized by averaging on the volume of the computational cell. 
Hence, the coupling term results strongly grid dependent unless the number 
of particles per cell $N_p/N_c$ exceeds a certain threshold, see e.g. the 
numerical results in \cite{jfm_2way}, the discussion by \cite{balachandar2009} 
and the comments in \cite{jenny}.

Alternative to the PIC approach, other methods which are able to work 
irrespective of particle number density do indeed exist. For instance \cite{pan} 
modeled the disturbance flow produced by each point particle in terms of 
the steady Stokeslet. Though interesting, this approach has several potential 
shortcomings. The disturbance 
flow decays in space away from the particle as slow as the inverse distance and 
the  perturbation induced by a single particle affects the whole domain.
In these conditions, any truncation is undoubtably bound to deeply alter the 
dynamics. Additionally, the disturbance flow presents the singularity associated 
with the steady Stokeslet. Moreover, the steady Stokes solution used to model 
the fluid-particle interaction is not uniformly valid and fails away from 
the particle. The Oseen correction consistently accounts for the unavoidable 
far field convective effects,  see classical textbooks like \cite{lamb,batchelor}. 
Numerical approaches based on  this improved modeling can be found, e.g., 
in \cite{subramanian2008,pignatel2011}.

In the present paper we propose a new approach able to provide a physically 
consistent and numerically convergent solution for the flow disturbance 
produced by a huge number of small, massive particles coupled to a generic, 
possibly turbulent, carrier flow. Hereafter this new formulation will be 
referred to as the Exact Regularized Point Particle (ERPP) method. 
As it will be shown in detail, this approach presents several advantages. 
The most significant one is related to the  physical accuracy of the momentum 
coupling modeling. In a nutshell, in the relative motion with respect to 
the fluid, the particle generates a vortical field. 
Even though the relative Reynolds number is small, the local flow is dominated by 
unsteady viscous effects as discussed by \cite{bruno2008}. Vorticity production is 
a localized process that takes a finite elapsed time $\epsilon_R$ since generation 
to reach the relevant hydrodynamic scales of the flow. It is indeed this transient 
process of localized generation and finite time diffusion that introduces the 
actual momentum coupling with the carrier flow. Indeed, the model here envisaged 
reproduces this physical process by addressing the velocity field, 
rather than vorticity. The finite time delay $\epsilon_R$ automatically provides 
the regularization of the disturbance field. Instead of being a purely 
mathematical  or numerical ingredient, the regularization featured by ERPP is 
intrinsically associated with the actual physical process of vorticity generation
and viscous diffusion. A distinguishing aspect of ERPP is that all the vorticity 
generated by the particle is properly transferred to the fluid phase, entailing 
momentum conservation. A crucial concern is the small scale component of the 
disturbance field associated with the instantaneously generated vorticity not yet 
diffused up-to the hydrodynamic scales. This localized, inner scale part of the 
disturbance exhibits a $1/r$ local singularity and vanishes altogether at 
the relevant hydrodynamic scales. Although, in principle, this field 
should contribute locally to the convective term of the Navier-Stokes equations,  
its effect is proportional to the (small) particle Reynolds number based on 
the slip velocity. Consistently, it negligibly contributes to the dynamics of 
the relevant hydrodynamic scales. 

Concerning the hydrodynamic force acting on the particles in the 
two way-coupling regime, the expression provided by \cite{maxril} is easily 
adapted to the present context. A crucial issue is the fluid-to-particle slip 
velocity appearing in the expression of the Stokes drag that should be 
understood as the undisturbed fluid velocity (i.e. the relative fluid-particle 
velocity in absence of the particle). In the ERPP the self-induced velocity 
disturbance can be evaluated in a closed form, allowing to explicitly remove its 
contribution. It follows a consistent evaluation of slip velocity and 
hydrodynamic force. 

Despite the underlying theoretical aspects may look complicated at first sight, 
the practical implementation of the ensuing algorithm is remarkably simple 
and efficient. In principle, the coupling algorithm can indeed be embedded in 
any available discretization scheme as implemented in one's favorite  
Navier-Stokes solver. This flexibility allows to easily handle hundred thousands 
particles at  affordable computational cost. 

The paper is organized as follows. The next section \S \ref{sec:methodology} 
forms the main theoretical body of the paper. Along with its subsections, it
introduces the physical model and discusses the inter-phase momentum coupling. 
In section \S \ref{sec:validation} the proposed approach is validated
against available analytical results. Section \S \ref{sec:turb_flow} reports 
preliminary results concerning a turbulent particle-laden shear flow.
Finally, the last section \S \ref{sec:final} summarizes the main findings. 
To smooth out the reading, several appendices are devoted to lengthy technical 
issues whose description inside the main text would have hampered a clear 
exposition of the main material.
\section{Methodology} \label{sec:methodology}
In this section we present the physical model used to achieve the momentum coupling 
between the carrier fluid and the disperse phase in view of describing the 
algorithm for the simulation of particle laden flows in the two way coupling regime.
In doing so, we assume to know the state of the system at time $t$ and propagate 
the solution for one time step $Dt$. Clearly, reiteration of the procedure allows 
to proceed in time, as in standard time integration algorithms. During the 
generic time step of length $Dt=t_{n+1}-t_n$ the state of the system 
will propagate from $t_n$ to $t_{n+1}$. For the sake of simplicity hereafter 
we shall often address the generic step as the step $n = 0$. In this case the 
running time will be $0 \le t \le Dt$ in all the differential equations to 
be addressed. In the discussion, a quantity $\epsilon_R$ with dimension of time 
and the role of a regularization parameter will play a central role.
In this case, having dubbed the current time as instant $t_0 = 0$, it could well 
happen that certain time-delayed variables  (i.e. $t-\epsilon_R$) could be 
negative: we like to assure the reader before hand that this will be no harm.
Integral representation of the solution may represent an exception to this rule. 
Indeed, such integral representation will be used to derive a systematic 
regularization procedure from which we obtain by differentiation the regularized 
pde's to be solved in the algorithm. In such cases the time extrema
will typically range in the interval $\left[0,t\right]$, with $t$ arbitrary, and 
typically larger than $\epsilon_R$.

In this framework, the short time evolution ($Dt \ll 1$) of the overall flow 
(fluid and particles) is conceptually split into a (modified) Navier-Stokes 
evolution of the carrier fluid and a superimposed disturbance flow produced by 
the relative motion of the particles, here assumed spherical, with respect to 
the fluid. Relying on the small Reynolds number of the particle-fluid relative 
motion, the disturbance flow is described by the linear unsteady Stokes equations. 
In fact, we will rearrange the equation in such a way that the exact solution 
of the particle disturbance field is consistently embedded into the carrier
phase Navier-Stokes solver allowing to reconstruct the actual fluid-particle 
coupled solution in the limit of vanishing time step and grid spacing for 
small particle Reynolds number.

The detailed derivation of the  coupling model needs a gradual illustration better 
achieved starting from a schematic description divided in five conceptual steps:
\begin{itemize}
\item[$i)$] Carrier flow-disperse phase interaction and disturbance flow 
equation (subsection \S \ref{sec:interaction})
\item[$ii)$] Solution of the disturbance flow equation 
(subsection \S \ref{sec:disturbance})
\item[$iii)$]Regularization (subsection \S \ref{sec:sing_reg})
\item[$iv)$] Embedding the disturbance flow into the Navier-Stokes equations  
(subsection \S \ref{sec:coupling_phases})
\item[$v)$] Evaluation of the hydrodynamic force on the particles in the
two-way coupling regime and removal of the self-induced velocity disturbance
(subsection \S \ref{sec:prtcl_motion_force}).
\end{itemize}

\subsection{Interaction between the two phases} \label{sec:interaction}
\begin{figure}
\centerline{
\includegraphics[scale=0.35,angle=-90]{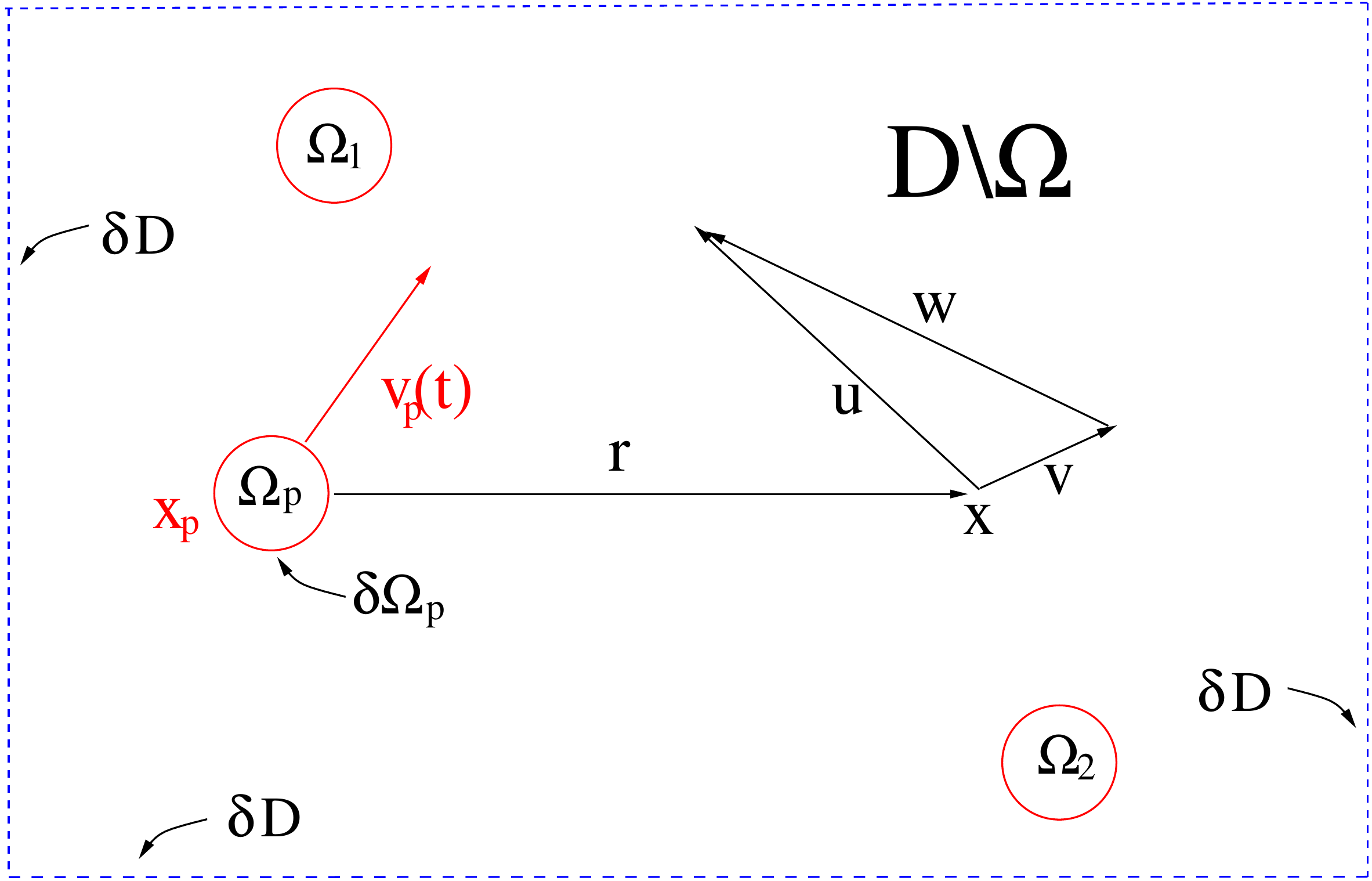}}
\caption{Sketch of the flow domain. The fluid fills the domain
${\cal D}\backslash \Omega$ with $\Omega(t) = \cup_p \Omega_p(t)$ the region 
occupied by the $N_p$ rigid particles and $\Omega_p(t)$ the time 
dependent domain of the $p$th particle. $\partial {\cal D}$ denotes
the boundary of $\cal D$. The fluid velocity at the generic point $\vx \in {\cal D}\backslash \Omega$ is decomposed as 
$\vu=\vw+\vv$, to be understood as the definition of $\vw$  given the fluid velocity $\vu$ and the solution $\vv$ of the linear, unsteady Stokes problem 
(\ref{eqn:unsteady_stokes}). 
\label{fig:domain_sketch}}
\end{figure}

In presence of a disperse phase, the carrier fluid fills  the domain 
${\cal D}\backslash \Omega$  where $\cal D$ is the flow domain and 
$\Omega(t) = \cup_p \Omega_p(t)$ denotes the region occupied by the collection of 
$N_p$ rigid particles, with $\Omega_p(t)$ the time dependent domain occupied by 
the $p$th particle, see the sketch in figure \ref{fig:domain_sketch}. 
The set theoretic notation $\cup_p$ denotes the union of 
sets indexed by $p$ and $A\backslash B$ denotes the complement in $A$ of set $B$.
The motion of the carrier fluid is assumed to be described  by the standard 
incompressible Navier-Stokes equations endowed with the no-slip condition at 
the particle boundaries
\begin{equation}
\label{eqn:ns_resolved}
\begin{array}{l}
\left.
\begin{array}{l}
\displaystyle \nabla \cdot \vu = 0 \\ \\
\displaystyle \frac{\partial \vu}{\partial t} + \vu \cdot \nabla \vu = 
-\frac {1}{\rho_f} \nabla {\rm p} + \nu \nabla^2 \vu 
\end{array} \right\} \qquad \vx \in {\cal D}\backslash \Omega(t)
\\
~\\
\displaystyle \vu\lvert_{\partial \Omega_p(t)} = \vV_p(\vx)\lvert_{\partial \Omega_p(t)} \qquad \qquad p =1,\ldots,N_p \\
\displaystyle \vu\lvert_{\partial {\cal D}} = \vu_{wall} \\
~\\
\displaystyle \vu(\vx,0)=\vu_0(\vx) \qquad \qquad \vx \in {\cal D}\backslash \Omega(0)\ .
\end{array}
\end{equation}
In equations (\ref{eqn:ns_resolved}), $\vu_0(\vx)$ is the velocity field at time 
$t=0$, $\rho_f$ denotes the fluid density, $\nu$ is the kinematic viscosity, 
$\partial \Omega_p$ is the boundary of the $p$th particle and $\partial {\cal D}$ 
is the boundary of the overall flow domain, see figure \ref{fig:domain_sketch}.
In this microscopic description, the 
particles affect the carrier fluid through the  no-slip condition at the moving 
particle surface $\partial \Omega_p(t)$ where the fluid matches the local rigid 
body velocity of the particle 
$\vV_p(\vx) = \vv_p + \vomega \times \left(\vx-\vx_p \right)$, with $\vv_p$  
the velocity of the particle geometric center $\vx_p(t)$
and $\vomega(t)$ the angular velocity. The equations of rigid body dynamics need 
be coupled to the equation for the fluid velocity field to determine the 
particle motions, where the fluid tension acting at the particle boundary provide 
the relevant forces and moments.

In principle the system (\ref{eqn:ns_resolved}) can be numerically integrated at 
the price of resolving all the particle boundaries on the computational grid. 
When the suspension is formed by a huge number of small particles their direct 
solution is unaffordable. In any case, equations (\ref{eqn:ns_resolved})  still 
provide the basic description of the flow in terms of the interaction between the 
two phases. Purpose of the present subsection is to manipulate and approximate the 
basic equations to derive a viable model for the suspension.

As a starting point, for small time intervals $0 \le t \le Dt \ll1$, the carrier flow velocity is decomposed  
into two parts, $\vu(\vx,t)=\vw+\vv$. The field $\vw(\vx,t)$  is assumed to satisfy the equations 
\begin{equation}
\label{eqn:ns_background}
\begin{array}{l}
\begin{array}{l}
\displaystyle \nabla \cdot \vw = 0 \\ \\
\displaystyle \frac{\partial \vw}{\partial t} + \vF = 
-\frac {1}{\rho_f} \nabla \pi + \nu \nabla^2 \vw
\end{array}
\\
\\
\displaystyle \vw\lvert_{\partial {\cal D}} = \vu_{wall} - 
\vv_{\partial {\cal D}} \\
\displaystyle \vw(\vx,0)={\bar \vu}_0(\vx)\, ,
\end{array}
\end{equation}
where $\vx \in {\cal D}$ and
\begin{equation}
\label{eqn:ns_F}
\vF = \left\{
\begin{array}{ll}
\vu \cdot \nabla  \vu & \qquad \mbox{for} \,\, \vx \in   {\cal D}\backslash\Omega(t)\\ \\
\vV_p \cdot \nabla \vV_p & \qquad \mbox{for}  \,\, \vx \in  \Omega(t)
\end{array}
\right.
\end{equation}
is a field reproducing the complete convective term of the Navier-Stokes equation 
in the carrier fluid domain ${\cal D}\backslash \Omega$ which is prolonged inside 
$\Omega$ using the solid particle velocity field. Other choices are possible, but 
the actual shape of the field inside the particle domains is irrelevant to our 
present purposes: under this respect, the solid body motion provides an elegant 
example given the continuity of the field $\vF$ at the particle boundaries. 
In problem (\ref{eqn:ns_background}), a part from the prolongation of the 
field $\vF$, the particles disappeared altogether from the domain and the 
convective term, retaining its complete nonlinear nature in the fluid domain, 
is treated as a prescribed forcing term. The initial field $\bar \vu_0$ is 
prolonged inside the particle domains by the same rule, i.e. as the solid body 
motion of relevant particle.

The field $\vv(\vx,t)$ exactly satisfies the \emph{linear} unsteady Stokes 
problem (the complete non-linear term has been retained in the equation for $\vw$) 
\begin{equation}
\label{eqn:unsteady_stokes}
\begin{array}{l}
\left.
\begin{array}{l}
\displaystyle \nabla \cdot \vv = 0 \\ \\
\displaystyle \frac{\partial \vv}{\partial t}
=-\frac{1}{\rho_f} \nabla {\rm q} +\nu \nabla^{2}{\vv}  
\end{array}
\right\} \qquad \vx \in {\cal D}\backslash \Omega(t)
\\
~\\
\displaystyle \vv\lvert_{\partial \Omega_p(t)}=\vV_p(\vx)\
\lvert_{\partial \Omega_p(t)} - 
\vw\lvert_{\partial \Omega_p(t)} \qquad p = 1, \ldots N_p \\
~\\
\displaystyle \vv(\vx,0)=0 \qquad \vx \in  {\cal D}\backslash \Omega(0)\, ,
\end{array}
\end{equation}
where boundary conditions are applied at the particle surfaces. It should be 
observed that no boundary condition are applied to the field $\vv$ at the 
flow domain boundary $\partial {\cal D}$. In other words, the field $\vv$ can 
be regarded as a free space solution in the whole $\Rset^3$ restricted the actual 
flow domain $\cal D$. Indeed the value of $\vv$ at the domain boundary is used 
to correct the boundary condition for $\vw$. It is worth calling the reader's 
attention to the initial conditions for the two complementary problems: the 
initial velocity field is assigned as initial condition for $\vw$, leaving 
homogenous initial data for $\vv$. As shown in a later section, the homogeneous 
initial conditions for the perturbation field $\vv$ will turn out to be a crucial 
feature of the decomposition.

The solution of equations (\ref{eqn:unsteady_stokes}) can be expressed 
in terms of the boundary integral representation of the unsteady Stokes equations 
that involves the unsteady Stokeslet $G_{ij}(\vx,\vxi,t,\tau)$, a second order 
Cartesian tensor, and the associated stresses in the form of the third order 
tensor  
${\cal T}_{ijk} (\vx,\vxi,t,\tau)$, see appendix \ref{app:unsteady_stokeslet}
and classical textbooks, \cite{zapryanov,kim_book}. 
The unsteady Stokeslet $G_{ij}(\vx,\vxi,t,\tau)$ is readily interpreted as
the fluid velocity ($i$th direction) at position  $\vx$ and time $t$ due to the
singular forcing $\delta(\vx-\vxi)\delta(t-\tau)$  ($j$th direction) applied at  
$\vxi$ at time $\tau$. Exploiting the vanishing initial condition,
the solution of  equations (\ref{eqn:unsteady_stokes}) is recast in the boundary  
integral representation
\begin{equation}
\label{eqn:unsteady_stokes_solution}
v_i(\vx,t)=\int_0^t d\tau \int_{\partial \Omega} t_j(\vxi,\tau) 
G_{ij}(\vx,\vxi,t,\tau) 
-v_j(\vxi,\tau) {\cal T}_{ijk} (\vx,\vxi,t,\tau) n_k(\vxi) \, dS_{\vxi}.
\end{equation}
Equation (\ref{eqn:unsteady_stokes_solution}) expresses $\vv(\vx,t)$ in terms of 
a boundary integral on $\partial \Omega = \cup_p \partial \Omega_p$ involving 
the (physical) tension $t_j(\vxi,\tau)$ and the boundary condition on the 
perturbation velocity $v_j(\vxi,\tau)$ at each particle boundary. 
In principle, the tension $t_j(\vxi,\tau)$ 
can be determined by solving the boundary integral equation (indeed a system of 
coupled boundary integral equations, one for each particle) associated with 
representation (\ref{eqn:unsteady_stokes_solution}). Once the tension is known at 
each  particle boundary, representation (\ref{eqn:unsteady_stokes_solution}) 
provides the perturbation field everywhere in the flow domain.  
Moreover the boundary integral of the tension $t_j$ would provide the forces 
acting on the particles.

Since the present aim is capturing the effects  of many small particles of 
diameter $d_p$, the interest is focused on the far field particle 
disturbance that can be approximated by a multipole expansion of equation 
(\ref{eqn:unsteady_stokes_solution}). Substituting in equation 
(\ref{eqn:unsteady_stokes_solution}) the first order truncation of the Taylor 
series of $G_{ij}(\vx,\vxi,t,\tau)$ and ${\cal T}_{ijk} (\vx,\vxi,t,\tau)$,
centered at the particle position $\vx_p$, leads to the far field expression for 
large $r_p/d_p$, where $r_p = |\vx -\vx_p |$,
\begin{equation}
\label{eqn:unsteady_stokes_far_field}
v_i(\vx,t)=-\sum_p \int_0^t D^p_j(\tau) G_{ij}(\vx,\vx_p,t,\tau) \, d\tau \, ,
\end{equation}
showing that the far field disturbance depends only on the hydrodynamic force 
$\vD_p(\tau)$, with Cartesian components $D^p_j$, which acts on the generic 
particle. Given the physical interpretation of the unsteady Stokeslet $G_{ij}$, 
the partial differential equation whose solution is given by 
(\ref{eqn:unsteady_stokes_far_field}) follows as
\begin{equation}
\label{eqn:unsteady_stokes_singular}
\frac {\partial \vv}{\partial t} - \nu \nabla^2 \vv + 
\frac{1}{\rho_f} \nabla {\rm q} =  - \frac{1}{\rho_f} \sum_p \vD_p(t)
\, \delta\left[ \vx - \vx_p(t) \right] \, ; \qquad \vv(\vx,0)=0 \, ,
\end{equation}
as is directly verified by combining the time derivative of equation 
(\ref{eqn:unsteady_stokes_far_field}) with its Laplacian. 
In equation (\ref{eqn:unsteady_stokes_singular}) the boundary condition at the
particle surfaces disappear altogether and the fluid-particle coupling  
occurs via the (singular) forcing term in the unsteady Stokes problem.
Given the linearity, hereafter we shall explicitly consider the single 
contribution of the generic particle $p$, keeping in mind that a final summation 
all over the particles is required.

It is also clear that as the particle diameter gets smaller and smaller, the 
term $\vF$ in equations (\ref{eqn:ns_background}) uniformly fills the
entire domain $\cal D$ and reduces everywhere to the standard convective term 
of the Navier-Stokes equation $\vu \cdot \nabla \vu$, where $\vu = \vw + \vv$. 

\subsection{Disturbance flow due to a small particle} \label{sec:disturbance}
The vorticity equation associated with (\ref{eqn:unsteady_stokes_singular})
is
\begin{equation}
\label{eqn:vort_stokes}
\frac {\partial \vzeta}{\partial t} - \nu \nabla^2 \vzeta
=  \frac{1}{\rho_f} \vD_p(t) \times \nabla 
\delta\left[\vx-\vx_p(t) \right]; \, \quad \vzeta(\vx,0) = 0 \, ,
\end{equation}
where $\vzeta = \nabla \times \vv$. The solution can be expressed as 
a convolution with the fundamental solution of the diffusion equation
$g(\vx-\vxi,t-\tau)$, given by (see appendix \ref{app:diffusion_eq})
\begin{equation}
\label{eqn:fundamental_fourier}
g(\vx-\vxi,t-\tau) = \frac{1}{\left[ 4 \pi \, \nu (t-\tau)\right]^{3/2}}
\exp\left[-\frac{\lVert \vx -\vxi \rVert ^2}{4 \nu (t-\tau)} \right] \, ,
\end{equation}
that is a Gaussian with time dependent variance 
$\sigma(t-\tau) = \sqrt{2 \nu (t-\tau)}$. Observe that $g$ is the fundamental 
solution of the diffusion equation in free-space, since $\vv$  is itself a free-space 
field, as noted when discussing eq.~(\ref{eqn:unsteady_stokes}).

By rearranging the forcing on the right hand side of equation 
(\ref{eqn:vort_stokes}) as a time-convolution,
\begin{equation}
\label{eqn:forcing_conv}
\vD_p(t) \times \nabla \delta\left[\vx-\vx_p(t) \right] = 
\int_0^{t^+} 
\vD_p(\tau) \times \nabla \delta\left[\vx-\vx_p(\tau) \right] 
\delta(t-\tau) d\tau \, 
\footnote{with $\displaystyle \int_0^{t^+} f(\tau) \, d \tau$ we intend 
$\displaystyle \lim_{\epsilon \to 0}\int_0^{t+\epsilon} f(\tau) \, d \tau$},
\end{equation}
the solution of equation (\ref{eqn:vort_stokes}) follows at once as
\begin{equation}
\label{eqn:sol_vort_forced}
\vzeta(\vx,t) = \frac{1}{\rho_f} \int_0^{t^+} 
\vD_p(\tau) \times \nabla g\left[ \vx-\vx_p(\tau),t-\tau\right] d\tau \ .
\end{equation}
The original  fluid velocity $\vv(\vx,t)$ can be reconstructed from the vorticity 
using the non-canonical decomposition
\begin{eqnarray}
\label{eqn:disturbance_decomp}
\vv(\vx,t) = \vv_{\vzeta}(\vx,t) + \nabla \phi(\vx,t) \, ,
\end{eqnarray}
where
\begin{eqnarray}
\label{eqn:pseudo_velo}
\vv_{\vzeta}(\vx,t) = - \frac{1}{\rho_f} \int_0^{t^+} \vD_p(\tau) 
g\left[ \vx-\vx_p(\tau),t-\tau\right] d\tau 
\end{eqnarray}
is a pseudo-velocity, such that its curl equals the vorticity,
$\nabla \times \vv_{\vzeta} = \vzeta$, and the gradient term is 
added to make the field solenoidal, as appropriate for incompressible flows,
\begin{eqnarray}
\label{eqn:velo_proj}
\nabla^2 \phi(\vx,t) = - \frac{1}{\rho_f}
\int_0^{t^+} \vD_p(\tau) \cdot \nabla g\left[ \vx-\vx_p(\tau),t-\tau\right] 
d\tau \ .
\end{eqnarray}
The pseudo-velocity $\vv_{\vzeta}$ obeys the equation
\begin{eqnarray}
\label{eqn:v_zeta_forced}
\frac{\partial \vv_{\vzeta}}{\partial t} - \nu \nabla^2 \vv_{\vzeta} = 
- \frac{1}{\rho_f} \vD_p(t) \delta\left[\vx - \vx_p(t) \right]\, , \qquad 
\vv_{\vzeta}(\vx,0) = 0 \ .
\end{eqnarray}

\subsection{Regularization of the disturbance field due to a small particle} 
\label{sec:sing_reg}
Both the velocity $\vv$ and the vorticity $\vzeta$ are apparently singular, 
with singularity arising from  the contribution to the integral  near the upper 
integration limit, $\tau \simeq t$, where  $g(\vx-\vxi,t-\tau)$ tends to 
behave as ``badly'' as the Dirac delta function. On the contrary away from 
the upper integration limit the integrand is nicely behaved since it involves a  
Gaussian or its gradient.

In this paragraph we define a regularization procedure based on a temporal cut-off 
$\epsilon_R$ such that the fields are additively split into a regular and a 
singular component. For instance the decomposition of the vorticity reads
\begin{eqnarray}
\label{eqn:vort_forced_decomp}
\vzeta(\vx,t) = \vzeta_R(\vx,t;\epsilon_R) + \vzeta_S(\vx,t;\epsilon_R) \, ,
\end{eqnarray}
with smooth and singular part respectively given by
\begin{eqnarray}
\label{eqn:vort_regular}
\vzeta_R(\vx,t) = \frac{1}{\rho_f} \int_0^{t - \epsilon_R}
\vD_p(\tau) \times \nabla g\left[ \vx-\vx_p(\tau),t-\tau\right] d\tau \, ,
\end{eqnarray}
and by
\begin{eqnarray}
\label{eqn:vort_singular}
\vzeta_S(\vx,t) = \frac{1}{\rho_f} \int_{t - \epsilon_R}^{t^+}
\vD_p(\tau) \times \nabla g\left[ \vx-\vx_p(\tau),t-\tau\right] d\tau \ .
\end{eqnarray}
%
As implied by the fundamental solution of the diffusion equation,
the regular part of the vorticity field is everywhere smooth and 
characterized by the smallest spatial scale 
$\sigma_R=\sigma(\epsilon_R) = \sqrt{2 \nu \epsilon_R}$. 
Thanks to the semigroup property of solutions of the diffusion equation,
the regular field $\vzeta_R(\vx,t)$ can be interpreted as the free 
diffusion from time $t-\epsilon_R$ to time $t$ of the complete
field at time $t-\epsilon_R$, $\vzeta(\vx,t-\epsilon_R)$, namely
\begin{eqnarray}
\label{eqn:vort_regular_diffused}
\vzeta_R(\vx,t) = \int \vzeta(\vxi,t-\epsilon_R)
g\left( \vx-\vxi,\epsilon_R\right) d\vxi \, ,
\end{eqnarray}
where the spatial convolution integral propagates the field from $t-\epsilon_R$ 
to $t$. Although physically obvious, equation~(\ref{eqn:vort_regular_diffused}) 
can be directly proved using the result 
\begin{eqnarray}
\label{eqn:group_green}
g(\vx,t) = \int g(\vxi,t-\epsilon_R) g\left( \vx-\vxi,\epsilon_R\right)
d\vxi \, ,
\end{eqnarray}
that is nothing more that a re-expression of the semigroup property for the 
free-space diffusion equation applied to the fundamental solution $g$. 
Actually, using the property (\ref{eqn:group_green}) and introducing 
eq.~(\ref{eqn:sol_vort_forced}) at time $t-\epsilon_R$ 
into eq.~(\ref{eqn:vort_regular_diffused}), after integration by parts, one 
readily gets
\begin{eqnarray}
\label{eqn:proof_vort_reg}
\vzeta_R(\vx,t) =  \frac{1}{\rho_f}
\int \left\{
\int_0^{t-\epsilon_R} 
\vD_p(\tau) \times \nabla_{\vxi} g\left[ \vxi-\vx_p(\tau),t-\epsilon_R-\tau\right] 
d\tau \right\} g\left( \vx-\vxi,\epsilon_R\right) d\vxi  = \nonumber \\
\frac{1}{\rho_f} \int_0^{t-\epsilon_R} 
\vD_p(\tau) \times \int \nabla_{\vxi} 
g\left[ \vxi-\vx_p(\tau),t-\epsilon_R-\tau\right]  
g\left( \vx-\vxi,\epsilon_R\right) d\vxi \,
d\tau = \nonumber \\
\frac{1}{\rho_f} \int_0^{t-\epsilon_R} 
\vD_p(\tau) \times \nabla \int 
g\left[ \vxi-\vx_p(\tau),t-\epsilon_R-\tau\right]  
g\left( \vx-\vxi,\epsilon_R\right) d\vxi \,
d\tau = \nonumber \\
\frac{1}{\rho_f} \int_0^{t-\epsilon_R} 
\vD_p(\tau) \times \nabla g\left[ \vx-\vx_p(\tau),t-\tau\right] \, 
d\tau \nonumber  \, ,
\end{eqnarray}
which is indeed equation (\ref{eqn:vort_regular}). 
The corresponding vorticity field $\vzeta_R$ at time $t$ obeys a forced diffusion 
equation where the forcing is applied at the slightly earlier time $t-\epsilon_R$,
\begin{eqnarray}
\label{eqn:pde_vort_regular}
\frac{\partial \vzeta_R}{\partial t} - \nu  \nabla^2 \vzeta_R = - \frac{1}{\rho_f}
\nabla \times \vD_p(t-\epsilon_R) g\left[\vx -\vx_p(t-\epsilon_R),\epsilon_R 
\right] \, ; \quad
\vzeta_R(\vx,0) = 0 \, ,
\end{eqnarray}
see appendix \ref{app:pde_vort_regular} for the detailed calculation. 
The velocity field $\vv_R$ associated with the regularized vorticity field 
$\vzeta_R$  can be expressed though the general decomposition 
(\ref{eqn:disturbance_decomp}),
\begin{equation}
\label{eqn:v_regular_decomp}
\vv_R(\vx,t) = \vv_{{\vzeta}_R} + \nabla \Phi_R \, ,
\end{equation}
where, by analogy with eq.~(\ref{eqn:v_zeta_forced}), the regularized 
pseudo-velocity $\vv_{{\vzeta}_R}$ is
\begin{eqnarray}
\label{eqn:v_zeta_regular}
\frac{\partial \vv_{{\vzeta}_R}}{\partial t} - \nu \nabla^2 \vv_{{\vzeta}_R} = 
- \frac{1}{\rho_f} \vD_p(t-\epsilon_R) 
g\left[\vx - \vx_p(t-\epsilon_R), \epsilon_R \right]\, ; \quad 
\vv_{\vzeta_R}(\vx,0) = 0 \, ,
\end{eqnarray}
and the potential correction follows from the equation
\begin{eqnarray}
\label{eqn:poisson_regular}
\nabla^2 \Phi_R = - \nabla \cdot \vv_{{\vzeta}_R} \ . 
\end{eqnarray}
It is worth noticing that the complete regularized field obeys instead the 
forced unsteady Stokes equation
\begin{eqnarray}
\label{eqn:velo_reg_eq}
\frac{\partial \vv_R}{\partial t} - \nu \nabla^2 \vv_R 
+\frac{1}{\rho_f}\nabla {\rm q}_R = - \frac{1}{\rho_f}
\vD_p(t-\epsilon_R) \, g\left[ \vx-\vx_p(t-\epsilon_R),\epsilon_R \right]
\end{eqnarray}
for the solenoidal field $\vv_R$. The crucial point to observe here is that the 
regularized component of the velocity disturbance $\vv_R(\vx,t)$ 
evolves according to a diffusion equation forced by the anticipating Stokes drag 
(i.e. evaluated at $t-\epsilon_R$) times the regular spatial distribution 
$g\left[ \vx-\vx_p(t-\epsilon_R),\epsilon_R \right]$. Equation~(\ref{eqn:v_zeta_regular}) 
can in principle be straightforwardly solved 
on a discrete grid, once the spatial scale $\sigma_R$ of the forcing is properly 
resolved by the grid. Once $\vv_{{\vzeta}_R}$ is known, the correction needed to 
make the field solenoidal calls for the solution of the Poisson 
equation~(\ref{eqn:poisson_regular}). 

For the future application to the full solver for the carrier phase in presence 
of the suspension, it is also worth mentioning that the field $\vv_{{\vzeta}_R}$ is 
rapidly decaying in space as far as the observation time $t$ is small, since it 
implies the short-time diffusion of a rapidly decaying forcing.
All the long-range effects are indeed confined to the potential 
correction $\nabla \Phi_R$.  As will be discussed in the forthcoming sections, 
the field $\vv_R$ does not need to be separately evaluated, since it will be 
embedded in the solution procedure for the single field $\vu$ 
which accounts for both the undisturbed carrier flows and the particle 
perturbation.

\begin{figure}
\centerline{
\includegraphics[scale=0.225,angle=-90]{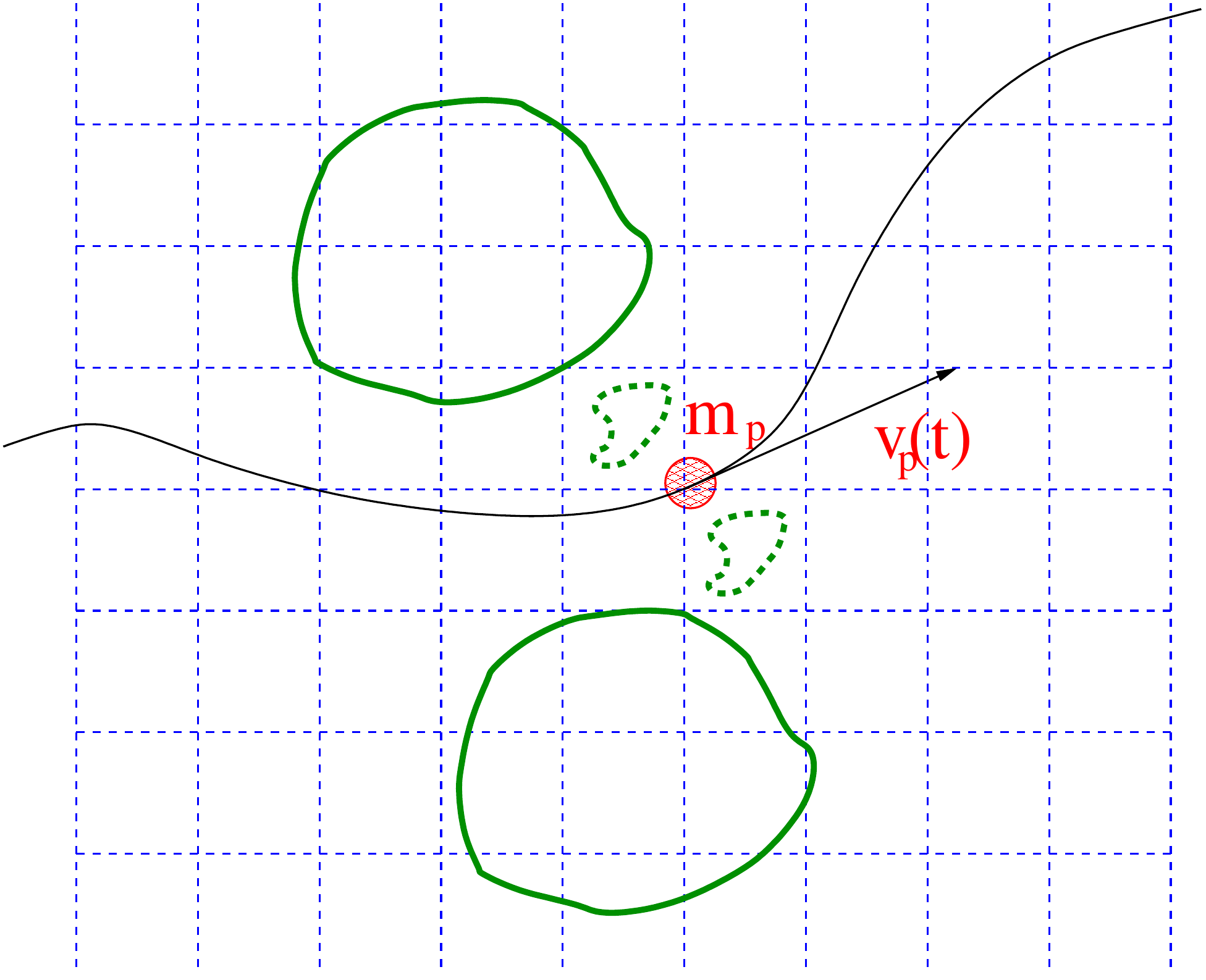}
\includegraphics[scale=0.225,angle=-90]{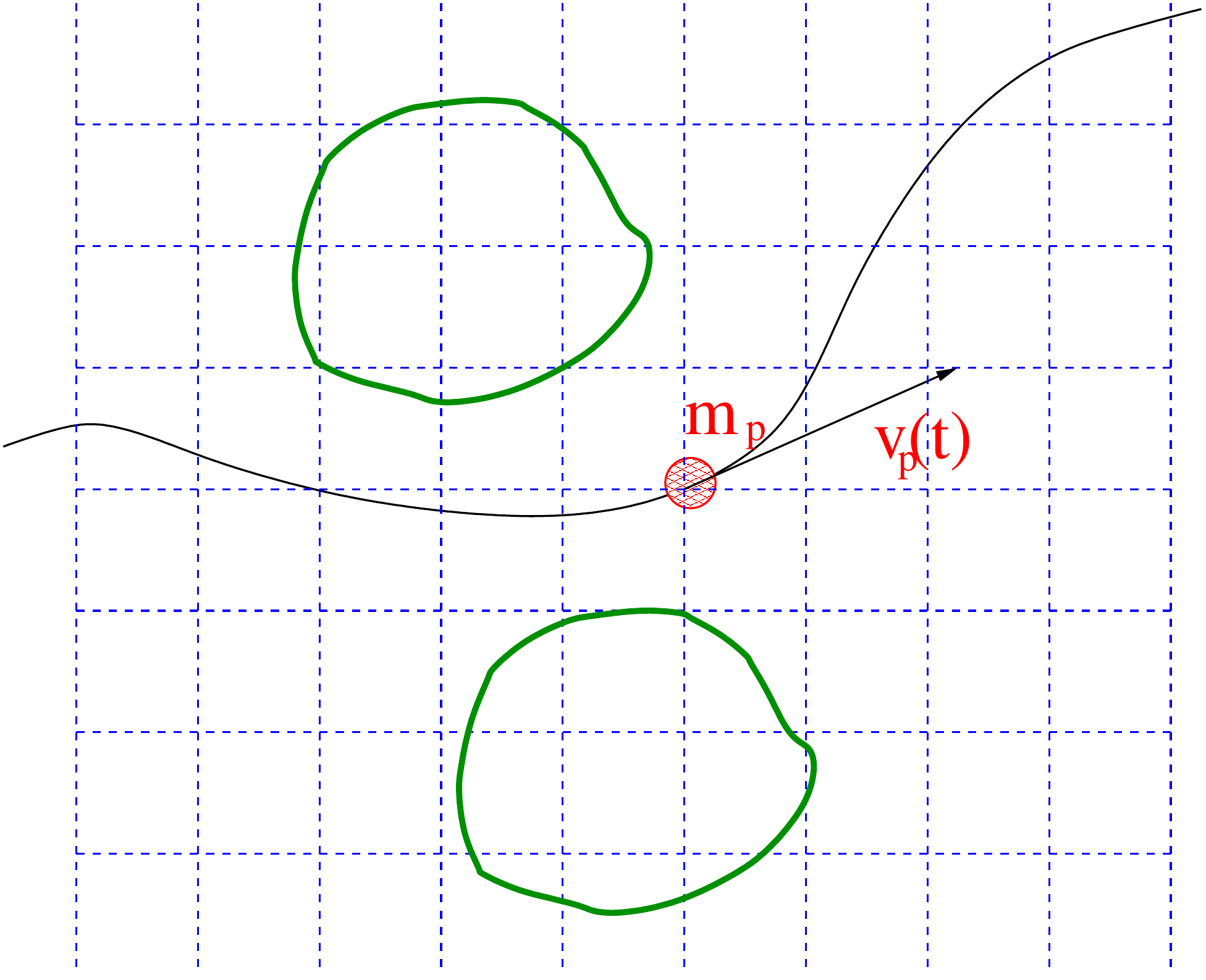}
\includegraphics[scale=0.225,angle=-90]{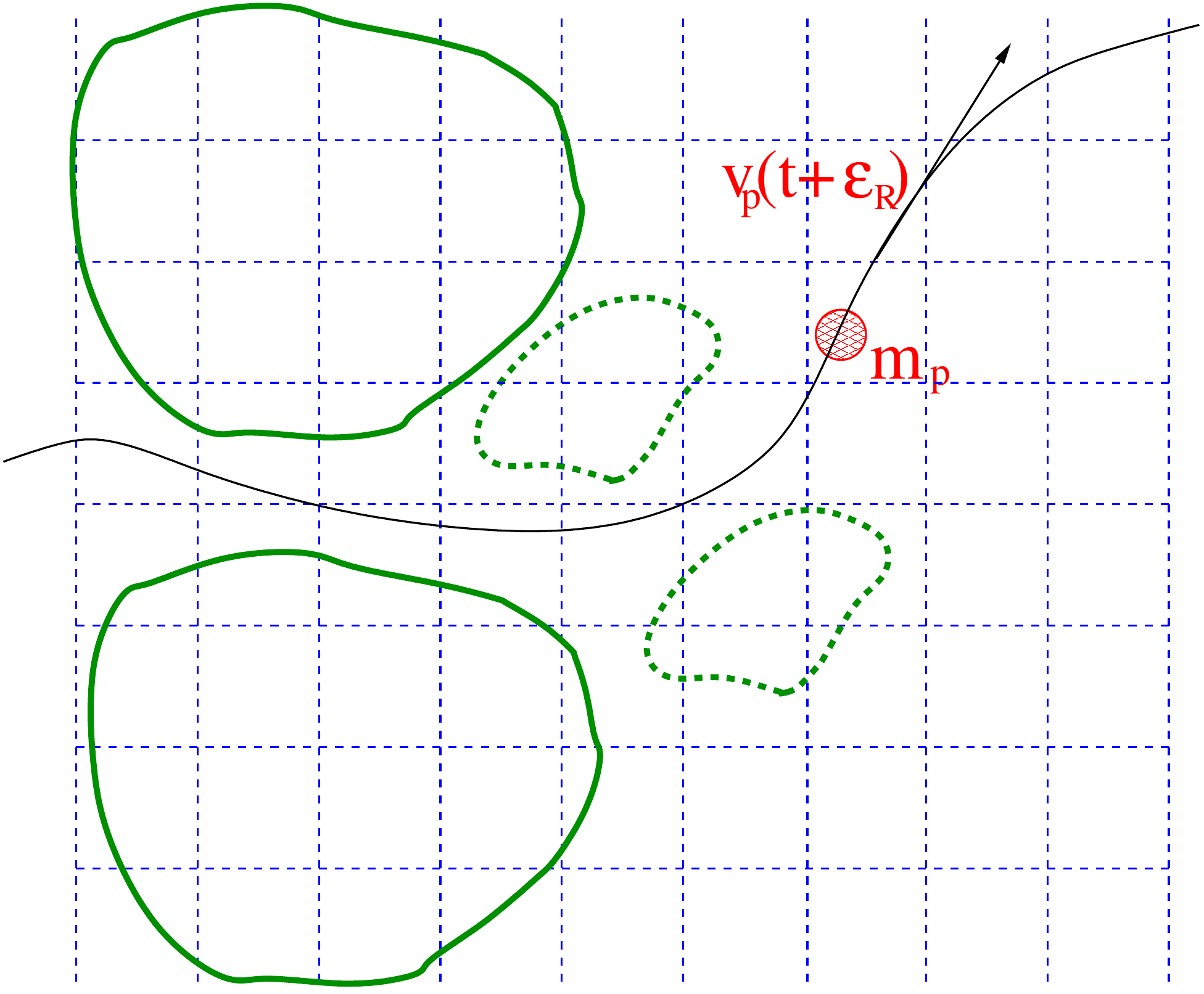}}
\caption{Coupling mechanism and regularization procedure. The green curves  
sketch the vorticity field. Left panel: the complete vorticity field generated by 
the particle at the current time $t$ is split into the regular $\vzeta_R(\vx,t)$ 
(solid green line) and singular $\vzeta_S(\vx,t)$ (dashed green line) components 
respectively. Central panel: only the regular component $\vzeta_R(\vx,t)$ can be 
represented by the computational grid with mesh size $Dx$ at the generic time $t$. 
Right panel: after the 
elapsed time $\epsilon_R$  (time $t+\epsilon_R$) the singular component of 
the vorticity field diffuses to scales large enough to be captured by the 
discrete grid. The momentum transfer towards the fluid occurs via viscous 
diffusion of the vorticity generated by the particle. When only the regularized 
field is considered a small error is incurred in  the exchanged momentum. However,
the successive diffusion of the singular field fully recovers the correct 
amount of vorticity at a successive time step. Thus the error does 
not accumulate in time and remains under control along the simulation.
\label{fig:vorticity_injection}}
\end{figure}

At variance with $\vv_R$, the singular contribution $\vv_S$ cannot be represented 
on a discrete grid. It can be decomposed as well into a vorticity related 
component plus a potential correction, according to the general 
representation~(\ref{eqn:disturbance_decomp}). The vortical component 
$\vv_{{\vzeta}_S}$ is an extremely fast decaying function of distance from the 
actual position of the particle, while its potential correction $\nabla \Phi_S$ 
is not. In order to address the error propagation of the algorithm that 
will be illustrated in the next section, it is instrumental to explicitly provide 
an estimate on the order of magnitude of the field $\nabla \Phi_S$.  
The singular part of the pseudo-velocity is given by
\begin{eqnarray}
\label{eqn:pseudo_velo_S}
\vv_{\vzeta_S}(\vx,t) = \frac{1}{\rho_f} \int_{t-\epsilon_R}^{t^+} \vD_p(\tau) 
g\left[ \vx-\vx_p(\tau),t-\tau\right] d\tau  \, ,
\end{eqnarray}
see eq.~(\ref{eqn:pseudo_velo}) for comparison. The equation for the potential 
correction is then
\begin{eqnarray}
\label{eqn:potential_S}
\nabla^2 \Phi_S  = - \nabla \cdot \vv_{\vzeta_S} =
\frac{1}{\rho_f} \int_{t-\epsilon_R}^{t^+} \vD_p(\tau)  \cdot 
\nabla g\left[ \vx-\vx_p(\tau),t-\tau\right] d\tau  \ .
\end{eqnarray}
It follows 
\begin{eqnarray}
\label{eqn:potential_S_solution}
\Phi_S  = - \frac{1}{\rho_f} \int_{t-\epsilon_R}^{t^+} d\tau  \vD_p(\tau)  \cdot 
\nabla \int_{\Rset^3}  
\frac{g\left[ \vy-\vx_p(\tau),t-\tau\right]}{4 \pi |\vx-\vy|} d^3\vy \, ,
\end{eqnarray}
where $-1/(4 \pi |\vx - \vy|)$ is the fundamental solution of the Laplace equation.
From the solution (\ref{eqn:potential_S_solution}) a rough estimate for the 
correction field $\nabla \Phi_S$ is immediately obtained as
\begin{eqnarray}
\label{eqn:estimate}
| \nabla \Phi_S | \le \frac{1}{\rho_f} \sup_{t-\epsilon_R \le \tau \le t^+} 
|\vD_p |
\Bigg|  \nabla \otimes \nabla \int_{\Rset^3}  \frac{\displaystyle 
\int_{t-\epsilon_R}^{t^+}  g\left[ \vy-\vx_p(\tau),t-\tau\right] d \tau}
{4 \pi |\vx-\vy|} d^3\vy \Bigg| \ .
\end{eqnarray}
Sufficiently away from the particle, i.e. $|\vx - \vx_p|/d_p \gg 1$, the above 
estimate is asymptotically expressed as
\begin{eqnarray}
\label{eqn:estimate_asy}
| \nabla \Phi_S | \le \frac{1}{\rho_f} \int_{\Rset^3} \int_{t-\epsilon_R}^{t^+}  
g\left[ \vy-\vx_p(\tau),t-\tau\right] d\tau d^3\vy    
\sup_{t-\epsilon_R \le \tau \le t^+} |\vD_p | \Bigg|  \nabla \otimes \nabla 
\frac{1}{4 \pi |\vx-\vx_p^*|}  \Bigg| \, ,
\nonumber \\
\end{eqnarray}
where $\vx_p^* = \vx_p(\tau^*)$, $t-\epsilon_R \le \tau^* \le t^+$, is the 
position along the portion of the particle trajectory closest to the point $\vx$.
Given the known integral
\begin{eqnarray}
\label{eqn:known_integral}
\int_{\Rset^3} \frac{1}{\left(2 \pi \sigma^2 \right)^{3/2} } 
e^{- r^2/(2 \sigma^2)} d^3 \vr =  1\, ,
\end{eqnarray}
one ends up with
\begin{eqnarray}
\label{eqn:estimate_asy2}
| \nabla \Phi_S | \le  \sup_{t-\epsilon_R \le \tau \le t^+} |\vD_p | 
\frac{\epsilon_R}{4 \pi \rho_f |\vx-\vx_p^*|^3} \, ,
\end{eqnarray}
where the norm of the double tensor 
$\nabla \otimes \nabla \left[1/(4 \pi |\vx -\vx_p^*)\right]$ is given by
\begin{eqnarray}
\label{eqn:tensor_norm}
\Bigg| \nabla \otimes \nabla \frac{1}{4 \pi |\vx - \vx_p^*|} \Bigg| =
\sup_{|\bf{ \hat e}  | = 1}  \left[ \left(\bf{ \hat e} \cdot \nabla\right) 
\nabla   
\frac{1}{4 \pi |\vx - \vx_p^* |} \right] = \frac{1}{4 \pi |\vx -\vx_p^*|^3} \ .
\end{eqnarray}
From the expression of the singular component of the pseudo-velocity 
(\ref{eqn:pseudo_velo_S}) it is clear that, far from the particle,
$\vv_{\vzeta_S}$ decays exponentially fast, hence the far field dominating 
component of $\vv_S$ is provided by the long range correction $\nabla \Phi_S$ 
that is order $\epsilon_R/r^3$. It is also clear that close to the particle the 
singular contribution is unbound. This singular near-field is however unessential 
as far as the relevant length scales of the system, either the smallest 
hydrodynamic scale $\eta$ or the inter-particle distance, are larger than 
$\sigma_R  = \sqrt{2 \nu \epsilon_R}$. For this reason, it will be neglected when 
advancing the solution of one time step in the actual algorithm illustrated in the 
following sections. However, this highly localized field will eventually diffuse 
to larger scales at later times. Hence, the singular contribution that is 
neglected 
during a single time step is successively reintroduced in the field as soon as 
it reaches the smallest physically relevant scales of the system. This procedure 
guarantees that the error does not accumulate in time, thereby maintaining the 
accuracy of the calculation. 
Figure~\ref{fig:vorticity_injection} sketches the decomposition into regular 
and singular fields, using the vorticity field to describe the process which 
is easier to visualize than the velocity field. The sketch highlight the 
singular production of vorticity by the particle, its diffusion, associated 
to momentum transfer the fluid, and the regularizing effects of viscosity.  
A crucial point is that the singular component of the field, which cannot be 
represented on a discrete mesh, is fully recovered at a successive time instant 
when its characteristic length-scale reaches the grid size. 
In the following section the convective effect of the 
singular field will be dealt with in more detail, to show that it is indeed 
negligible when the dynamics is observed at the relevant hydrodynamical scale.

\subsection{Coupling with the carrier flow} \label{sec:coupling_phases}

The regularized fluid velocity of the carrier flow in presence of the perturbing 
particles is obtained by aggregating  the two contributions of the velocity 
decomposition $\vu_R=\vw+\vv_R$ described in subsection \S \ref{sec:interaction}. 
The resulting field obeys the equations
\begin{equation}
\label{eqn:ns_regularized}
\begin{array}{l}
\displaystyle \nabla \cdot \vu_R = 0 \\ \\
\displaystyle \frac{\partial \vu_R}{\partial t} + \vu_R \cdot \nabla \vu_R +
\left\{ \vv_S \cdot \nabla \vu_R + \vu_R \cdot \nabla \vv_S + \vv_S \cdot \nabla 
\vv_S \right\} =  \\
 \displaystyle
-\frac {1}{\rho_f} \nabla p + \nu \nabla^2 \vu_R 
- \frac{1}{\rho_f} \sum_p^{N_p}
\vD_p(t-\epsilon_R) \, g\left[ \vx-\vx_p(t-\epsilon_R),\epsilon_R \right] \\
\end{array}
\end{equation}
with boundary and initial conditions given by
\begin{equation}
\label{eqn:ns_regularized_bc}
\displaystyle \vu_R\lvert_{\partial {\cal D}} = \vu_{wall}  - 
\vv_S\vert_{\partial {\cal D}}  \, , \qquad 
\displaystyle \vu_R(\vx,0)=\vu_0(\vx)\, ,
\end{equation}
where we have added the contributions arising from all the $N_p$ particles
transported by the fluid. It should be stressed that the boundary condition for the 
regularized velocity $\vu_R$ at  $\partial {\cal D}$ needs taking the singular 
contribution $\vv_S$ into account.

An interpretation of equation (\ref{eqn:ns_regularized}) could now be helpful. 
Along its motion the particle experiences the hydrodynamic force. In the 
formulation here proposed, the force is naturally regularized by viscous 
diffusion, 
hence the mollified Dirac delta functions takes the form of the fundamental 
solution of the diffusion equation. The effect of the hydrodynamic force is the 
generation of the regularized vorticity, (\ref{eqn:vort_regular}), that is 
characterized by the smallest length-scale $\sigma_R=\sqrt{2 \nu \epsilon_R}$
where $\epsilon_R$ is the regularization diffusion time scale. 
A crucial point to be stressed again is that the hydrodynamic forcing acting on 
the regularized solution at time $t$ is the one experienced by the particles at  
a slightly previous time $t-\epsilon_R$ when their position were 
$\vx_p(t-\epsilon_R)$. The net effect of the disperse phase on the regularized 
carrier flow field is then accounted for by  the extra forcing term corresponding 
to the time-delayed hydrodynamic force expressed as the Gaussian 
$g\left[ \vx-\vx_p(t-\epsilon_R),\epsilon_R \right]$ with variance $\sigma_R$.

The total field $\vu$ will involve a singular part that is concentrated on scales 
smaller than the physically relevant ones. As such the singular contribution is 
actually neglected since only the regularized field needs to be considered.
However contributions from the singular disturbance field $\vv_S$ appear
in the term in curly brackets in eq.~(\ref{eqn:ns_regularized}).
Indeed in the far field of the particles $\vv_S$ was already shown to be of the 
order of $\epsilon_R/r^3$, which is negligible in comparison with the other terms 
in the equation. It is worth reminding that, at the successive time step, the 
corresponding contribution is re-introduced in the field, giving rise to no 
error accumulation in the long run. The crucial point here is that 
eq.~(\ref{eqn:ns_regularized}) is taken to hold everywhere in ${\cal D}$, also 
close and inside the domains which are actually occupied by the particles. 
In this near field the curly bracket term needs to be treated with some care. 
In fact, due to the scale separation between $\vu_R$ and $\vv_S$, the 
filtering of the fields on a scale $\Delta$ which is order of the smallest 
hydrodynamic scale does not alter $\vu_R$, i.e. denoting by ${\hat \vu}_R$ the 
filtered field one has $\vu_R = {\hat \vu}_R$. In such conditions the equations 
for the regularized field follows by applying the filter to 
system~(\ref{eqn:ns_regularized}). As a result 
of scale separation, the filter is actually acting only on the terms in curly 
brackets which involve the singular contribution $\vv_S$. As explicitly shown in 
appendix \ref{app:singular_velocity}, a few detailed calculations show that 
the filtered terms give contribution of the order
\begin{eqnarray}
\label{eqn:h_estimate_text}
\widehat{\vv_S \cdot \nabla \vv_S} \sim 
Re_p \left(\frac{\sigma_R}{\Delta}\right)^3 \frac{\vD_p}{\rho_f}g_{max} 
\nonumber \\
\\
\widehat{\vv_S \cdot \nabla \vu_R} \sim  \widehat{\vu_R \cdot \nabla \vv_S} \sim
Re_p \left(\frac{\sigma_R}{\Delta}\right)^2 \frac{\vD_p}{\rho_f}g_{max} \, ,
\nonumber
\end{eqnarray}
where $g_{max} = 1/(2 \pi \sigma_R^2)^{3/2}$ is the maximum of the mollified delta 
function. Clearly the above filtered convective terms are an order $Re_p$ smaller 
than the forcing term on the right hand side of eq.~(\ref{eqn:ns_regularized}). 
Under the assumption of small particle Reynolds number they can be safely 
neglected in the evolution equation of the regularized field.

We like to stress  the simplicity of the final equations that have to 
be solved,
\begin{equation}
\label{eqn:ns_regularized_filtered}
\begin{array}{l}
\displaystyle \nabla \cdot \vu_R = 0 \\ \\
\displaystyle \frac{\partial \vu_R}{\partial t} + \vu_R \cdot \nabla \vu_R
 =  
 \displaystyle
-\frac {1}{\rho_f} \nabla p + \nu \nabla^2 \vu_R 
- \frac{1}{\rho_f} \sum_p^{N_p}
\vD_p(t-\epsilon_R) \, g\left[ \vx-\vx_p(t-\epsilon_R),\epsilon_R \right]  \ .
\end{array}
\end{equation}
The effects of the disperse phase on the carrier fluid is taken 
into account by an extra term in the Navier-Stokes 
equations.  Under this point of view, any standard Navier-Stokes solver 
can be easily equipped with such extra term which is known in closed form. 
Furthermore each particle will produce an active 
forcing on the fluid localized in a sphere of radius order $\sigma_R$ 
centered at the particle position. In presence of many particles only the 
few grid points in the sphere of influence of the particle will receive the 
disturbance produced by the particle itself. Finally, the forcing term is 
grid independent in the sense that, once the grid spacing is refined 
($Dx$ progressively getting smaller at fixed $\sigma_R$), any successively finer 
grid will only provide a better numerical approximation of the same forcing. 

\subsection{Evaluation of the hydrodynamic force \& removal of self-interaction}
\label{sec:prtcl_motion_force}
The dynamics of a point particle of mass $m_p$ in the relative motion with 
respect to a Newtonian fluid is described by the equation of motion, 
\begin{equation}
\label{eqn:particle_motion}
\frac{d \vx_p}{dt}= \vv_p(t), \quad m_p \frac{d \vv_p}{dt} = \vD_p(t) 
+\left( m_p -m_f\right)\vg,
\end{equation}
where $m_f$ is the  displaced mass of fluid, $\vD_p(t)$ is the hydrodynamic force, 
and $\vg$ the acceleration due to gravity. 
Clearly, for the accurate evaluation of the particle trajectories and of the 
inter-phase momentum coupling, an accurate and efficient expression for the hydrodynamic 
force is mandatory. To obtain such expression one should reconsider the equation 
for the perturbation field $\vv$ addressed in \S~\ref{sec:interaction}.

As shown there, the perturbation due to the presence of a particle obeys the 
unsteady Stokes equation~\eqref{eqn:unsteady_stokes} where, noteworthy, the 
initial condition for the perturbation field $\vv$ is homogeneous. Indeed, in our 
scheme, the solution of the unsteady Stokes equation for $\vv$ at the generic time 
step provides the stress at the fluid-particle interface and ultimately yields the 
drag force. Luckily there is no need to work out the details, since \cite{maxril} 
already provided the expression for the general unsteady drag force of a spherical 
particle when the field has homogenous initial condition, which is the case of 
interest here. Their solution can in fact be fully exploited to provide
the drag force for the perturbation flow that is asymptotically expressed 
as \eqref{eqn:unsteady_stokes_far_field} during the generic time step.

Following \cite{maxril} the force $\vD_p(t)$ can be  evaluated as
\begin{eqnarray}
\label{eqn:force_particle}
\vD_p(t) &=&   6\pi\mu a_p \left[\tilde \vu(\vx_p,t) 
         +\frac{a_p^2}{6}\nabla^2 \tilde \vu(\vx_p,t)-\vv_p(t) \right] \nonumber \\
         &+&m_f \frac{D \tilde \vu}{Dt}\bigg|_{x_p} +\frac{1}{2} m_f\frac{d}{dt} 
         \left[\tilde \vu(\vx_p,t) + \frac{a_p^2}{10}\nabla^2 \tilde \vu(\vx_p,t)
         -\vv_p(t) \right] \nonumber \\
         &+&6\pi\mu a_p^2 \int_0^t d\tau \, 
         \frac{1}{\left[\pi\nu\left(t-\tau\right)\right]^{1/2}}
         \frac{d}{d\tau}\left[ \tilde \vu(\vx_p,\tau)+
         \frac{a_p^2}{6}\nabla^2 \tilde \vu(\vx_p,\tau)-\vv_p(\tau)\right]
\end{eqnarray}
where $a_p=d_p/2$ is the particle radius. Expression (\ref{eqn:force_particle}) 
involves the steady Stokes drag (first line), the added mass terms (second line), 
and the Basset history force (third line). In all terms the Faxen correction 
associated with spatial non-uniformity of the flow is included and,
following the original derivation by \cite{maxril}, the velocity 
$\tilde \vu(\vx_p,t)$ must be interpreted as the fluid velocity, at the particle 
position, in absence of the particle self-interaction, i.e. $\tilde \vu_p$ should  
account for the background -- possibly turbulent -- flow  and for the disturbance 
generated by all the other particles except the $p$th one, see also \cite{boivin}
and \cite{gatignol}. In the regime of 
our interest, where the particle back-reaction modifies the carrier flow, the 
correct calculation of $\tilde \vu_p$ is crucial and calls for an effective 
procedure to deprive the field $\vu(\vx,t)$ evaluated at the particle position 
from the particle self-interaction contribution.
Luckily, the (regularized) disturbance flow generated by each particle is known in 
closed form and can thus be easily removed from the complete field in computing  
the hydrodynamic force, at least for numerical algorithms using explicit time 
integration schemes. As an illustration, let us consider the simple case of 
heavy small particles, $\rho_p \gg \rho_f$, where the 
hydrodynamic force (\ref{eqn:force_particle}) reduces to the Stokes drag
\begin{eqnarray}
\label{eqn:force_stokes_drag}
\vD_p(t) &=&  6\pi\mu a_p \left[\tilde \vu(\vx_p,t) 
         -\vv_p(t) \right].
\end{eqnarray}
The explicit calculation of the velocity $\vv_R(\vx-\vx_0,t)$ induced at time $t$  
and position $\vx$ by a particle located at $\vx_0$ is provided
in appendix \ref{app:self_velocity}. This result can be exploited to remove the 
self-interaction term in the illustrative case of an explicit Euler time 
advancement algorithm. Indeed, in this case, to correctly evaluate the right 
hand side of equation~(\ref{eqn:force_stokes_drag}) it suffices to subtract from 
$\vu(\vx_p,t)$ the value $\vv_R[\vx_p(t)- \vx_p(t-Dt), Dt]$ induced at time $t$ 
at the current particle position $\vx_p(t)$ by the same particle when
it was placed at $\vx_p(t -Dt)$. The same kind of reasoning can be 
straightforwardly extended to other explicit time integration schemes, e.g.  
to each intermediate step of a Runke-Kutta algorithm and to the different 
contributions in the general expression of the force 
(\ref{eqn:force_particle}), e.g. bubbly flows \cite{climent2006dynamics}.

Before closing this section devoted to force evaluation, a final note is in 
order concerning the Basset force: it represents the effects on the 
force due to the particle-fluid interaction during the particle previous motion 
before the actual time $t$. In cases where the particle do not modify the 
carrier flow, see the derivation by \cite{maxril}, this interaction is modeled 
by a memory convolution integral which mimics the particle vorticity production 
and its viscous diffusion occurring from the initial time $t=0$ up to the actual 
time $t$. In our case, the carrier fluid is perturbed step by step by the
particle motion (two-way coupling regime) and the diffusion of the 
vorticity produced by the particle during the past motion before the actual time $t$ 
is captured without any modeling by equations (\ref{eqn:ns_regularized_filtered}).
Hence, the time integral must model the only vorticity production occurring 
during the last time step $Dt$, i.e. the memory integral is limited to a single 
time step of the eventual integration algorithm. Actually, the effects of the 
previous history come in through the boundary condition of equations 
(\ref{eqn:unsteady_stokes}) where the field $\vw$ must be interpret as the 
background velocity acting on the particle, i.e. as the carrier flow velocity field 
that would occur at the particle boundary during the last time step
in absence of the particle. For small particles such field reduces to the
value at particle center plus a Faxen-like correction accounting for spatial flow 
variations on the scale of the particle.
\section{Algorithm validation} \label{sec:validation}

The methodology illustrated in the previous sections needs to be validated.
We will address several test cases where analytical data can be employed for 
comparison. To better focus our attention on the interaction between the fluid 
and the disperse phase, we will consider  a periodic  box ${\cal D}$ free from solid 
boundaries which may hinder the analysis.  The numerical solution of equation 
(\ref{eqn:velo_reg_eq}) and (\ref{eqn:ns_regularized}) for the carrier fluid is based 
on a pseudo-spectral Fourier-based spatial discretization where the non linear terms are 
calculated by the standard $3/2$ dealiasing procedure. Time advancement is achieved 
by a low-storage semi-implicit Runge-Kutta method with a fourth-order Adams-Bashforth 
formulation for the convective terms and an implicit Crank-Nicholson formula for the 
diffusive terms. The details on the implementation are described elsewhere, see 
\cite{pof_shear}.

\subsection{Response to a localized force}
\label{sec:spatio_temporal_convergence}
We start by addressing a simple case where a known small amplitude constant force 
$\vF_0$ is applied at a fixed point $\vx_p$ to the fluid which is initially at rest in 
the domain ${\cal D}$. Due to the small amplitude of the forcing, the flow is assumed to 
obey the linear, unsteady Stokes equations. Equation (\ref{eqn:velo_reg_eq}) is 
suitable of an analytical solution which, in terms of vorticity, is provided by 
equation (\ref{eqn:vort_regular}). This reference solution allows to verify that the 
algorithm correctly transfers the proper impulse to the fluid, a crucial aspect in view 
of simulations in the two-way coupling regime.

\begin{figure}
\centerline{
\includegraphics[scale=0.30]{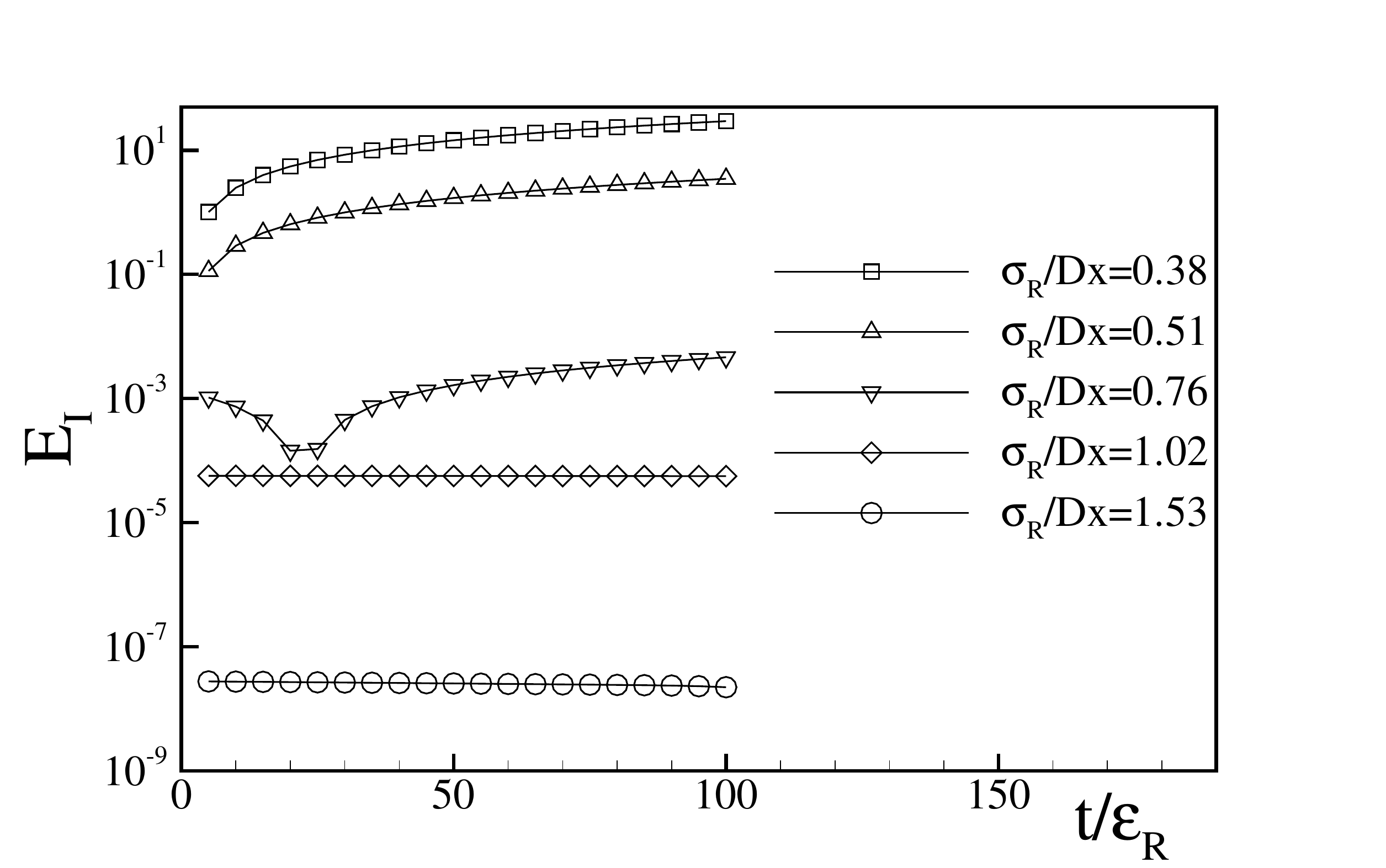}
\includegraphics[scale=0.30]{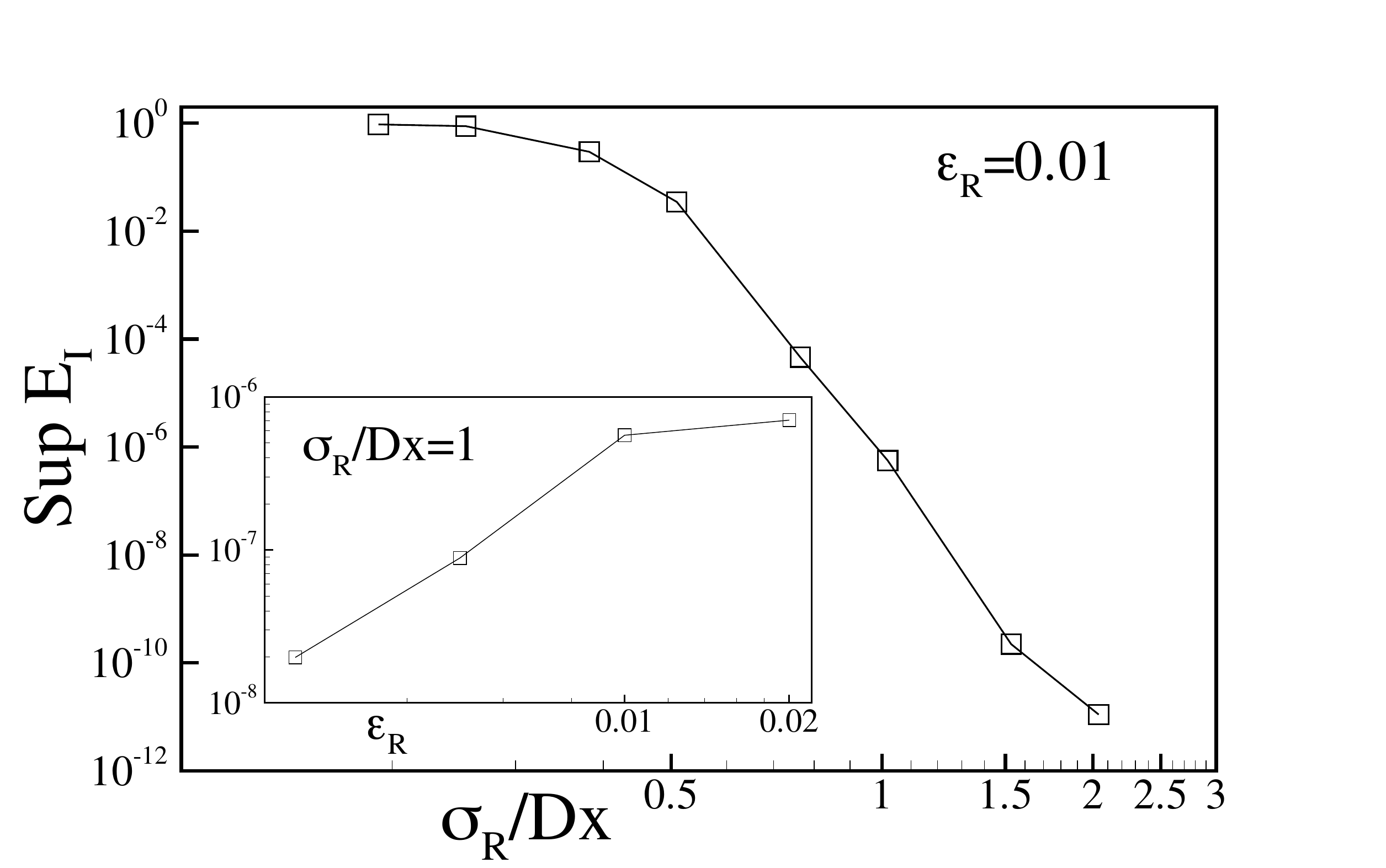}}
\caption{Convergence study in the case of a constant force $\vF_0=(1,0,0)$
applied in a fixed point $\vx_p$ to the fluid initially at rest in a periodic box
$\cal D$.
Left: the error $E_I$ of the total impulse, see text for definition,
is plotted against the normalized time $t/\epsilon_R$ for a fixed regularization 
timescale $\epsilon_R=0.01$ and different values of the spatial resolution $\sigma_R/Dx$.
Right: the supreme $\sup_{t \ge 0} E_I$ is plotted against the ratio $\sigma_R/Dx$ for 
$\sigma_R=0.01$. The inset reports the supreme $\sup_{t \ge 0} E_I$ for
different values of $\epsilon_R$ and a fixed spatial resolution 
$\sigma_R/Dx=1$.
\label{fig:conv_imp} }
\end{figure}

Due to periodicity, the exact impulse,  
$\vI_E = \vF_0 t = \int_{\cal D} \rho_f \vu(\vx,t) d^3 \vx$, can be expressed in terms 
of the vorticity moment  (\cite{saffman1992vortex}),
\begin{equation}
\label{eqn:impulse_def}
\vI_E(t)=\frac{1}{2} \rho_f \int_\Omega \vx \times \vzeta(\vx,t) \, d^3 \vx.
\end{equation} 
The error $E_I = \lvert \vI_E(t)-\vI_N(t) \rvert$, where  
$\vI_N(t)$ is the estimate of (\ref{eqn:impulse_def}) from the numerical solution,
is shown in semi-logarithmic scale in the left panel of figure \ref{fig:conv_imp} as
a function of the normalized time $t/\epsilon_R$ for a fixed value of the 
regularization timescale $\epsilon_R=0.01$ and different spatial resolutions, namely
the ratio $\sigma_R/Dx$. In the unresolved cases ($\sigma_R/Dx < 1$),
the error is order one and increases in time. In contrast, when a proper spatial 
resolution is adopted, i.e. $\sigma_R/Dx > 1$, the error becomes progressively 
smaller as the resolution is increased and stays constant in time. 
In other words, as the simulation advances in time 
$E_I$ does not accumulate. The right panel of figure \ref{fig:conv_imp} reports the 
supreme $\sup_{t \ge 0} \lvert \vI_E(t)-\vI_N(t) \rvert$ as a function of the
ratio $\sigma_R/Dx$. This plot emphasized the convergence rate of the impulse
against the spatial resolution at a fixed value of $\epsilon_R$. The inset
shows $\sup_{t \ge 0} E_I$ in a different manner. Here the spatial resolution
is fixed, $\sigma_R/Dx=1$, and the regularization timescale is progressively
reduced denoting convergence also with respect to the parameter $\epsilon_R$.

\begin{figure}
\centerline{
\includegraphics[scale=0.30]{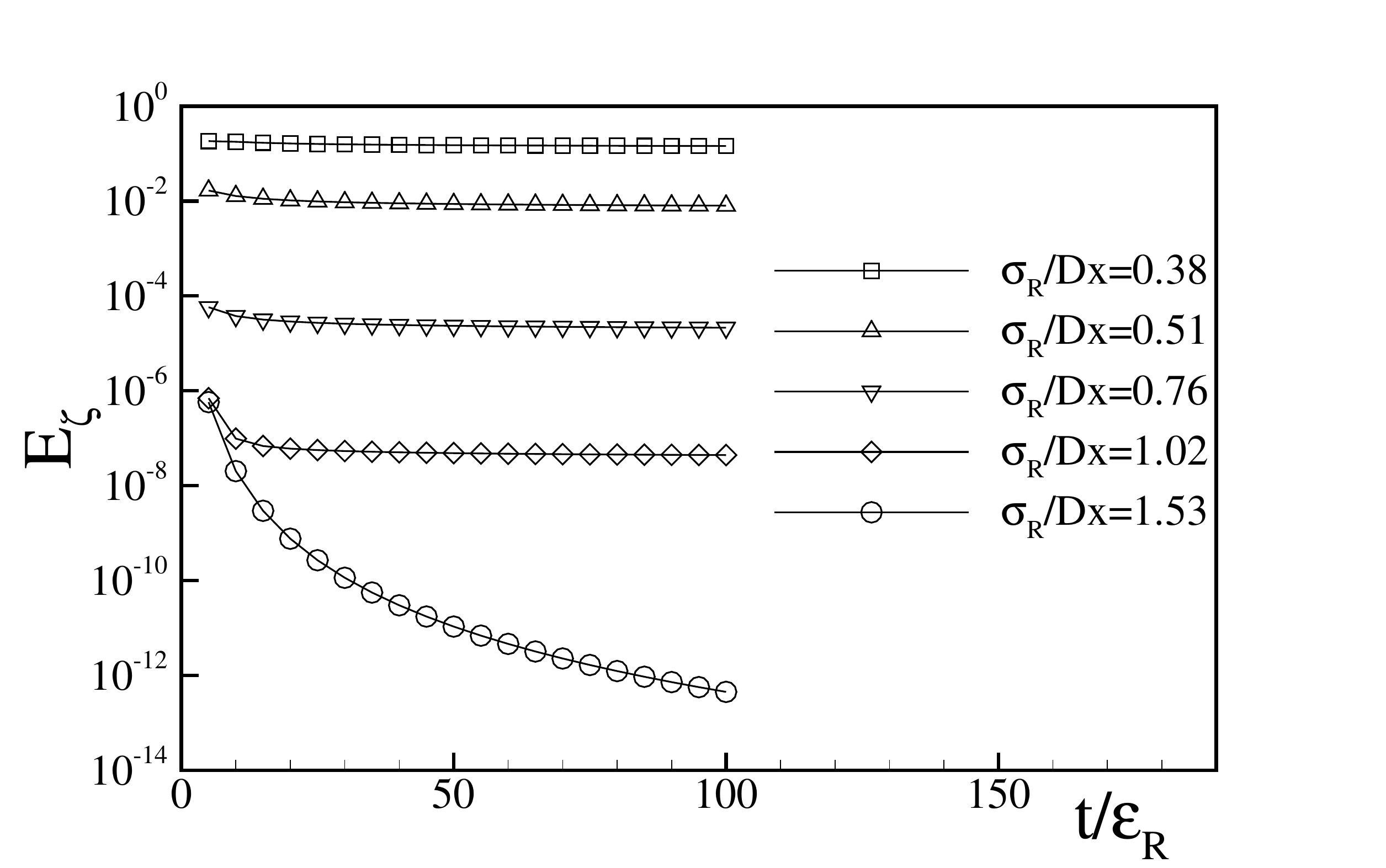}
\includegraphics[scale=0.30]{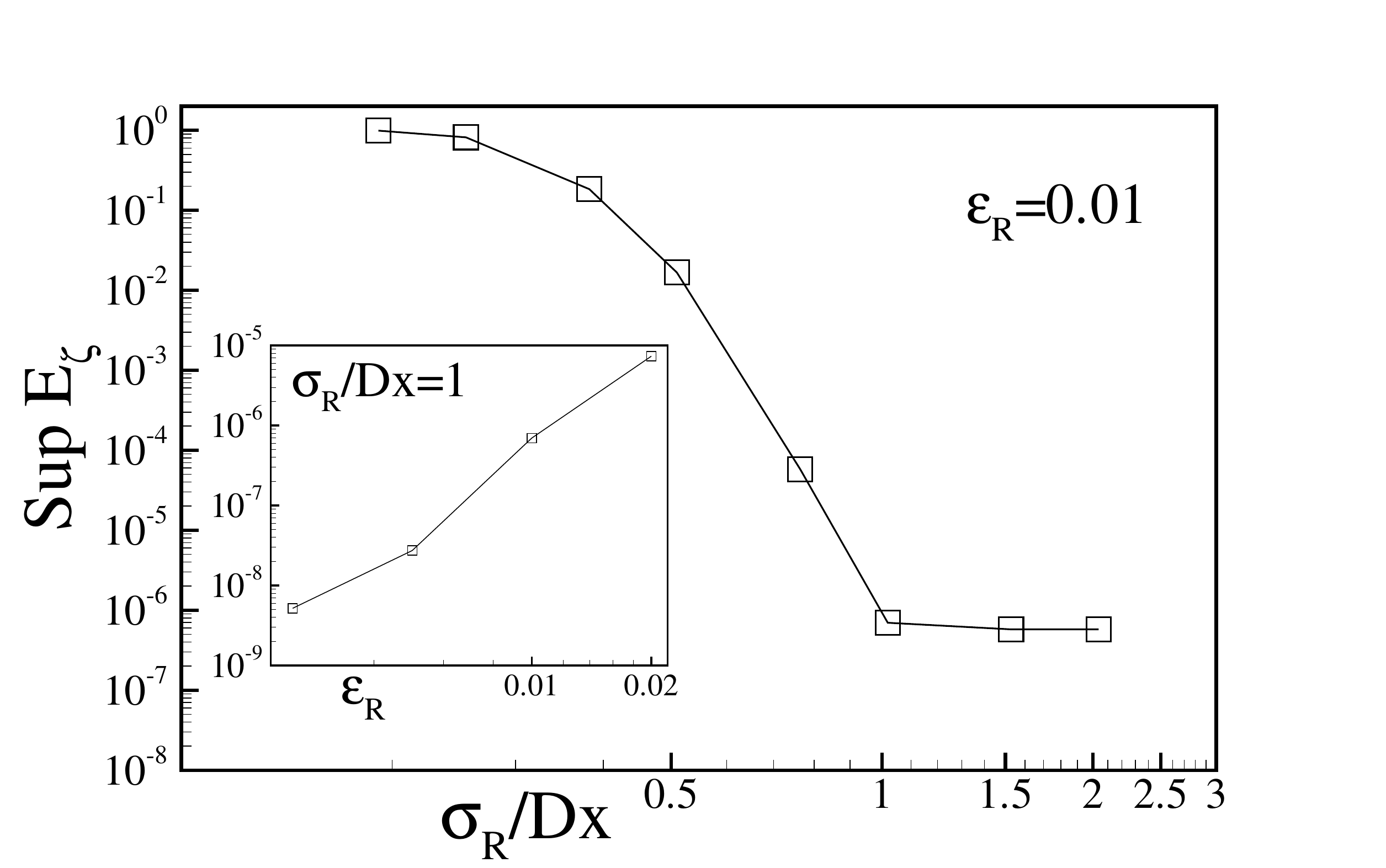}}
\caption{Convergence study in the case of a constant force $\vF_0=(1,0,0)$
applied in a fixed point $\vx_p$ to the fluid initially at rest in a periodic box
$\cal D$.
Left: the error $E_{\vzeta}$, see text for definition,
is plotted against the normalized time $t/\epsilon_R$ for a fixed regularization timescale 
$\epsilon_R=0.01$ and different values of the spatial resolution $\sigma_R/Dx$.
Right: the supreme $\sup_{t \ge 0} E_{\vzeta}$ is shown as a function of 
the ratio $\sigma_R/Dx$ for $\sigma_R=0.01$. The inset reports the supreme 
$\sup_{t \ge 0} E_{\vzeta}$ versus the regularization timescale
$\epsilon_R$ for a fixed spatial resolution $\sigma_R/Dx=1$.
\label{fig:conv_vort} }
\end{figure}

The impulse, though a fundamental quantity, does not retain any information 
concerning the spatial structure of the fluid field. To go more in depth into 
the convergence analysis we have addressed the vorticity field. 
The error is now defined by using the standard $L^2$ norm 
as $E_{\vzeta}=\lVert \vzeta_E-\vzeta_N \rVert_2$ where the subscripts refer to the
exact regularized solution (\ref{eqn:vort_regular}) and its numerical counterpart.
The error $E_{\vzeta}$ is shown in the left panel of figure \ref{fig:conv_vort} 
as a function of time for $\epsilon_R=0.01$ and several values of the ratio
$\sigma_R/Dx$.  When a proper spatial resolution is adopted, also the error  $E_{\vzeta}$  
 stays constant in time or decreases. Note that the largest error is
achieved at the early stages of the simulation when the force is applied to
the fluid at rest. In any case, the supreme $\sup_{t \ge 0} E_{\vzeta}$ 
converges with respect to the refinement of the spatial resolution as
shown in the right panel of the figure. The inset
reports $\sup_{t \ge 0} E_{\vzeta}$ against the regularization timescale
$\epsilon_R$ for a fixed spatial resolution documenting the convergence of 
$\sup_{t \ge 0} E_{\vzeta}$ also with respect to $\epsilon_R$.

\begin{figure}
\centerline{
\includegraphics[scale=0.30]{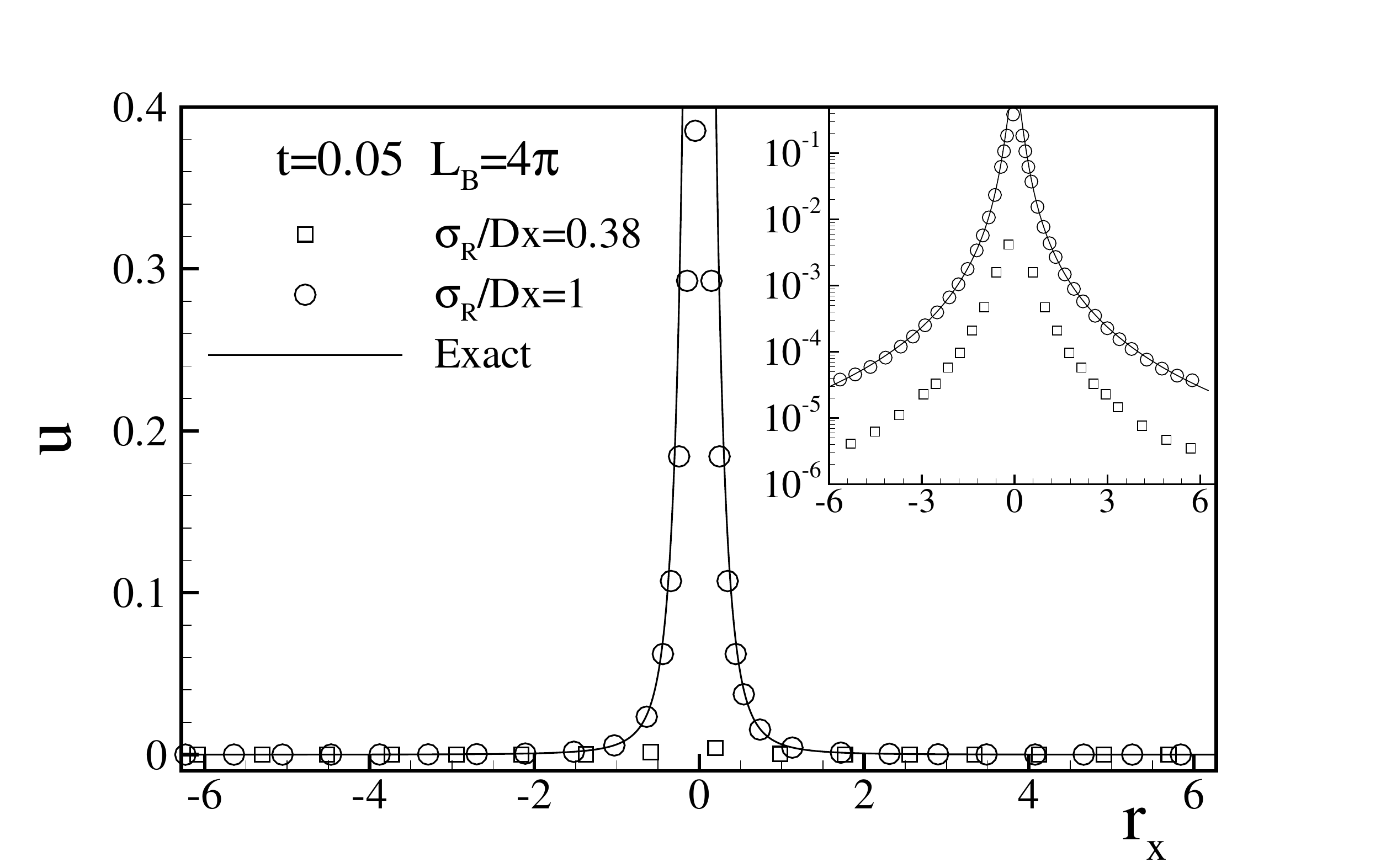}
\includegraphics[scale=0.30]{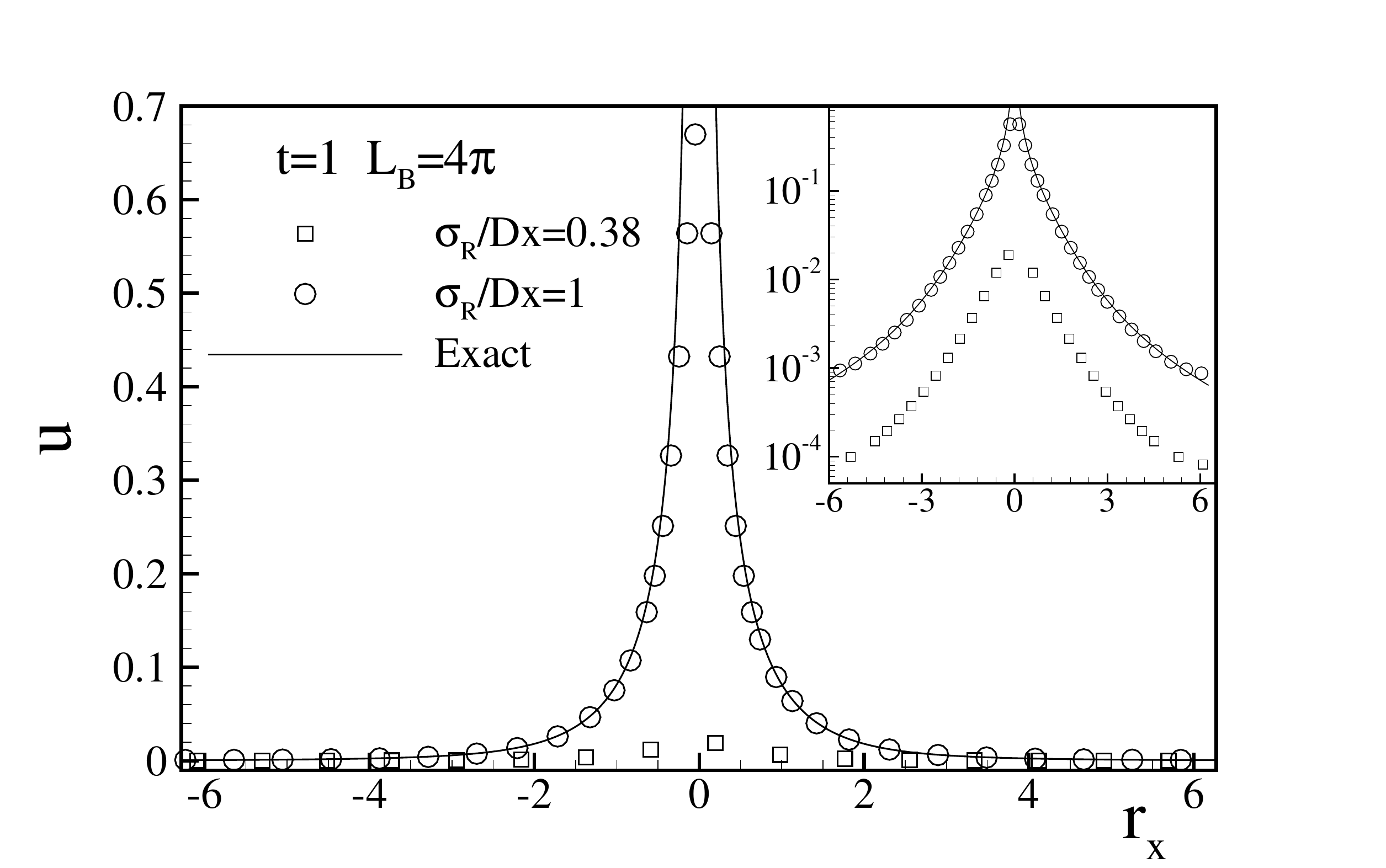}}
\caption{Fluid velocity disturbance generated by a fixed constant force 
$\vF_0=(1,0,0)$ on an initially motionless fluid contained in a
periodic box $L_B=4\pi$. The $1D$ profile representing the fluid velocity 
in the direction of the force (symbols) is compared against the exact solution 
(solid line). The velocity component $u(\vr,t)$ in the $x-$direction is plotted 
against the separation $r_x$ for $\epsilon_R=0.01$ and two spatial resolutions,
namely $\sigma_R/Dx=0.38$ ($\square$) and 
$\sigma_R/Dx=1$ ($\bigcirc$). Left: velocity disturbance at $t=0.05$. 
Right: velocity disturbance at $t=1$. In the insets of the two panels 
the data are plotted in a semi-logarithmic scale.
\label{fig:u_vs_rx_sigmaR_f=1} }
\end{figure}

A more detailed insight concerning the ensuing fluid motion generated by the fixed 
force is achieved by a direct inspection of the flow field.
Figure \ref{fig:u_vs_rx_sigmaR_f=1} reports the fluid velocity in a one dimensional
cut across the complete three-dimensional field. It is 
useful to fix the notation: the constant force has cartesian components 
$\vF_0=(F_x,F_y,F_z)$, the corresponding velocity vector is $\vu=(u,v,w)$ and the 
distance from $\vx_p$ is measured by the vector $\vr=\vx-\vx_p$ whose cartesian 
components are $\vr=(r_x,r_y,r_z)$. 
In the plots of figure \ref{fig:u_vs_rx_sigmaR_f=1} the force is $\vF_0=(1,0,0)$. 
The figure presents the velocity disturbance 
$u$ in the direction of the force as a function of the separation $r_x$ along a one dimensional cut aligned
with the x-axis passing through the point where the force is applied.
The different profiles pertain to simulations which share the same regularization 
timescale $\epsilon_R=0.01$ but differ for the spatial resolution. Namely,
a typical unresolved case $\sigma_R/Dx=0.38$ and a resolved simulation
$\sigma_R/Dx=1$ are compared against the exact solution at two different times
along the simulation. In fact, when a fixed constant force is applied to the 
fluid, an exact solution can be easily determined in closed form 
by evaluating the time convolution integral between the unsteady Green tensor, see 
appendix equation (\ref{eqn:green_tensor}), and the force $\vF_0$. After some algebra, 
the fluid velocity disturbance in the direction of the force reads
\begin{equation}
\label{eqn:ref_velo_f_fixed}
u(\vr,t)=\frac{1}{4 \pi \mu r}
\left[\frac{1}{2 \eta_t^2}\mbox{erf}\left(\eta_t\right) - 
\mbox{erf}\left(\eta_t\right)-
\frac{1}{\sqrt{\pi} \eta_t}\exp\left(-\eta_t^2\right)+1\right]
\end{equation}
where $\eta_t=r/\sqrt{4 \nu t}$ and $r=\sqrt{r_k \, r_k}$. 
As shown in the  figure \ref{fig:u_vs_rx_sigmaR_f=1},
when $\sigma_R/Dx \ge 1$ the present algorithm well reproduces 
the exact solution. Note that insufficient spatial resolution results in a
clear underestimate of the fluid velocity disturbance. This is emphasized 
by the plots reported in the insets figure \ref{fig:u_vs_rx_sigmaR_f=1}
where the data are represented in a semi-logarithmic scale. 

\begin{figure}
\centerline{
\includegraphics[scale=0.30]{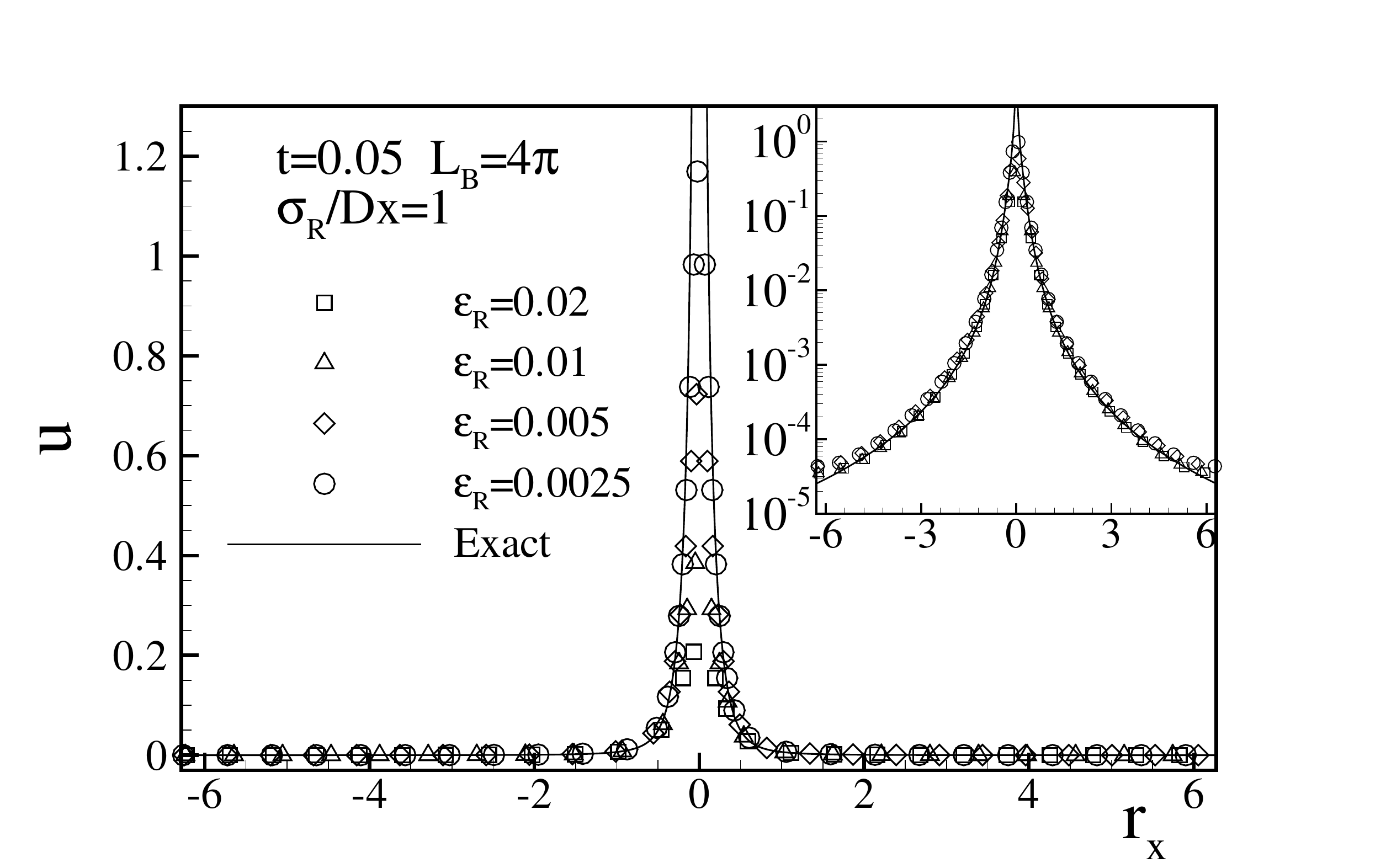}
\includegraphics[scale=0.30]{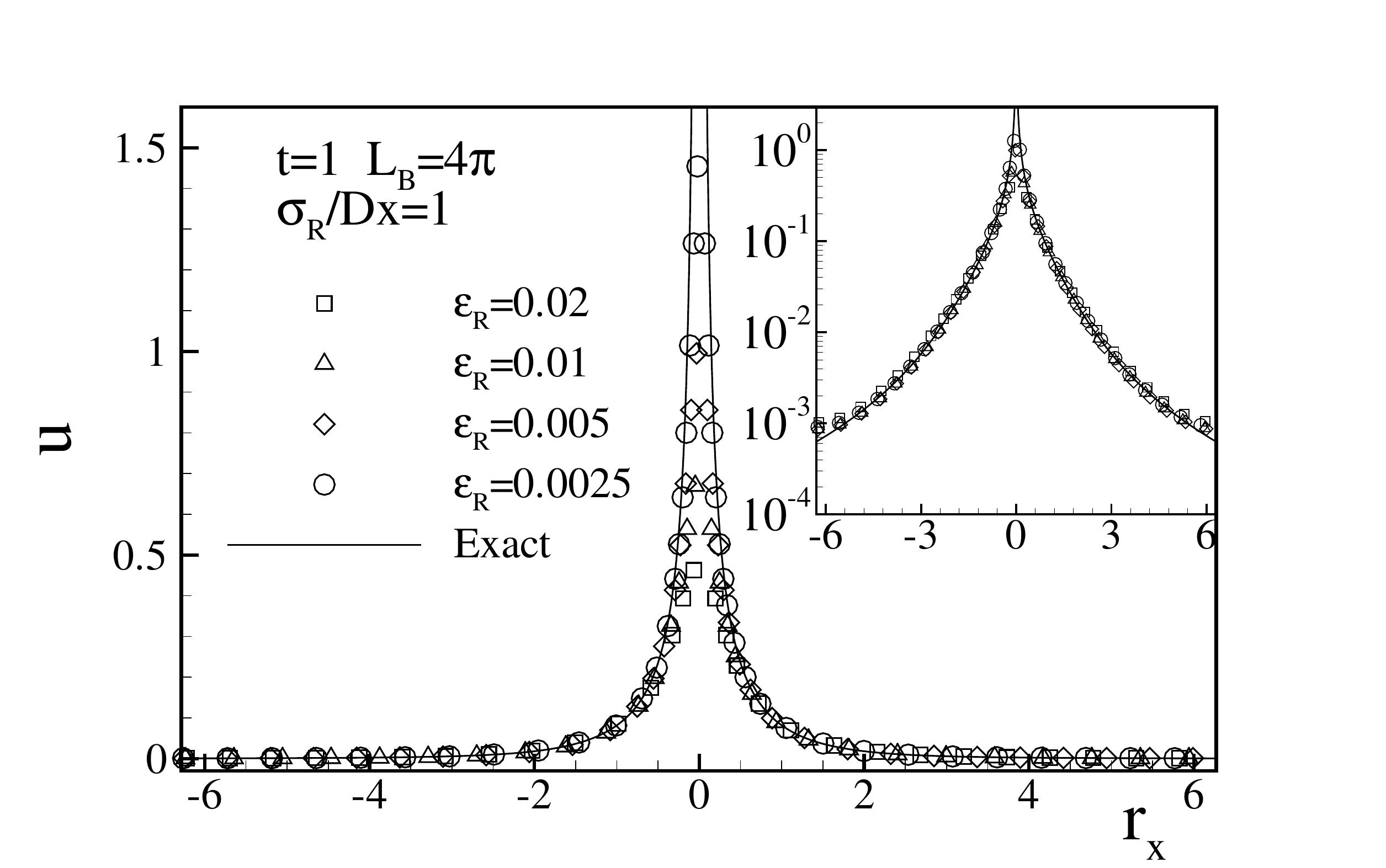}}
\caption{Fluid velocity disturbance generated by a fixed constant force 
$\vF_0=(1,0,0)$ on an initially motionless fluid contained in a
periodic box $L_B=4\pi$. The $1D$ profile representing the fluid velocity 
in the direction of the force (symbols) is compared against the exact solution 
(solid line). The velocity component $u(\vr,t)$ in the $x-$direction is plotted 
against the separation $r_x$ for several values of the regularization 
timescale, $\epsilon_R=0.01$ ($\square$); $\epsilon_R=0.02$ ($\triangle$);
$\epsilon_R=0.005$ ($\diamond$); $\epsilon_R=0.0025$ ($\bigcirc$), at a fixed 
spatial resolution $\sigma_R/Dx=1$.
Left: velocity disturbance at $t=0.05$. 
Right: velocity disturbance at $t=1$. The insets of the two panels show
the data plotted in a semi-logarithmic scale.
\label{fig:u_vs_rx_epsR_f=1} }
\end{figure}

Figure \ref{fig:u_vs_rx_epsR_f=1} documents the behavior of the ERPP method
when the spatial resolution $\sigma_R/Dx$ is kept fixed and the regularization
timescale is progressively reduced. In fact, as $\epsilon_R$ is decreased,
the numerical solution describes a progressively wider range of the exact solution 
avoiding in all cases the occurrence of the singularity at the point $\vx_p$ 
where the force is applied, $r_x=0$ in the plot. 
The different cases share the same far field behavior 
away from $\vx_p$ irrespective of the value of $\sigma_R$ as
emphasized by the plots in the insets of figure \ref{fig:u_vs_rx_epsR_f=1} 
where the velocity disturbance is represented in a semi-logarithmic scale.
In summary, the solution provided by the ERPP retains the relevant features 
of the exact solution and avoids the occurrence of the singularity at $\vx_p$ 
which is clearly an unwanted trait in any numerical solution. 

\begin{figure}
\centerline{
\includegraphics[scale=0.30]{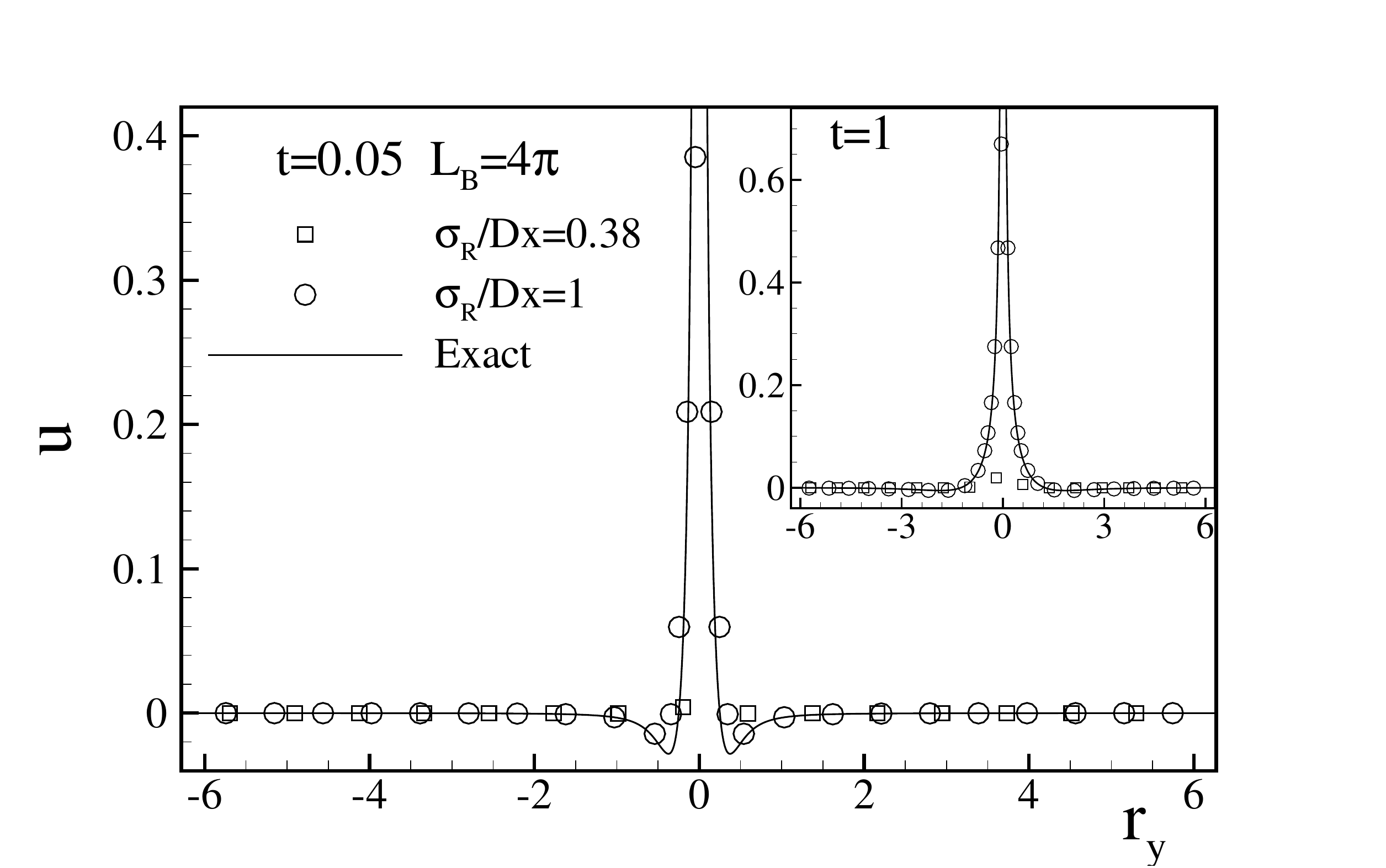}
\includegraphics[scale=0.30]{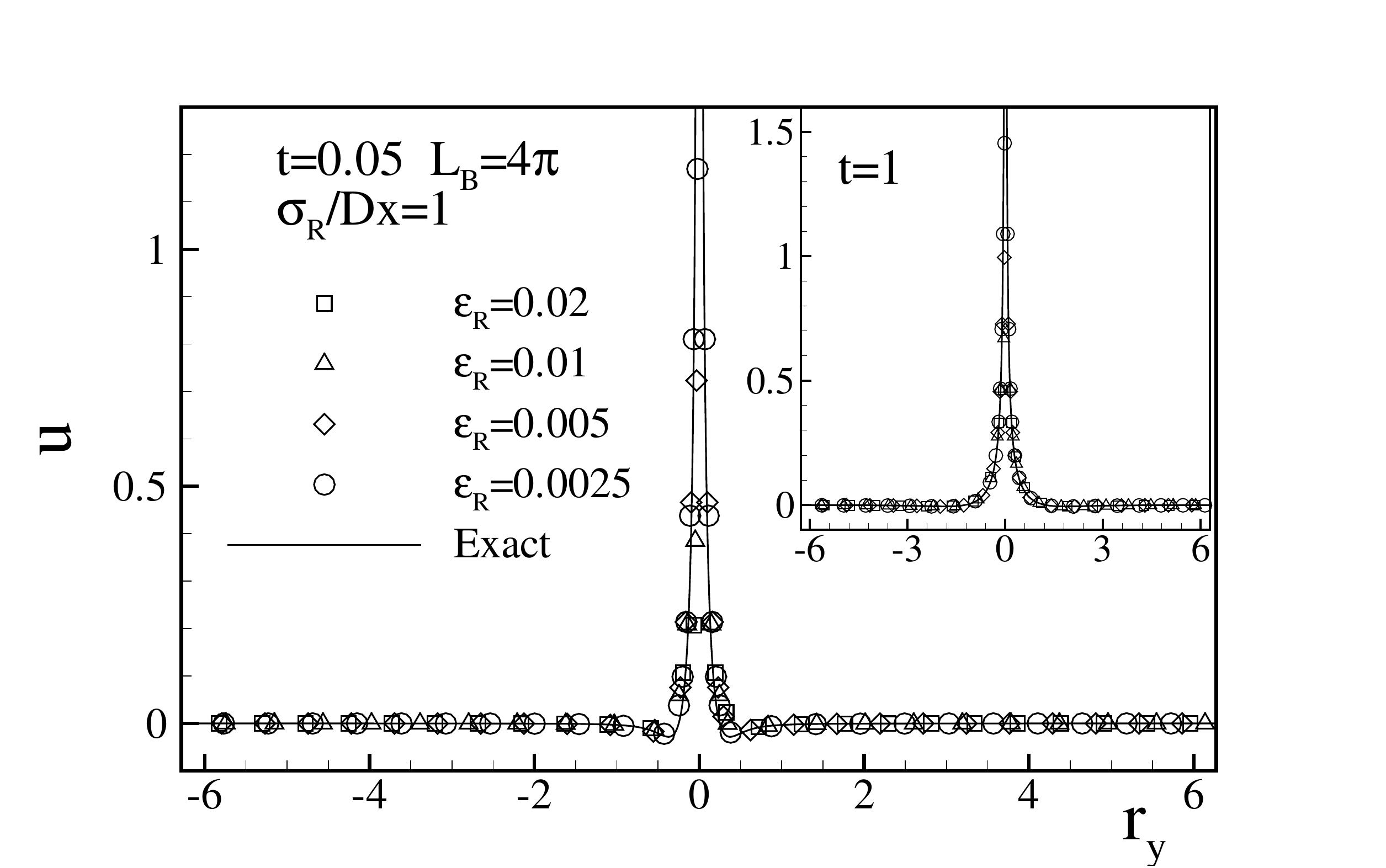}}
\caption{Fluid velocity disturbance generated by a fixed constant force 
$\vF_0=(1,0,0)$ on an initially motionless fluid contained in a
periodic box $L_B=4\pi$. The $1D$ profile representing the fluid velocity 
component $u(\vr,t)$ in the $x-$direction (symbols) 
is plotted against the separation $r_y$. The exact solution (solid line) 
is reported for comparison. 
Left: fluid velocity disturbance at $t=0.05$ (main panel) and
$t=1$ (inset) plotted for a fixed value of $\epsilon_R=0.01$ and two spatial resolutions
namely $\sigma_R/Dx=0.38$ and $\sigma_R/Dx=1$. 
Right: fluid velocity disturbance at $t=0.05$ (main panel) and
$t=1$ (inset) plotted for several values of $\epsilon_R$ and a fixed spatial
resolution $\sigma_R/Dx=1$. 
\label{fig:u_vs_ry}}
\end{figure}

\begin{figure}
\centerline{
\includegraphics[scale=0.30]{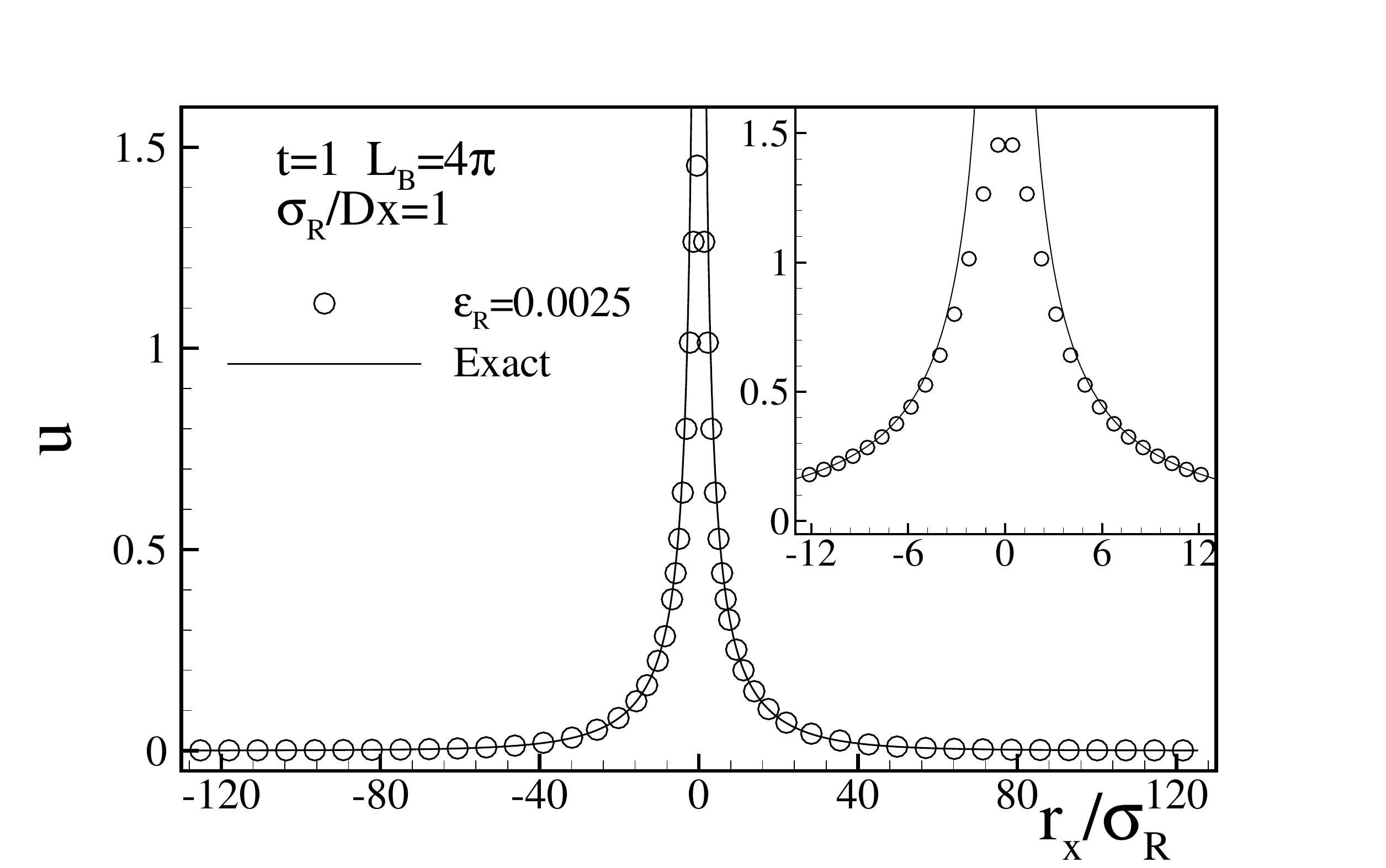}
\includegraphics[scale=0.30]{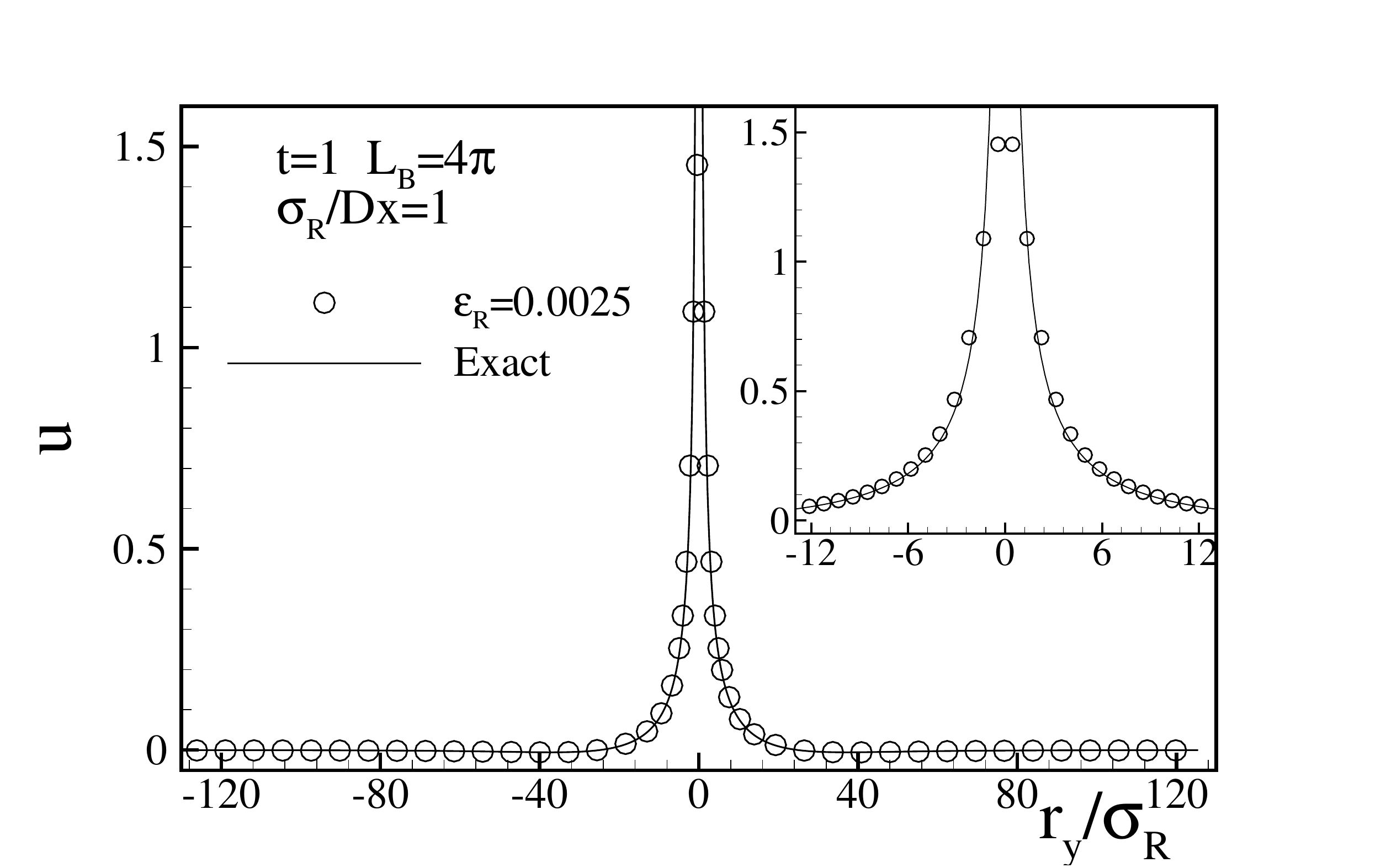}}
\caption{Fluid velocity disturbance generated by a fixed constant force 
$\vF_0=(1,0,0)$ on an initially motionless fluid contained in a
periodic box $L_B=4\pi$. The velocity component $u(\vr,t)$ in the $x-$direction 
(direction of the force, symbols) is plotted against the normalized 
separation $r_x/\sigma_R$ (left panel) and $r_y/\sigma_R$ (right panel) for 
$\epsilon_R=0.0025$ and $\sigma_R/Dx=1$. The inset of the two panels
provides a close-up view of the solution at $\vx_p=0$.
\label{fig:u_vs_r_normalized}}
\end{figure}

Figure \ref{fig:u_vs_ry} reinforces  the conclusion of the previous analysis by showing the fluid velocity 
component in the direction of the force $u(r_y)$ as a function of the distance $r_y$ in a transversal one-dimensional
cut through the point of application of the force.
Finally figure \ref{fig:u_vs_r_normalized} reports the fluid velocity component
in the direction of the force plotted against the normalized distance
$r_x/\sigma_R$ and $r_y/\sigma_R$. The discussion of these plots requires some care.
The ERPP model has been conceived to describe the far field effect produced on the 
fluid by a point-like particle avoiding the occurrence of singularities in the
point $\vx_p$ where the particle is located. In fact, the disturbance produced by the
particle is described by retaining only the first term in the multipole
expansion of the general solution of the unsteady Stokes flow and the regularization
timescale $\epsilon_R$ accounts for the viscous diffusion process which naturally 
regularizes the solution. Hence, the solution provided by the ERPP has un intrinsic 
inner cut-off provided by $\sigma_R$ and is expected to reproduce the disturbance flow
generated by a point-particle in the far field. This is indeed what happens and what is
documented by the plots in figure \ref{fig:u_vs_r_normalized}. The regularized 
solution stays on top of the exact solution everywhere, see e.g. the main panels of 
figure \ref{fig:u_vs_r_normalized} which, on the scale of the complete computational
domain $\cal D$, reports the fluid velocity disturbance produced by the particle. The
insets of figure \ref{fig:u_vs_r_normalized} show the same data in proximity of the
origin where the particle is placed. Such representation emphasizes that,
after a distance of a few $\sigma_R$, the ERPP solution falls on top of the exact solution. 
The threshold $3\sigma_R$ can be safely assumed as an inner cut-off for the
disturbance flow field produced by the particle.

\subsection{Unsteady motion of an isolated particle} 
\label{sec:unsteady_motion}
The following subsections address more realistic cases where the point $\vx_p$ 
is allowed to move according to equations (\ref{eqn:particle_motion}) with 
the initial conditions $\vx_p(t=0)=\vx_p^0; \, \vv_p(t=0)=0$. 
In order to proceed gradually we will first discuss a series of tests where the 
particle motion is not affected by the fluid velocity disturbance that the particle 
generates. In such decoupled cases the fluid disturbance field is still amenable
of exact solutions which can be employed for further comparisons.
Successively, we will consider the fully coupled case where the dynamics of the
particles and of the fluid are intertwined.

\subsubsection{Imposed particle motion} 
We consider the motion of a small particle subjected to an external force, 
e.g. gravity, and to the Stokes drag. The particle velocity is given by the 
solution of the equation
\begin{equation}
\label{eqn:Drag_one_way}
m_p \frac{d \vv_p}{dt}= m_p\vg - 6\pi\mu a_p \vv_p(t) \, ,
\end{equation}
namely
\begin{equation}
\label{eqn:uncoupled_solution}
v_p(t)=v_t\left(1-e^{-t/\tau_p}\right)
\end{equation}
where $v_p(t)$ denotes the particle velocity in the direction of the gravity
acceleration, say the $x-$direction, $v_t=\tau_p\,g$ is the particle terminal 
velocity and $\tau_p=\rho_p d_p^2/{18 \mu}$ the Stokes relaxation timescale. 
In this framework the motion of the particle is imposed and its dynamics
is decoupled from the dynamics of the carrier fluid.

\begin{figure}
\centerline{
\includegraphics[scale=0.30]{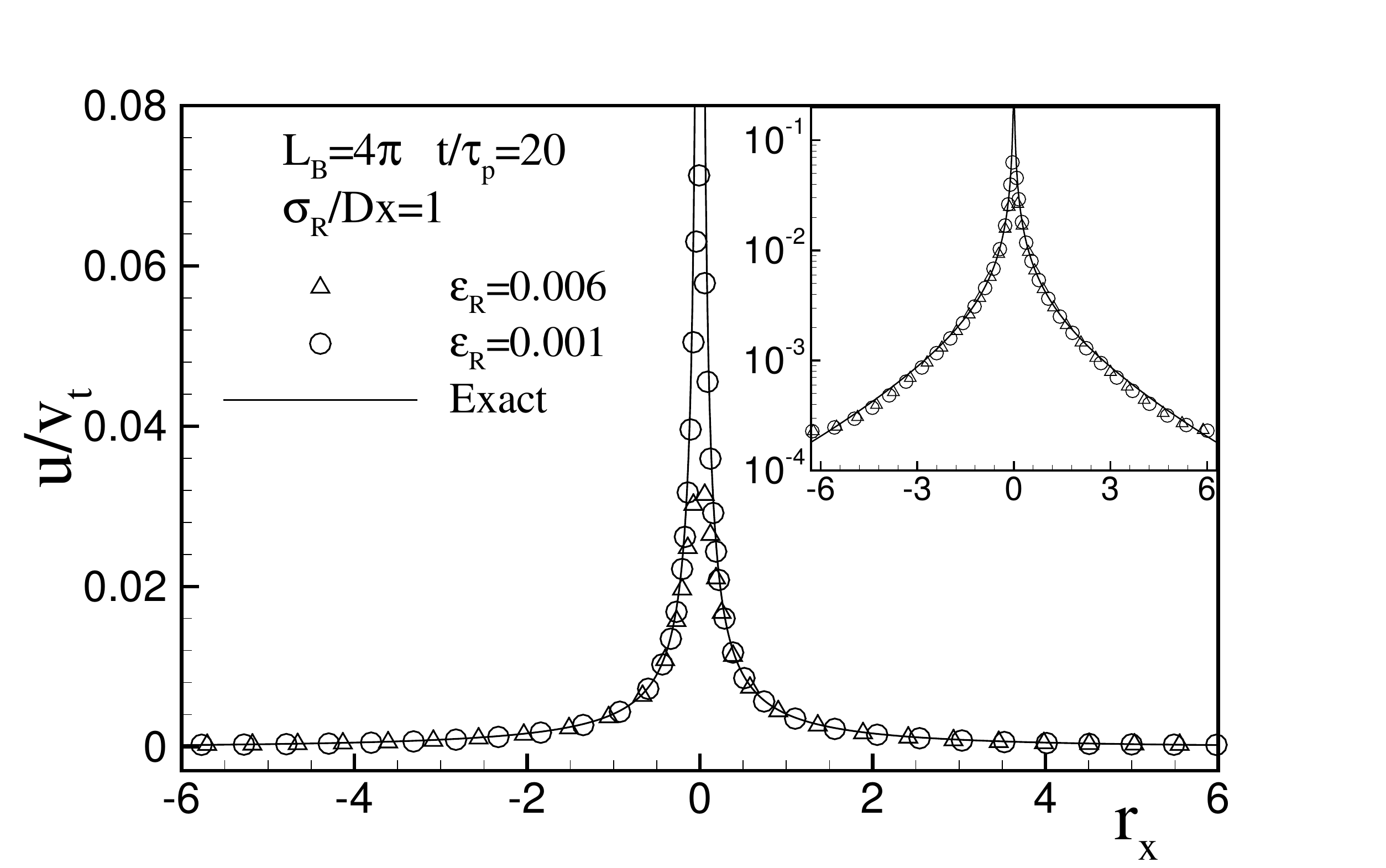}
\includegraphics[scale=0.30]{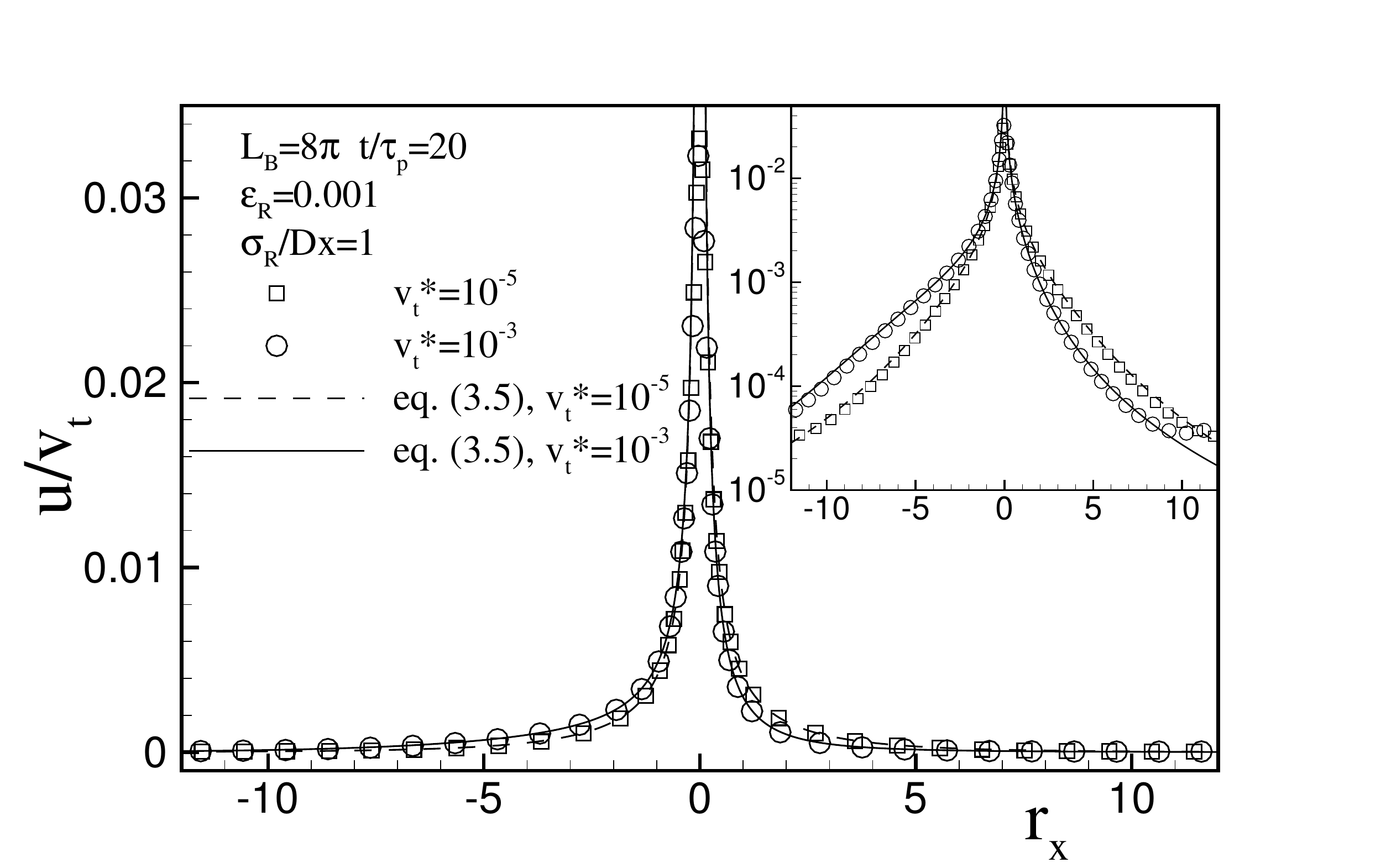}}
\caption{Normalized velocity disturbance produced by a particle moving with
velocity $v_p(t)=v_t\left(1-e^{-t/\tau_p}\right)$ in the $x-$direction in a 
periodic box $L_B=4\pi$.
The normalized velocity disturbance $u/v_t$ in the direction of
the particle motion, is plotted against the separation $r_x$ at time 
$t/\tau_p=20$ when the particle has reached the terminal velocity $v_t$.
Left panel: data obtained for different values of the regularization timescale 
$\epsilon_R=0.006$ ($\triangle$); $\epsilon_R=0.001$ ($\bigcirc$) 
at a fixed spatial resolution $\sigma_R/Dx=1$, are
compared against the exact solution (solid line) given by equation
(\ref{eqn:ref_velo_disturbance}). The inset reports 
the data of the main panel plotted in a semi-logarithmic scale.
Right panel: data pertaining particles with
with different terminal velocities, $v_t^*= v_t d_p/\nu=10^{-5}$ ($\square$);
$v_t^*=10^{-3}$ ($\triangle$), are compared against the corresponding exact
solution (\ref{eqn:ref_velo_disturbance}), dashed and solid line respectively,
in a periodic box $L_B=8\pi$. The inset reports the date of the main panel 
plotted in a semi-logarithmic scale.
\label{fig:sigmaR_Re} }
\end{figure}

The left panel of figure \ref{fig:sigmaR_Re} reports
the fluid velocity disturbance produced by the moving particle
at $t/\tau_p=20$ when the particle has reached its terminal velocity $v_t$. 
The fluid velocity profile is plotted for two cases which differs for
the value of the regularization timescale $\epsilon_R$. Once again
the value of $\epsilon_R$ controls the regularized near field and does not
affect the far field, see the inset of the figure. 
In this case the fluid velocity has an explicit solution given by
the time convolution integral of 
the unsteady Stokeslet, equation (\ref{eqn:green_tensor}), and the 
the hydrodynamic force $D_p(t)$ where the Stokeslet is placed at the instantaneous position of
the particle evolving as specified by equation (\ref{eqn:uncoupled_solution}). Performing the time convolution integral is now a little more
tricky than in the previous example since $\vr = \vx-\vx_p(t)$ where $\dot{x}_p=v_p(t)$. 
The integration of the expression
\begin{equation}
\label{eqn:ref_velo_disturbance}
u_i(\vx,t)=\int_0^t G_{ik}\left[\vx-\vx_p(\tau),t-\tau\right] D^p_k(\tau) 
\quad d\tau \,
\end{equation}
in a closed form becomes  cumbersome even though the integral can be evaluated 
numerically by a quadrature formula. Indeed, such numerical approximation 
of the exact solution can be still used for useful comparisons.
The data produced by the ERPP algorithm are  compared with this reference 
solution in figure \ref{fig:sigmaR_Re}.
The plots show that the ERPP approach is able to capture 
the expected solution and provides a consistent regularization of the 
singularity which occurs at $\vx_p$. The large fluid domain, note that the box size 
is $L_B=4\pi$, allows a good approximation of the unbounded domain where the reference
solution (\ref{eqn:ref_velo_disturbance}) holds. The right panel of figure 
\ref{fig:sigmaR_Re} provides the fluid velocity disturbance produced by  
particles with different terminal velocities $v_t$. 
The effect of increasing the terminal  velocity is worth discussing. As $v_t$ is increased the 
front-aft symmetry in the disturbance flow is progressively broken. 
This symmetry breaking is indeed easily explained in terms of vorticity released along the path of
the moving particle. In fact in the body-fixed frame, the  convection term
$\vv_p\cdot \nabla \vu$ is responsible of the constant velocity advection of the
vorticity even in a Stokes regime. In order to capture the tails of the disturbance flow in the far field and
to avoid confinement effects introduced by  the periodic boundary conditions, the computational box is now $L_B=8\pi$ wide. 
%
%

\begin{figure}
\centerline{
\includegraphics[scale=0.50]{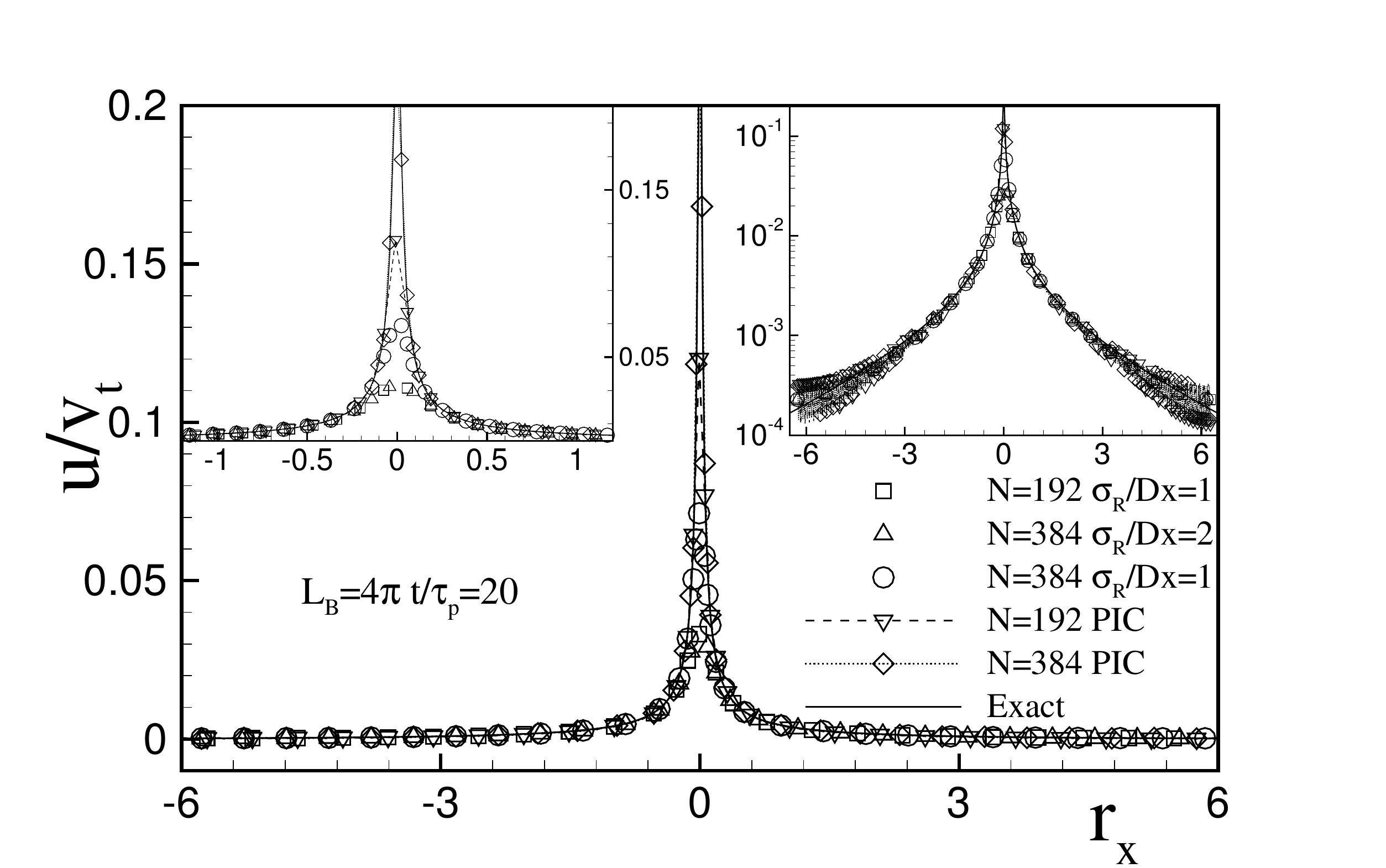}}
\caption{Normalized velocity disturbance at $t/\tau_p=20$ produced by a particle 
moving with velocity $v_p(t)=v_t\left(1-e^{-t/\tau_p}\right)$ in the $x-$direction
in a periodic box $L_B=4\pi$. The velocity profile calculated with
the ERPP for different values of the regularization timescale
$\epsilon_R=0.006$, $N=192^3$ Fourier modes, $\sigma_R/Dx=1$ ($\square$); 
$\epsilon_R=0.006$, $N=384^3$ Fourier modes, $\sigma_R/Dx=2$, ($\triangle$); 
$\epsilon_R=0.001$, $N=384^3$ Fourier modes, $\sigma_R/Dx=1$, ($\bigcirc$);
is compared against the exact solution (solid line) and corresponding results obtained 
by the PIC approach,  $N=192^3$ Fourier modes, ($\triangledown$); 
$N=384^3$ Fourier modes, ($\diamond$).
Top left panel: close-up view of the velocity disturbance near
the singular point $\vx_p$. Top right inset: velocity profile plotted in 
semi-logarithmic scale.
\label{fig:cfr_PIC_uncoupled} }
\end{figure}

We conclude the discussion by presenting in figure \ref{fig:cfr_PIC_uncoupled}
the comparison between the ERPP solution and what one would obtain by using the 
classical Particle In Cell (PIC) approach. As expected, the solution
provided by the PIC method is grid dependent, as demonstrated by comparing the disturbance
velocity profile of two simulations which share the same physical parameters but 
differ for the grid resolution, namely $N=192^3$ and $N=384^3$ Fourier modes. 
As the grid is refined
a singular-like behavior occurs at $\vx_p$ and the field is characterized by 
numerical aliasing, see e.g. the top right inset where the velocity profile 
is plotted in a semi-logarithmic scale. In contrast, once the regularization timescale
$\epsilon_R$ is fixed, the ERPP approach provides a numerically convergent, asymptotically
grid-independent solution. This behavior 
can be better appreciated in the top left inset where a close-up view of the 
velocity disturbance is reported. In a nutshell the ERPP retains all the features
of the physical solution produced by a small point-like particle except for the (undesired)
singularity which unavoidably occurs at $\vx_p$. The regularization 
of the solution is controlled by the timescale $\epsilon_R$ which is 
related to a diffusive lengthscale $\sigma_R$. Indeed  $\sigma_R$ is naturally 
introduced by the process of vorticity diffusion and can be fixed on a physical ground. 
For instance, in a 
turbulent flow, velocity fluctuations are physically irrelevant below the Kolmogorov 
lengthscale $\eta$. At the same time  the effects that a swarm of point-like particles generates on length scales larger
than $\eta$ is physically relevant.  In such framework the regularization lengthscale $\sigma_R$ 
is naturally selected as $\sigma_R=\eta$. 
%
%
\subsubsection{Particle motion in the coupled regime} 
This subsection addresses the unsteady motion of a particle which
settles from rest under the action of gravity in the coupled regime
where the  particle induces a disturbance in the surrounding fluid and such 
disturbance enters the expression of the hydrodynamic force. 
For simplicity we will consider small particles much heavier than the surrounding 
fluid, i.e. $\rho_p\gg\rho_f$, where the only relevant force is  the Stokes drag. 
The general expression of the force \eqref{eqn:force_particle} simplifies to
\begin{equation}
\label{eqn:drag_coupled}
\vD_p(t)=m_p\vg + 6\pi\mu a_p \left[\tilde\vu(\vx_p,t) 
         -\vv_p(t) \right] \,.
\end{equation}
Following the discussion of section \ref{sec:prtcl_motion_force} the velocity
$\tilde\vu(\vx_p,t)$ must be interpreted as the background fluid velocity in absence of the 
p$th$ particle, e.g. turbulent fluctuations plus the disturbance flow generated by all 
the other particles. This makes the calculation of the 
hydrodynamic force particularly challenging in the two-way coupling regime. 
In the particular case where only one particle is considered the value 
$\tilde\vu(\vx_p,t)$ should be set to zero. However this way of proceeding is unfeasible 
in the general case where many particles are present since the value of 
$\tilde\vu(\vx_p,t)$ must also account for the velocity disturbance generated by all the 
other particles and the background flow. This conundrum can be disentangled in 
the context of the ERPP approach since the disturbance flow produced by the $p$th 
particle on itself is known in a closed  form and thus can be removed from 
the background fluid velocity $\vu(\vx_p,t)$ even in presence of many other particles. 

\begin{figure}
\centerline{
\includegraphics[scale=0.30]{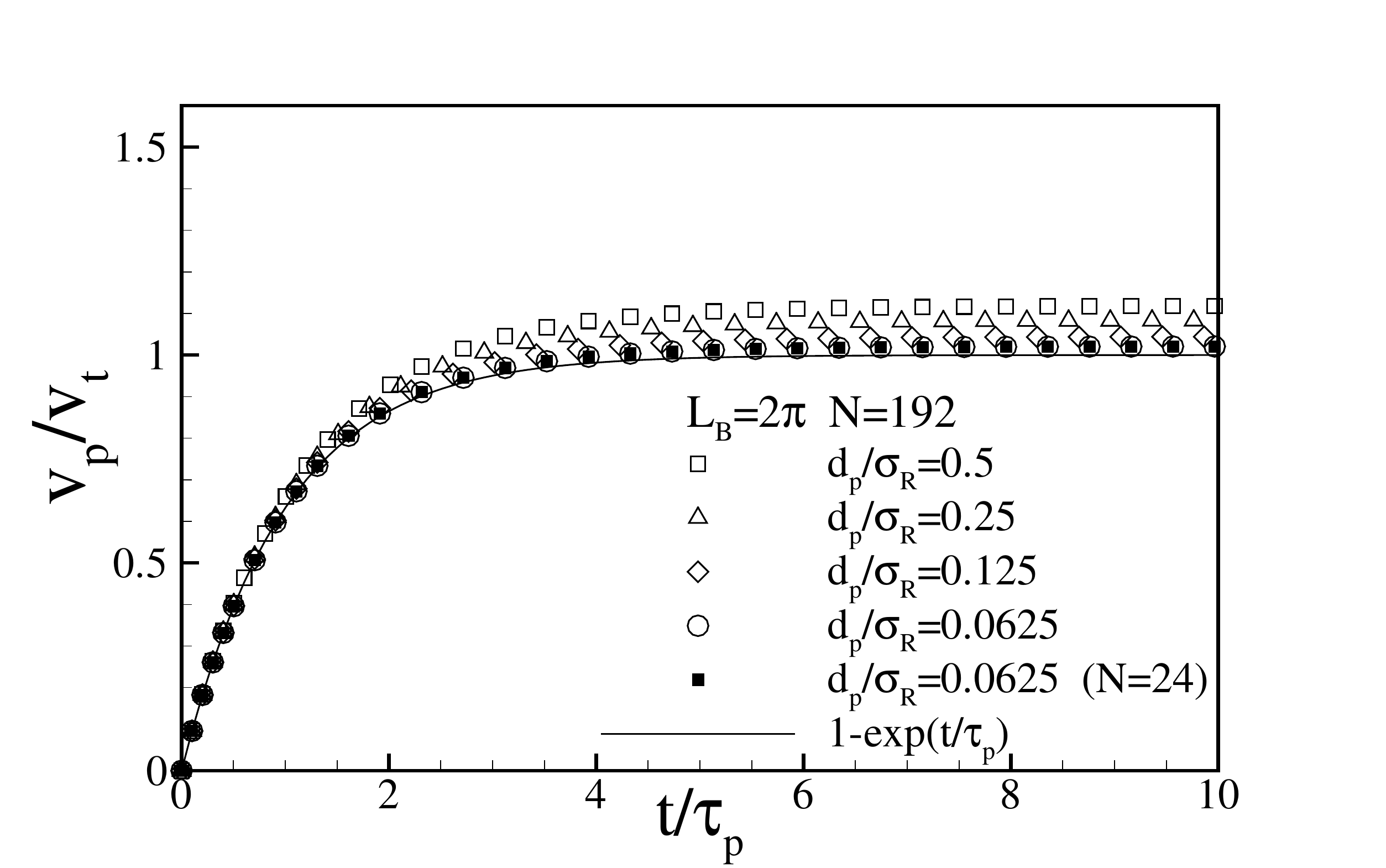}
\includegraphics[scale=0.30]{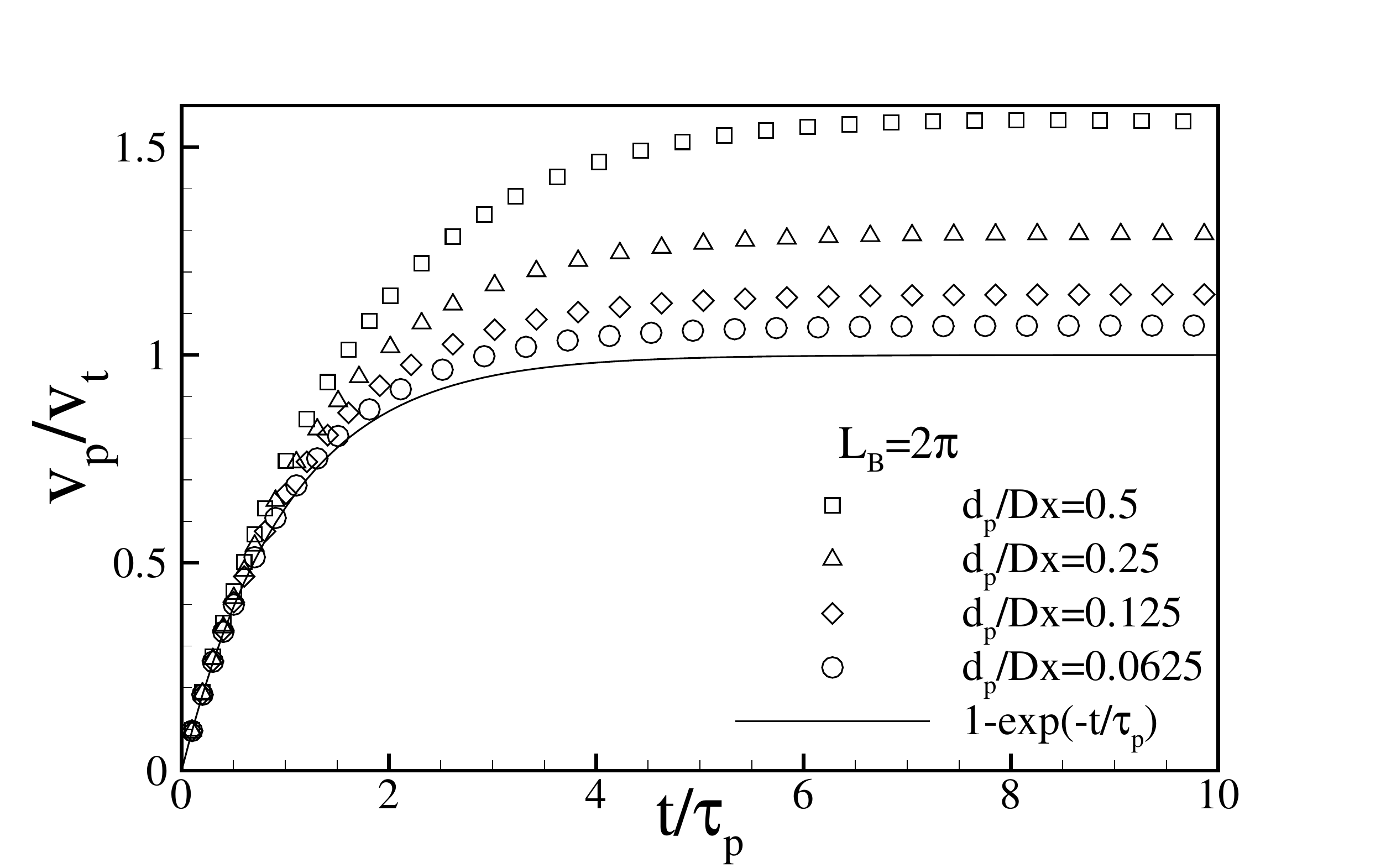}}
\caption{Normalized particle velocity $v_p/v_t$ as a function of the dimensionless
time $t/\tau_p$. Left panel: particle velocity in the ERPP simulations for
different values of the ratio; $d_p/\sigma_R=0.5$, ($\square$); 
$d_p/\sigma_R=0.25$, ($\triangle$); $d_p/\sigma_R=0.125$, ($\diamond$);
$d_p/\sigma_R=0.0625$, ($\bigcirc$). The fluid field is discretized with $N=192^3$ 
Fourier modes (open symbols) in a periodic box $L_B=2\pi$. In the case 
$d_p/\sigma_R=0.0625$, ($\blacksquare$) the fluid field is discretized with $N=24^3$
Fourier modes to check grid-independence.
Right panel: particle velocity provided by the PIC approach in comparable conditions 
as in the ERPP. $d_p/Dx=0.5$, ($\square$); $d_p/Dx=0.25$, ($\triangle$); 
$d_p/Dx=0.125$, ($\diamond$); $d_p/Dx=0.0625$, ($\bigcirc$). 
In both panels the reference solution (\ref{eqn:uncoupled_solution}) is reported for
comparison (solid line).
\label{fig:velo_vs_t} }
\end{figure}

The two panels of figure \ref{fig:velo_vs_t} provide evidence to the above 
considerations.
The plots report the particle velocity normalized with the settling velocity
$v_t$ as a function of the dimensionless time $t/\tau_p$ both for the ERPP
calculation and for the PIC approach. The particle trajectory should be compared with
the reference solution given by equation (\ref{eqn:uncoupled_solution}).
The left panel shows the particle velocity calculated by
the ERPP method for different values of the ratio $d_p/\sigma_R$. We recall
that our approach is designed to model the disturbance flow produced by point-like 
particles, i.e. particles whose diameter $d_p$ is much smaller that any other 
lengthscale in the system. In the ERPP approach, in absence of any other length-scales
introduced by the background flow, the only significant lengthscale is the 
diffusive scale $\sigma_R$. Hence, the nominal diameter of the particle should be 
smaller than $\sigma_R$. Indeed, as the ratio $d_p/\sigma_R$ decreases the particle 
velocity rapidly approaches the reference curve provided by 
equation (\ref{eqn:uncoupled_solution}).  When the scale $\sigma_R$ has been fixed 
the ERPP gives a grid-independent solution as can be appreciated in 
figure \ref{fig:velo_vs_t} where two trajectories which share the same $\sigma_R$ 
but have different grids, namely $N=192^3$ and $N=24^3$ Fourier modes, 
give practically undistinguishable results. It's worth noting that the error in the 
particle velocity is already below $10\%$ for the relatively large ratio 
$d_p/\sigma_R=0.5$ we have considered. The right panel of figure \ref{fig:velo_vs_t} 
reports the particle velocity calculated with the PIC method. The solution 
presents now larger deviations from the exact 
result (\ref{eqn:uncoupled_solution}). This is due to a poor estimate of the
hydrodynamic force. In fact, in the PIC approach the self-induced 
disturbance produced by the $p$th particle is unknown or, if eventually modeled by 
the steady Stokeslet, is singular at the particle position $\vx_p$. In both cases 
it can not be removed from the particle-to-fluid slip velocity resulting in 
an inaccurate prediction of the hydrodynamic force and, consequently, of the 
particle trajectory. For instance, for $d_p/Dx=0.5$  the error on the terminal velocity is $50\%$ for the PIC approach 
compared to a much lower $10\%$ for the ERPP. Clearly the error reduces as 
the ratio $d_p/Dx \to 0$.

\begin{figure}
\centerline{
\includegraphics[scale=0.30]{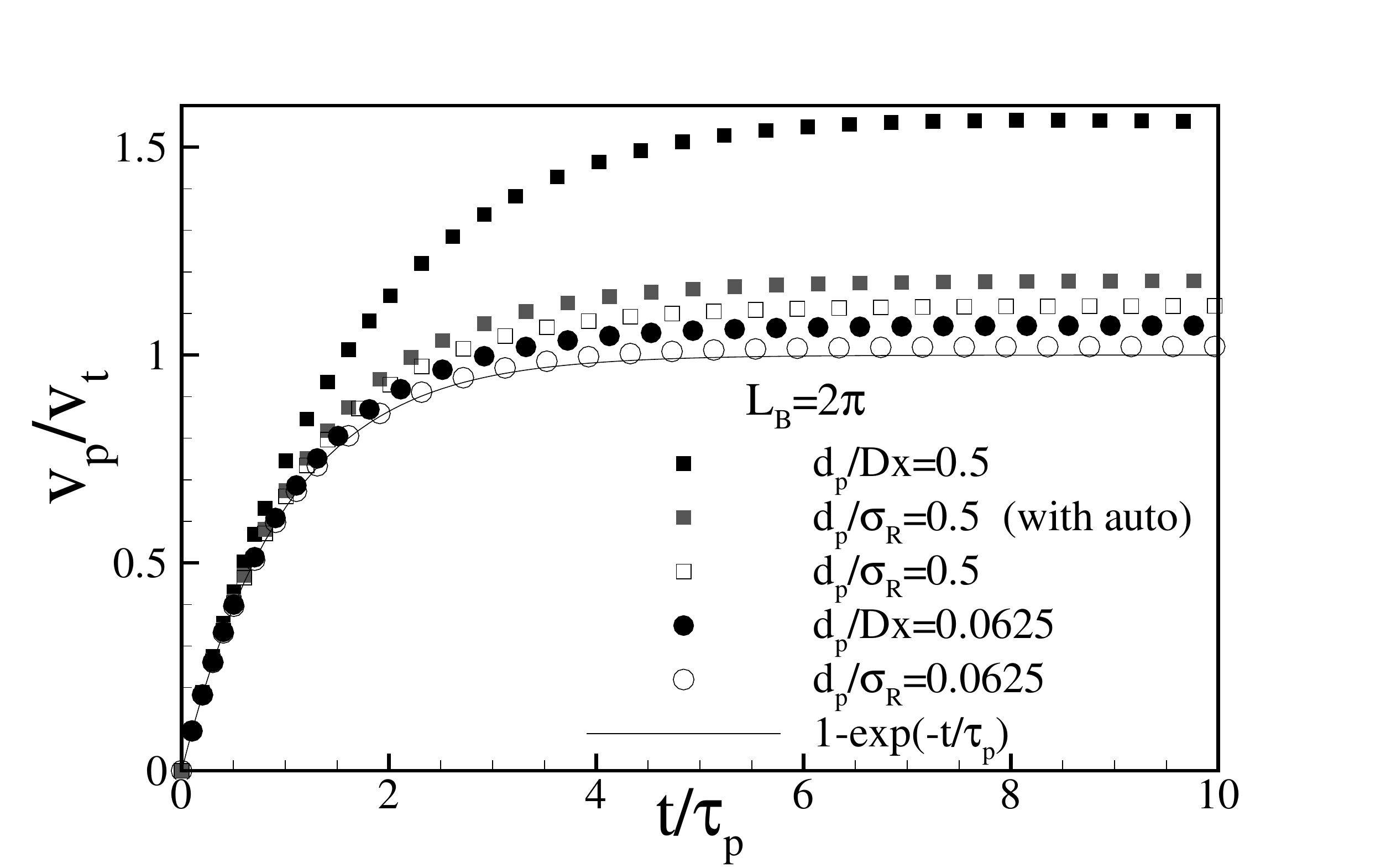}
\includegraphics[scale=0.30]{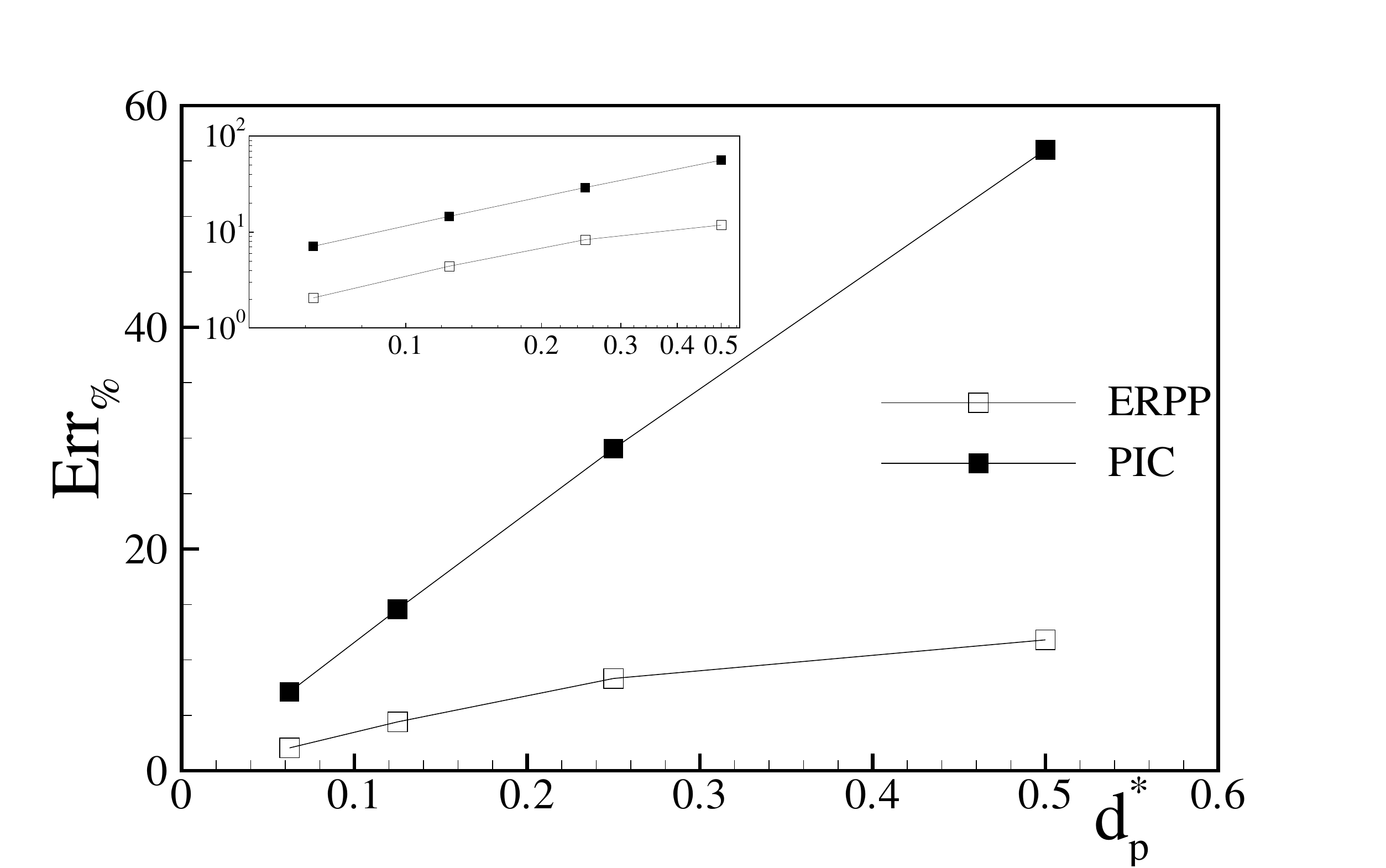}}
\caption{Normalized particle velocity $v_p/v_t$ as a function of the dimensionless
time $t/\tau_p$. Left panel: direct comparison of the ERPP results 
against the PIC approach and the reference solution (solid line) 
for different values of the ratio $d_p/\sigma_R$ or equivalently $d_p/Dx$. 
$d_p/\sigma_R \div d_p/Dx =0.5$ (squares), $d_p/\sigma_R \div d_p/Dx =0.0625$
(circles). Black filled symbols refer to the PIC approach, open symbols to the ERPP 
method. The grey square refers to an ERPP calculation where intentionally we
did not remove the self-induced velocity disturbance.
Right panel: relative error committed on the evaluation of the
terminal velocity as a function of the dimensionless parameter $d_p^*=d_p/\sigma_R$
for the ERPP and the PIC. In the inset data plotted in log-log scale.
\label{fig:velo_terminal} }
\end{figure}

\begin{figure}
\centerline{
\includegraphics[scale=0.30]{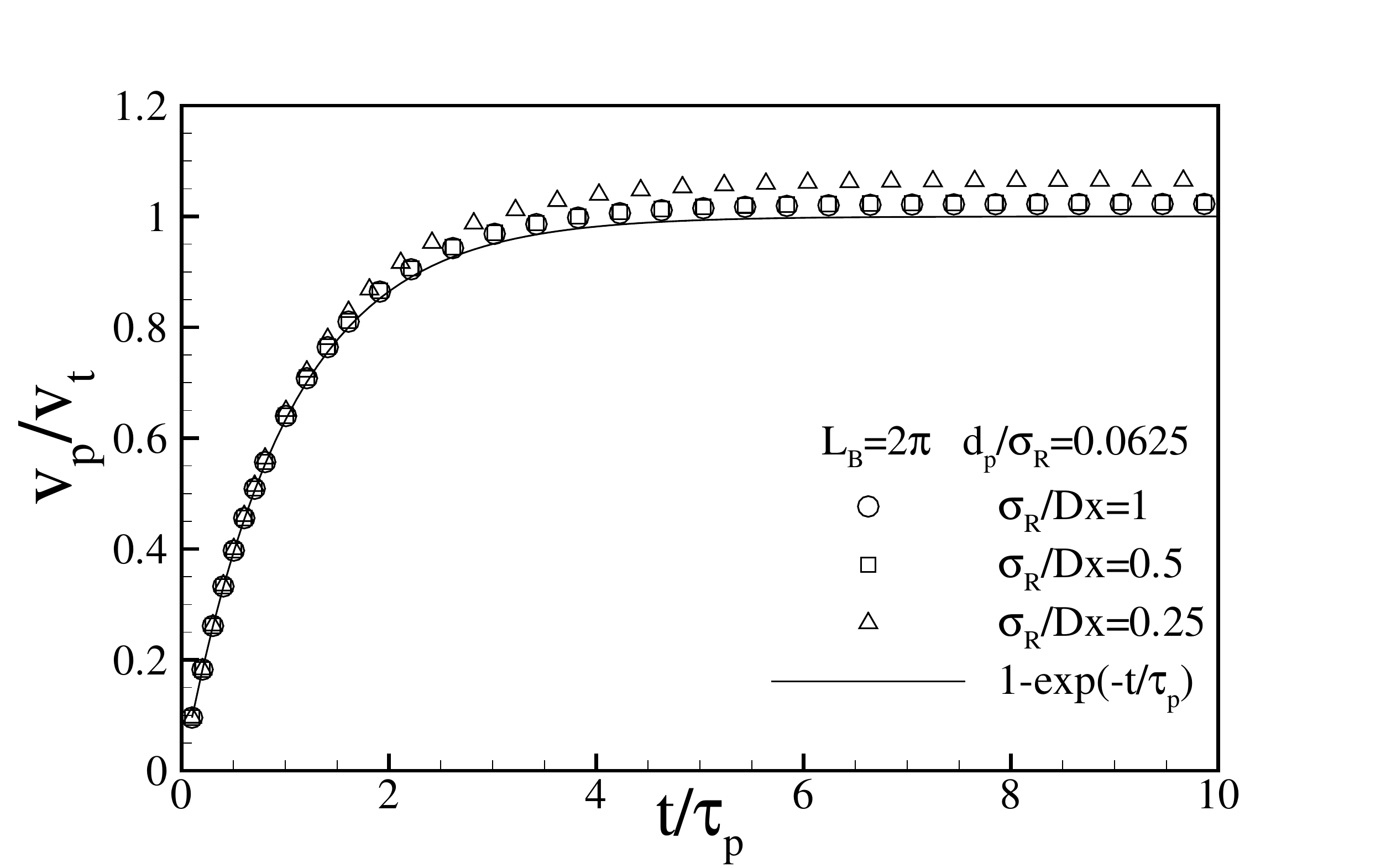}}
\caption{Normalized particle velocity $v_p/v_t$ as a function of the dimensionless
time $t/\tau_p$. The trajectories obtained in two unresolved cases, namely
$\sigma_R/Dx=0.5$ and $\sigma_R/Dx=0.25$ are compared against 
a resolved case at $\sigma_R/Dx=1$ for a given value of the ratio
$d_p/\sigma_R=0.0625$.
\label{fig:velo_unresolved}}
\end{figure}

A more direct comparison between the two approaches is provided in the
left panel of figure \ref{fig:velo_terminal} where we plot the particle velocity 
for the largest ratio $d_p/\sigma_R=0.5$ and the smallest one $d_p/\sigma_R=0.0625$
both for the ERPP and the PIC. For comparison in the ERPP calculation we 
have reported the particle velocity in a case where we did not subtract 
from $\vu(\vx_p,t)$ the self-induced disturbance. The right panel of the figure 
presents the relative error committed in the estimate of the terminal velocity 
as a function of $d_p/\sigma_R$.  Although the error scaling with grid resolution is comparable (see inset),
the error pertaining to the ERPP approach is substantially smaller.
A last issue concerns the sensitivity of the ERPP
method in poorly resolved cases where $\sigma_R < Dx$. In figure 
\ref{fig:velo_unresolved} we compare the particle velocity for 
three different resolutions at fixed  $d_p/\sigma_R$. As expected, the method loses accuracy as the 
regularization  kernel is not resolved on the computational grid. 
\section{Application to turbulent flows} \label{sec:turb_flow}

\begin{figure}
\begin{center}
\includegraphics[scale=0.35]{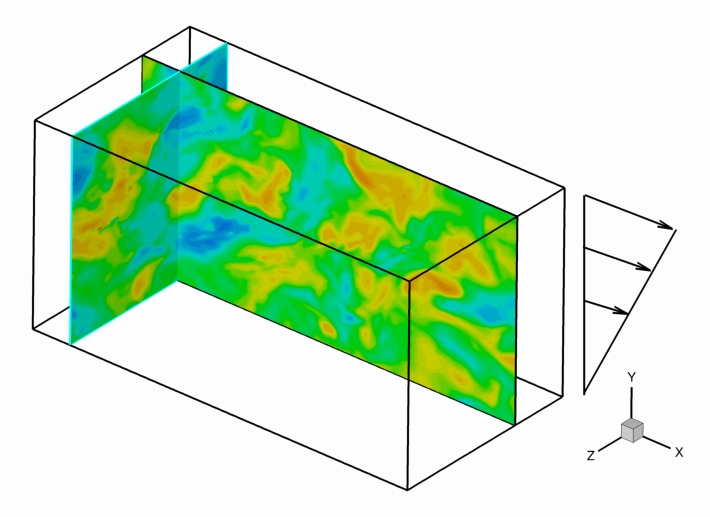}
\end{center}
\caption{Sketch of the flow configuration. The flow domain is represented by
a periodic box of length $L_x=4\pi$, $L_y=2\pi$ and $L_z=2\pi$ in the
steamwise, shear and span wise directions respectively. The mean flow $S y \ve_x$
is in the $x$-direction and linearly changes at a rate $S$ along the 
$y$-direction. The contour plot shows the intensity of the velocity fluctuations
in selected coordinate planes.
\label{fig:flow_config}}
\end{figure}

In order to discuss the feasibility of turbulent, particle-laden  flow simulations,
in this section we present preliminary results obtained by the ERPP method
for a homogeneous shear flow  at moderate  Reynolds number. The mean velocity profile 
in the $x$-direction (streamwise direction) is imposed and is characterized 
by a constant velocity gradient $S$ along the $y$-direction (shear direction). 
The third coordinate is denoted with $z$ (spanwise direction). The Reynolds 
decomposition $\vu = S y \ve_x + \vu^\prime$ allows to write the Navier-Stokes
equations for the turbulent fluctuating component $\vu^\prime$ which
are solved in a reference frame advected by the mean flow, 
see e.g. \cite{rogallo}. The Rogallo's transformation allows to restore
the spatial homogeneity of the fluctuations in the convected frame. A sketch 
of the flow domain is reported in figure \ref{fig:flow_config}.

In the homogeneous shear flow the turbulent fluctuations are sustained by the 
off-diagonal component of the Reynolds shear stress $- \langle u \, v\rangle$ 
resulting in a neat turbulent kinetic energy production rate 
${\cal P}=-S \langle u \, v \rangle$.
The large scale anisotropic forcing feeds the energy cascade operated by the
non-linear terms of the Navier-Stokes equations which eventually restore 
isotropy at smaller scales. The so called shear scale $L_S$ ideally
separates the production range $L_S < \ell < L_0$ ($L_0$ is the
integral scale), from the isotropy recovery range $\eta < \ell < L_S$.
It follows that the nature of turbulent fluctuations is parametrized
by two dimensionless parameter,
the shear intensity $S^*=\left(L_0/L_S\right)^{2/3}$ and the Corsin parameter
$S_c=\left(\eta/L_S\right)^{2/3}$. The latter can be recast in terms of the 
inverse of the classical turbulent Reynolds number $Re_\lambda$ based on the 
Taylor length-scale.

In the conditions discussed above the transport of inertial particles is non 
trivial. In fact, the disperse phase is characterized by small scales 
aggregates (clusters) which preserve a spatial preferential orientation induced 
by the large scale motions, up to the smallest scales where, in contrast, 
turbulent fluctuations recover isotropy, see e.g. \cite{jfm_part,jfm_2way}.
%
%

The first result concerns a flow at a Taylor based Reynolds number of
$Re_\lambda=60$ and shear intensity $S^*=7$. The carrier phase is resolved by 
using $N_x\times N_y\times N_z = 256 \times 256 \times 128$ Fourier modes in a 
$4\pi \times 2\pi \times 2\pi$ periodic box. Such spectral-based 
discretization corresponds to $384 \times 384 \times 192$ collocation
points in physical space due to the $3/2$ dealiasing procedure required for
the calculation of the non linear terms. The spatial discretization
fully resolves the Kolmogorov length scale with $Dx/\eta \sim 1.07$.
Time integration is performed
by the low-storage Runge-Kutta method already mentioned in section 
\S \ref{sec:validation}. The carrier fluid is laden with
$N_p=2200000$ inertial particles. The particle to fluid density ratio is 
$\rho_p/\rho_f=1800$ corresponding to a Stokes number 
$St_\eta=\tau_p/\tau_\eta=1$ where 
$\tau_p=\left(\rho_p/\rho_f\right) d_p^2/18 \nu$ is the Stokes relaxation 
time and $\tau_\eta$ is the Kolmogorov time scale. In such conditions
the particle diameter $d_p$ is much smaller than the Kolmogorov length,
namely $d_p/\eta=0.1$. The mass load $\Phi$ defined as the 
ratio between the mass of the disperse phase and the carrier fluid 
is $\Phi=0.4$. 

\begin{figure}
\begin{center}
\includegraphics[scale=0.45]{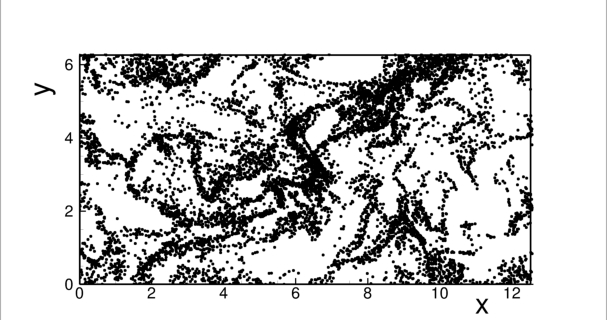} 
\includegraphics[scale=0.45]{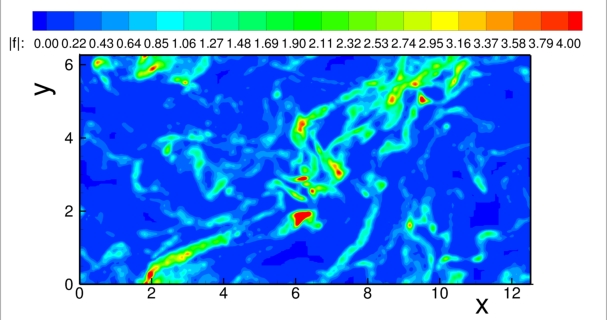} 
\includegraphics[scale=0.45]{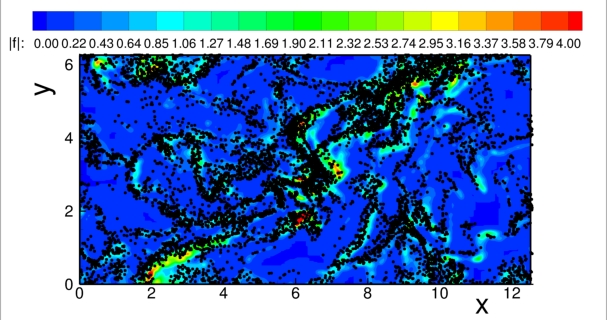}
\end{center}
\caption{
Snapshot of the instantaneous particle configuration (top) and corresponding 
intensity of the forcing on the fluid (center) in a thin slice along the 
$xy$ plane. The mean flow $S\, y$ is in the $x$ direction from left to right.
The two top panels are superimposed in the bottom panel to provide a visual 
correlation between the instantaneous particles configuration and the corresponding 
forcing field.
\label{fig:snapshots}}
\end{figure}

In figure \ref{fig:snapshots} we present a snapshot of the particle position
in a $xy$ plane containing the mean flow (from left to right). 
As expected, particles with unitary  Stokes number 
are characterized by small scale clusters, i.e. the particles concentrate in 
narrow regions, the clusters, separated by voids where neither a particle 
can be found. The preferential alignment of the aggregates
along the principal strain direction of the mean flow is also evident from 
the snapshot. This is the signature 
of the persistent anisotropy of the clusters at small scales.
In the context of the ERPP methodology we are able to compute
in a closed  form the forcing operated by the particles on the
fluid. In the middle panel of figure \ref{fig:snapshots} we report 
the intensity of the forcing term of equations
(\ref{eqn:ns_regularized}) which accounts for the back-reaction on the 
fluid. The pattern of the back-reaction field is strongly correlated to the 
cluster structure and inherit from the latter its characteristic 
multi-scale nature. The forcing is actually active in a broad range
of scales up to the smallest scales where intense peaks occur. Note 
however, that the forcing field is everywhere smooth and can be successfully 
represented on the discrete grid by virtue of the regularization naturally 
operated by the viscosity. The highest forcing intensity is 
localized in the spatial regions where the particles concentrate 
while in the void regions the forcing  vanishes. 
The correlation between the instantaneous particles spatial 
configuration and the corresponding back-reaction on the fluid can be 
visually appreciated in the bottom panel of the figure where the two fields are 
superimposed. 

As anticipated, this short paragraph is aimed at the clear illustration of 
the potential of the ERPP in dealing with
actual turbulent flows laden with millions of particles. Clearly a complete 
analysis of the turbulence modulation in the two-way coupling regime
would require  a more complete statistical analysis which is however beyond 
the intents of the present work and is postponed to future investigation.
\section{Final remarks}  \label{sec:final}
In this paper we have presented a new methodology, dubbed the ERPP method,  
able to capture the
inter-phase momentum exchange between a carrier flow and a disperse phase modeled
as lumped massive points. The coupling mechanism is designed on the physical ground
provided by the unsteady Stokes flow around a small sphere. In short, along its 
trajectory the particle continuously generates a highly localized vorticity field 
that can be evaluated in a closed form. Successively, because of viscous 
diffusion, the vorticity field reaches the physically significant length-scales of 
the flow field. When this occurs, the newly generated vorticity can be injected 
on the computational grid 
where the Navier-Stokes equations for the carrier flow are solved thus 
achieving the inter-phase momentum coupling. Under this respect
the viscous diffusion naturally regularizes the disturbance flow produced by
each particle without requiring any ``ad hoc'' numerical artifact. 
The proposed approach can be implemented in a highly efficient computational 
algorithm since the disturbance produced by a particle is strongly 
compact in space and localized around the actual particle position. 
This means that at each time step only few grid points perceive the particle 
disturbance which decays exponentially fast  in space. As a consequence the ERPP 
method can handle millions of particles at an affordable computational cost as 
proved by preliminary results of a particle laden turbulent homogeneous shear flow 
in the two-way coupling regime.

The ERPP overcomes several drawbacks of established methods, like the Particle In 
Cell method. Indeed, the regularization of the back-reaction field  provided by
the viscous diffusion allows numerically convergent solutions, preventing the 
strong grid-dependence which spoils singularity-based approximations.
Even more important, the  ERPP  method solves the intrinsic difficulty of 
numerical simulations
in the two-way coupling regime associated with the calculation of the correct 
particle-to-fluid slip velocity. Actually, in the ERPP method the disturbance flow
produced by the particles at each time step can be evaluated in a closed form.
This allows to remove from the particle-to-fluid slip velocity the
spurious self-induced velocity disturbance allowing for a correct evaluation of
the hydrodynamic force.

The preliminary results concerning a turbulent particle-laden shear flow presented 
in the last section demonstrate the potential of the ERPP method in the 
simulation of turbulent flows in the two-way coupling regime. It is known that the 
dynamics of the two phase system is fully characterized by a given set of 
dimensionless parameters, namely 
$\left\{Re_0, \, St_\eta, \, \rho_p/\rho_f, \, d_p/\eta, \, \Phi, \, N_p \right\}$. 
To comment on the effectiveness of the ERPP method in modeling turbulent suspensions,
let us assume that the turbulence characteristics are prescribed i.e. the 
the turbulent Reynolds number $Re_0$, the integral scale $L_0$, the
Kolmogorov scale $\eta$ or timescale $\tau_\eta$ are fixed, and let us 
consider small particles, i.e. $d_p/\eta \ll 1$. In these conditions
the Stokes number  $St_\eta=\displaystyle \tau_p / \tau_\eta$
controls the dynamics of the disperse phase in terms of its preferential spatial
accumulation, either small scale clustering in homogeneous 
flows 
\cite{toschi2009lagrangian,meneguz2011statistical,jfm_bec,monchaux2010preferential,reade2000effect} 
or turbophoresis in wall bounded flows \cite{pica,marsol}.
Once the Stokes number is fixed, the mass load $\Phi$ follows as 
\begin{equation}
\label{eqn:phi_constraint}
\Phi=\frac{\pi}{6} \, N_p \, \frac{\left(18 \, St_\eta\right)^{3/2}}
{\left(\rho_p/\rho_f\right)^{1/2} \, Re_0^{9/4}} \, .
\end{equation}
The value of $\Phi$ can be adjusted by means of the density ratio $\rho_p/\rho_f$
and the number of particles $N_p$. 
However, in an actual experiment the ratio $\rho_p/\rho_f$ must fall in the 
range of the 
available materials and the most straightforward way to achieve the desired mass 
load consists in adjusting the number of particles $N_p$. Although rather 
easy in experiments, adjusting the number of particles turns out to be a big 
issue in numerical simulations since, most often, the momentum coupling model 
is unable to handle an arbitrary number of particles while providing 
grid-independent and physically consistent results. At variance with most 
available methods, both these requirements are fulfilled by the ERPP approach.
Indeed, the number of  particles can be freely changed since disturbance flow 
and back-reaction of each particle are smooth fields. This implies that the 
solution is correctly  reproduced also in flow regions where the particles are 
extremely dilute, like it happens for the exterior region of
turbulent jets or spatially evolving boundary layers. 
For comparison, classical approaches like the PIC method intrinsically 
suffer of spurious numerical oscillations in the back-reaction field 
when too few particles per computational cell are available, leading to  
strong limitations in the achievable mass load $\Phi$.
Indeed, in any Direct Numerical Simulation of a turbulent flow the number of 
computational cells scales with the Reynolds number like $N_c \sim Re_0^{9/4}$ 
suggesting through equation (\ref{eqn:phi_constraint}) that no room is available  
to adjust the mass load if  the additional constraint $N_p/N_c \sim 1$
needs to be enforced.  This limitation is overcome in the approach proposed here 
by relaxing the request on the particle density to allow for the
modeling freedom needed to reproduce any physically relevant condition.
\section{Ackonwledgements}
The results of this research have been achieved using the 
PRACE-2IP project (FP7 RI-283493) resource {\em Zeus} based in Poland at 
Krakow. The authors are gratefull to the COST Action MP0806 Particles in 
Turbulence.
\appendix \section{} \label{app:appendix}
\subsection{Fundamental solution of the diffusion equation} 
\label{app:diffusion_eq}

The fundamental solution of the diffusion equation $g(\vx-\vxi,t-\tau)$ 
can be found by solving the following singularily forced diffusion problem
\begin{equation}
\label{eqn:green_diffusion}
\frac {\partial g}{\partial t} -\nu \nabla^2 g 
= \delta\left(\vx-\vxi\right) \delta(t-\tau),
\end{equation}
with $\lim_{t \rightarrow \tau^-} g(\vx-\vxi,t-\tau) = g(\vx-\vxi,0^-) = 0$
expressing the causality principle.
By integrating equation (\ref{eqn:green_diffusion}) in the interval 
$\left[\tau-\epsilon, \tau+\epsilon\right]$ and letting $\epsilon$ approach 
zero, the singularly forced diffusion equation is recast into an initial value 
problem
\begin{equation}
\label{eqn:green_initial}
\begin{array}{l}
\displaystyle \frac {\partial g}{\partial t} -\nu \nabla^2 g
= 0 \qquad t > \tau \\ \\
\displaystyle \lim_{t \rightarrow \tau^+} g(\vx-\vxi,t-\tau) = 
g(\vx-\vxi,0^+) = \delta(\vx-\vxi) \,  ,
\end{array}
\end{equation}
whose solution is immediate in Fourier space. By denoting with 
$\fg(\vk,t-\tau)$ the Fourier transform of $g(\vx-\vxi,t-\tau)$, 
equation (\ref{eqn:green_initial}) reads 
\begin{equation}
\label{eqn:green_iv_fourier}
\begin{array}{l}
\displaystyle \frac {\partial \fg}{\partial t} + \nu \lVert \vk \rVert^2 \fg 
= 0 \qquad t > \tau \\ \\
\displaystyle \lim_{t \rightarrow \tau^+} \fg(\vk,t-\tau) = \fg(\vk,0^+) = 
\frac{1}{(2 \pi)^3} \, .
\end{array}
\end{equation}
The solution in Fourier space is 
\begin{equation}
\label{eqn:fundamental_fourier}
\fg(\vk,t-\tau) =  \frac{1}{(2 \pi)^3}
\exp\left[-\nu \lVert \vk \rVert^2 (t-\tau)\right] \, ,
\end{equation}
which, after inverse Fourier transformation, yields the fundamental solution
\begin{equation}
\label{eqn:fundamental_phys}
g(\vx-\vxi,t-\tau) = \frac{1}{\left[ 4 \pi \, \nu (t-\tau)\right]^{3/2}}
\exp\left[-\frac{\lVert \vx -\vxi \rVert ^2}{4 \nu (t-\tau)}
\right] 
\end{equation}
as a Gaussian function with time dependent variance 
$\sigma(t-\tau) = \sqrt{2 \nu (t-\tau)}$.

\subsection{Fundamental solution of the unsteady Stokes equations}
\label{app:unsteady_stokeslet}
The fundamental solution of the unsteady Stokes operator can be found by
solving the singularly forced unsteady Stokes equations, namely
\begin{equation}
\label{eqn:stokes_singular}
\begin{array}{l}
\displaystyle \nabla \cdot \vv=0 \\ \\
\displaystyle \rho_f \frac{\partial \vv}{\partial t}
=-\nabla p +\mu \nabla^{2}{\vv} 
+\hat{\ve} \delta\left(\vx-\vxi\right)\delta(t-\tau) \\ \\
\lim_{t \rightarrow \tau^-} \vv(\vx-\vxi,t-\tau) = \vv(\vx-\vxi,0^-) = 0
\end{array}
\end{equation}
where the sigular forcing $\delta\left(\vx-\vxi\right)\delta(t-\tau)$ is 
applied at the point $\vx=\vxi$ at time $t=\tau$ along the direction 
$\hat{\ve}$. The solution of equations (\ref{eqn:stokes_singular}) is more 
easily found in terms of the associtaed vorticity field 
$\vzeta=\nabla \times \vv$. By taking the curl of equations 
(\ref{eqn:stokes_singular}) it follows
\begin{equation}
\label{eqn:vorticity_singular}
\rho_f \frac{\partial \vzeta}{\partial t}
=\mu \nabla^{2}{\vzeta} 
-\hat{\ve} \times \nabla \delta\left(\vx-\vxi\right)\delta(t-\tau)
\end{equation}
with the corresponding initial condition $\vzeta(\vx-\vxi,0^-) = 0$.
Equation (\ref{eqn:vorticity_singular}) can be reconducted to the standard
scalar diffusion equation (\ref{eqn:green_diffusion}) by the ansatz
\begin{equation}
\label{eqn:vorticity_solution}
\vzeta = - \frac{1}{\rho_f} \hat{\ve} \times \nabla g.
\end{equation}
Equations (\ref{eqn:vorticity_solution}) and (\ref{eqn:fundamental_phys})
provides the solution of the singularly forced unsteady Stokes problem in terms
of vorticity. The solution in terms of velocity can be found by introducing
a divergence free vector potential $\vA$, namely $\vv = \nabla \times \vA$
and $\nabla^2 \vA = - \vzeta$. The Laplace equation for the vector potential 
can be transformed into a scalar equation by looking for solutions for the vector 
potential in the form
\begin{equation}
\label{eqn:A_solution}
\vA = \frac{1}{\rho_f} \hat{\ve} \times \nabla G \, ,
\end{equation}
where the scalar function $G$ satisfy the standard Laplace equation 
$\nabla^2 G = g$. The solution for $G$ reads
\begin{equation}
\label{eqn:G_solution}
G = -\frac{1}{4 \pi r} \mbox{erf}\left( \frac{r}{\sqrt{4 \nu (t-\tau)}}\right)
\end{equation}
where $r = \lVert \vx - \vxi \rVert$. The velocity field can be readily detemined
by substituting the expressions (\ref{eqn:A_solution}) and (\ref{eqn:G_solution})
into $\vv=\nabla\times \vA$. After some algebra the velocity reads
\begin{equation}
\label{eqn:velo_solution}
\vv=\left( g\, \vI - \nabla \otimes \nabla G \right)\hat{\ve} \, .
\end{equation}
The solution (\ref{eqn:velo_solution}) is usually written in terms of the
Green tensor $G_{ik}(\vx-\vxi,t-\tau)$. In fact by using 
(\ref{eqn:fundamental_phys}) and (\ref{eqn:G_solution}), equation 
(\ref{eqn:velo_solution}) can be written as
\begin{equation}
\label{eqn:velo_green}
v_i = G_{ik} \, \hat{e}_k
\end{equation}
where the Green tensor is given by the expresion
\begin{equation}
\label{eqn:green_tensor}
G_{ik}(\vx-\vxi,t-\tau)=\frac{1}{\rho_f}\left[ 
\left( 1 +\frac{\sigma^2}{r^2} \right) g
+\frac{G}{r^2} \right] \delta_{ik}
-\frac{1}{\rho_f}\left[ \left(1+ \frac{3\sigma^2}{r^2}\right) g
+\frac{3G}{r^2} \right] \frac{r_i r_k}{r^2} \, .
\end{equation}
The solution of the singularly forced Stokes problem is completed by the
expression of the pressure and of the viscous stressed. 
The pressure field associeted to the original problem
(\ref{eqn:stokes_singular}) can be computed by taking the divergence of the 
momentum equation, namely 
\begin{equation}
\label{eqn:p_unsteady_stokes}
\nabla^2 p =
\hat{\ve} \cdot \nabla \delta\left(\vx-\vxi\right)\delta(t-\tau) \, .
\end{equation}
The Laplace equation for the pressure can be readily solved by the substitution
$p=\hat{\ve} \cdot \nabla q \, \, \delta(t-\tau)$. In fact, the function $q$ 
satisfies the Laplace problem $\nabla^2 q = \delta\left(\vx-\vxi\right)$, i.e.
$q=-\frac{1}{4 \pi r}$. The pressure field then follows at once as
\begin{equation}
\label{eqn:p_solution}
p =\frac{\hat{\ve} \cdot \vr}{4 \pi r^3} \delta(t-\tau) \,.
\end{equation}
The stress tensor associated to the singularly forced unsteady Stokes problem
can be computed as
\begin{equation}
\label{eqn:stress_def}
T_{ij} = -p \delta_{ij} +\mu \left( \frac{\partial v_i}{\partial x_j}
+ \frac{\partial v_j}{\partial x_i} \right)
\end{equation}
where the pressure $p$ is given by (\ref{eqn:p_solution}) and the velocity
$\vv$ by (\ref{eqn:velo_green}). Usually the stress tensor is expressed via
a third rank tensor
\begin{equation}
\label{eqn:stress_tensor}
T_{ij} = {\cal T}_{ijk}\hat{e}_k
\end{equation}
where ${\cal T}_{ijk}$ is the Green stress tensor defined as
\begin{equation}
\label{eqn:stress_tensor_def}
{\cal T}_{ijk} = -\frac{r_k}{4 \pi r^3} \delta(t-\tau) \delta_{ij} +
\mu \left( \frac{\partial G_{ik}}{\partial x_j}
+\frac{\partial G_{jk}}{\partial x_i} \right)\, .
\end{equation}
The expression of the Green tensor (\ref{eqn:green_tensor}) can be substituted
into the definition (\ref{eqn:stress_tensor_def}) and, after some algebra,
the final expression for ${\cal T}_{ijk}$ reads
\begin{eqnarray}
\label{eqn:stress_tensor_final}
{\cal T}_{ijk} &=& -\frac{r_k}{4 \pi r^3} \delta(t-\tau) \delta_{ij} \nonumber \\
&+& \nu \left[ -2 \frac{B}{r^2} \delta_{ij} r_k + 
\left(\frac{1}{r} \frac{dA}{dr} -\frac{B}{r^2}\right)
\left(\delta_{jk} r_i + \delta_{ik} r_j \right)
-2 \left( \frac{2}{r} B + \frac{dB}{dr}\right) \frac{r_i r_j r_k}{r^3}
\right]
\end{eqnarray}
where the functions $A(r)$ and $B(r)$ are defined as
\begin{equation}
\label{eqn:A_B_def}
\begin{array}{l}
\displaystyle A(r) = \left( 1 +\frac{\sigma^2}{r^2} \right) g +\frac{G}{r^2} \\ \\
\displaystyle B(r) = \left(1 +\frac{3\sigma^2}{r^2} \right) g +\frac{3}{r^2} G
\end{array}
\end{equation}
\subsection{Evolution of the regular vorticity field, 
proof of equation (\ref{eqn:pde_vort_regular})}
\label{app:pde_vort_regular}

Let's first differentiate equation (\ref{eqn:vort_regular}) with respect to time,
\begin{eqnarray}
\label{eqn:eq23_bis}
\frac{\partial \vzeta_R}{\partial t} &=& \frac{1}{\rho_f}
\vD_p(t-\epsilon_R) \times \nabla_x g\left[\vx -\vx_p(t-\epsilon_R), 
\epsilon_R \right] 
\nonumber \\
&+&
\frac{1}{\rho_f} \int_0^{t-\epsilon_R} \vD_p(\tau) \times \nabla_x 
\frac{\partial g }{\partial t} 
\left[ \vx - \vx_p(\tau),t-\tau \right] d\tau \, , 
\end{eqnarray}
take the Laplacian,
\begin{eqnarray}
\label{eqn:eq23_tris}
\nabla^2 \vzeta_R = \frac{1}{\rho_f}
\int_0^{t-\epsilon_R} \vD_p(\tau) \times \nabla_x \nabla^2 g 
\left[ \vx - \vx_p(\tau),t-\tau \right]  d\tau \, ,
\end{eqnarray}
and recombine the two results with the kinematic viscosity yielding
\begin{eqnarray}
\label{eqn:eq23_quat}
\frac{\partial \vzeta_R}{\partial t} - \nu  \nabla^2 \vzeta_R &=& 
\frac{1}{\rho_f} \vD_p(t-\epsilon_R) \times \nabla_x g
\left[\vx -\vx_p(t-\epsilon_R), \epsilon_R \right] 
\nonumber \\
&+&
\frac{1}{\rho_f} \int_0^{t-\epsilon_R} \vD_p(\tau) \times \nabla_x \left\{ 
\frac{\partial g }{\partial t}  -\nu \nabla^2 g \right\}
\left[ \vx-\vx_p(\tau),t-\tau \right] d \tau 
\nonumber \\
&=&
\frac{1}{\rho_f} \vD_p(t-\epsilon_R) \times \nabla_x 
g\left[\vx -\vx_p(t-\epsilon_R), \epsilon_R \right] 
\nonumber \\
&+&
\frac{1}{\rho_f} \int_0^{t-\epsilon_R} \vD_p(\tau) \times \nabla_x 
\delta\left[ \vx-\vx_p(\tau) \right] \delta(t-\tau) d \tau 
\nonumber  \\
&=&
\frac{1}{\rho_f} \vD_p(t-\epsilon_R) \times \nabla_x 
g\left[\vx -\vx_p(t-\epsilon_R), \epsilon_R \right]  \ .
\nonumber 
\end{eqnarray}
\subsection{Singular part of the velocity field}
\label{app:singular_velocity}
\begin{figure}
\begin{center}
\includegraphics[scale=0.35,angle=-90]{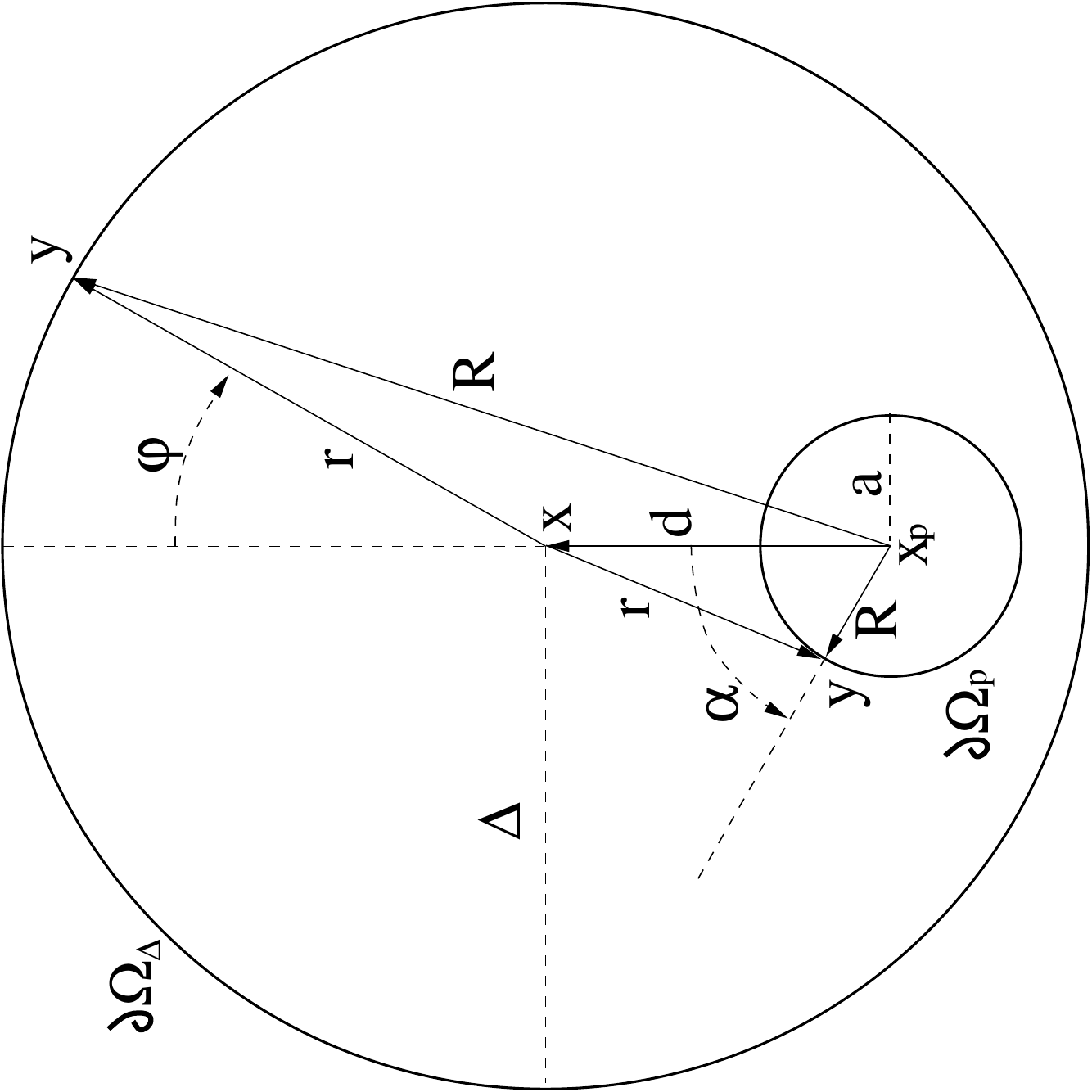} \hspace*{0.5cm}
\includegraphics[scale=0.35,angle=-90]{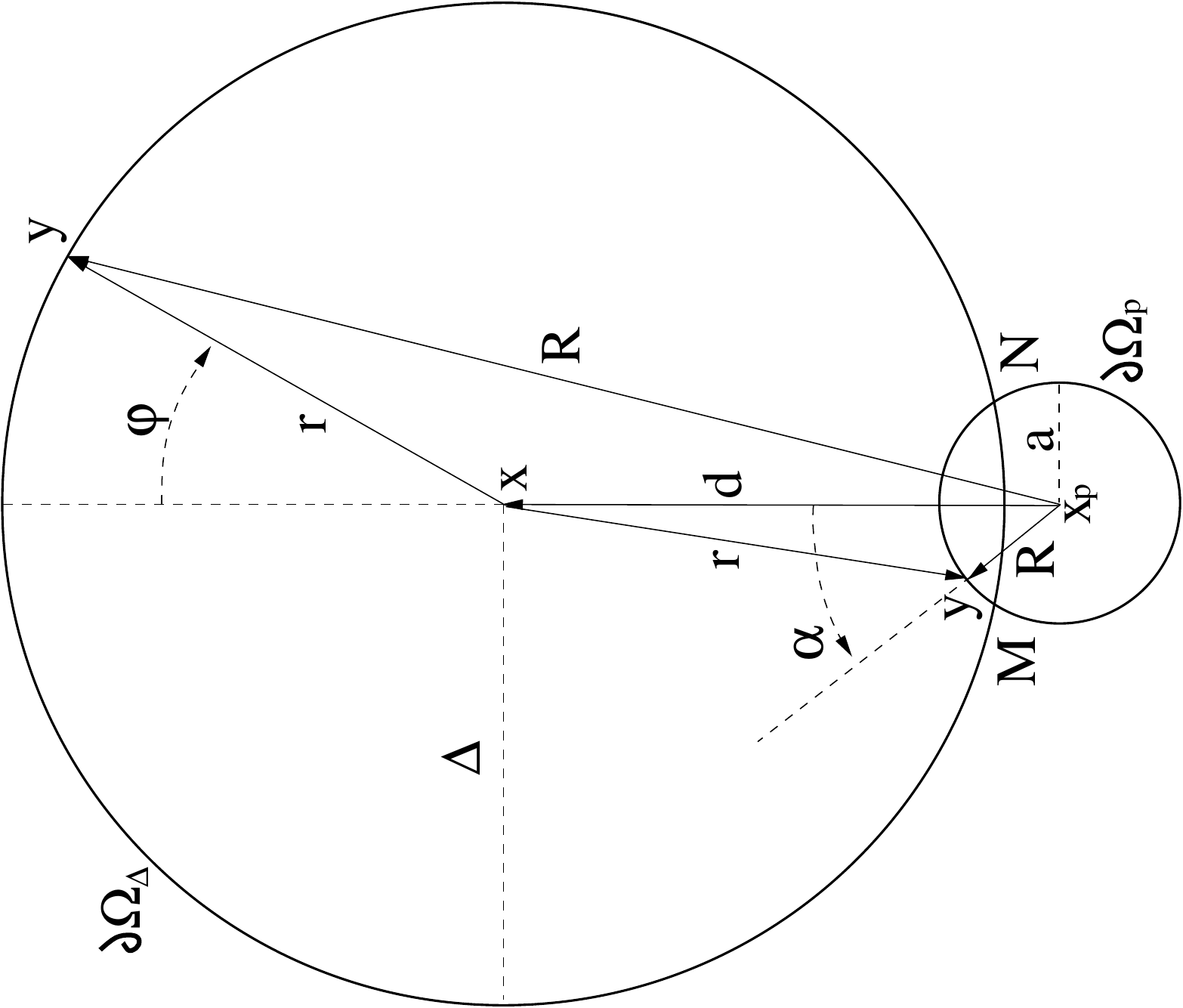} 
\end{center}
\caption{\label{fig:sketch_palle} 
Sketch of the coarse-graining procedure in the neigbourhood of the Eulerian point
$\vx$. The boundary $\partial \Omega_\Delta$ denotes the fileter kernel of width
$\Delta$ and $\partial \Omega_p$ is the particle spherical surface of
radius $a<\Delta$. Two cases are possible. The particle entirely lays inside the
region where the filter kernel is non vanisshing (left), or the particle lays 
partially ouside the filter radius $\Delta$.
}
\end{figure}

The contribution of the singular component of the velocity disturbance $\vv_S$ 
due to the particles can be estimated starting from the expression of the 
associated singular vorticity field given in equation (\ref{eqn:vort_singular}) 
that we report here for convenience, namely
\begin{eqnarray}
\label{eqn:vort_singular_def}
\vzeta_S(\vx,t) = \frac{1}{\rho_f} \int_{t - \epsilon_R}^{t^+}
\vD_p(\tau) \times \nabla g\left[ \vx-\vx_p(\tau),t-\tau\right] d\tau \ .
\end{eqnarray}
%
The time integral for small values of $\epsilon_R$ can be approximated as
\begin{eqnarray}
\label{eqn:vort_singular_approx}
\vzeta_S(\vx,t) = \frac{\vD_p^*}{\rho_f} \times \nabla \int_{t - \epsilon_R}^{t^+}
g\left[ \vx-\vx_p(\tau^*),t-\tau\right] d\tau \ .
\end{eqnarray}
where $\displaystyle \vD_p^*=\sup_{t-\epsilon_R < \tau < t^+} \vD_p(\tau)$ and $\tau^*$ 
is the time corresponding to the minimum distance between the actual particle 
position $\vx_p(\tau^*)$ and the point $\vx$. The time integral in equation 
(\ref{eqn:vort_singular_approx}) can be explicitly computed 
leading to the following expression of the singular vorticity field
\begin{equation}
\label{eqn:vort_singular_final}
\vzeta_S=\vD_p^* \times \nabla H \qquad \mbox{with} \qquad
H=\frac{1}{4\pi\mu r}\left[1-\mbox{erf} \left(\frac{r}{\sqrt{2} \sigma_R}\right) 
\right] \,, 
\end{equation}
where $r=\lvert \vx - \vx_p(\tau^*)\rvert$.
Given the vorticity, the corresponding velocity can be found in terms of the 
associated divergence free vector potential, namely $\vv_S=\nabla \times \vA_S$, by solving the 
Poisson problem $\nabla^2 \vA_S=-\vzeta_S$. The solution can be found in the
form $\vA_S=-\vD_p^* \times \nabla \psi$ where $\psi$ is the solution of the 
scalar problem $\nabla^2 \psi = H$. The singular velocity field is then expressed as
\begin{eqnarray}
\label{eqn:velo_singular_compact}
\vv_S(\vx,t) = \left( \nabla \otimes \nabla\psi -\nabla^2 \psi \right) \vD_p^*\,
\end{eqnarray}
and $\psi$ is given by 
\begin{eqnarray}
\label{eqn:psi_solution}
\psi = \frac{\sqrt{2} \sigma_R}{8 \pi \mu}
\left[ \eta-\eta \mbox{erf}\left(\eta\right)-\frac{1}{2\eta}\mbox{erf}
\left(\eta\right)
-\frac{1}{\sqrt{\pi}}\exp{\left(-\eta^2\right)}\right] \,,
\end{eqnarray}
in terms of the dimensionless variable $\eta=r/\sqrt{2} \sigma_R$.
The explicit expression of $\vv_S$ is only a matter of successive derivation 
of expression (\ref{eqn:psi_solution}). After calculations, the 
singular velocity field can be finally expressed as
\begin{eqnarray}
\label{eqn:velo_singular_explicit}
\vv^S_i(\vx,t) = \frac{{D^p_j}^*}{8 \pi \mu}\left[ 
\left( \frac{\partial^2 \psi}{\partial \eta^2}-
\frac{1}{\eta}\frac{\partial \psi}{\partial \eta}\right) \frac{r_i r_j}{r^2}
-\left( \frac{\partial^2 \psi}{\partial \eta^2}+
\frac{1}{\eta}\frac{\partial \psi}{\partial \eta}\right) \delta_{ij}
\right] \, .
\end{eqnarray}
Expression (\ref{eqn:velo_singular_explicit}) is amenable of further manipulation to
extract the near field behavior of the singular velocity field, i.e. the
expression of $\vv_S$ in the limit of $\eta \to 0$. In fact, for small values of 
$\eta$, the error function which appears in $\psi$ and in its first and second 
derivatives can be expanded in McLaurin series. After some algebra, equation 
(\ref{eqn:velo_singular_explicit}) can be recasted in the form
\begin{eqnarray}
\label{eqn:velo_singular_close}
\vv^S_i(\vx,t) = -\frac{{D^p_j}^*}{8 \pi \mu r}\left(\delta_{ij} +\frac{r_i r_j}{r^2}
\right) \, ,
\end{eqnarray}
which express the behavior of the singular field for small distances $r$ from the 
particle when compared with the diffusion length-scale $\sigma_R$.
From equation (\ref{eqn:velo_singular_close}) it appears that the singular
velocity field still presents a singularity which diverges as $1/r$ in the
neighborhood of the actual particle position $\vx_p$. In principle, the singular
velocity field gives a finite contribution to the convective terms of the 
Navier-Stokes equations, see e.g. equations (\ref{eqn:ns_regularized}).
By coarse graining the equations on a scale $\Delta$, small with respect to the hydrodynamic scale but larger than the particle size
one can show that the contribution arising from $\vv_S \cdot \nabla \vv_S$, $\vv_S \cdot \nabla \vu_R$ and $\vu_R \cdot \nabla \vv_S$
are negligibly small and can be neglected. 
In performing the coarse graining, one has to consider that the convolution integral should be performed in the region occupied by the fluid, i.e.
outside the particles. For instance, let us refer to the sketch in figure \ref{fig:sketch_palle} where
the particle placed at $\vx_p$ induces a  velocity disturbance in $\vy$
and the coarse grained velocity field is evaluated at point $\vx$.
In the relative position of the particle with respect to the point $\vx$
we will discuss two typical cases. The particle can lay entirely inside the 
region where the filter kernel is non vanishing or it can partially lay outside 
the filter kernel radius. Lastly the particle might lay completely outside the 
filter kernel. In all cases the coarse grained advective terms 
at point $\vx$ can be computed as a convolution of the relevant part of the 
convective term with a filter Kernel $K$, e.g.
\begin{eqnarray}
\label{eqn:advective_filtered}
\vh_{SS}(\vx,t) = \int_{\Omega_\Delta \backslash \Omega_p} \vv_S(\vy,t) 
\cdot \nabla_{\vy} \vv_S(\vy,t)\,  K(\vy-\vx) \, d^3\vy
\end{eqnarray}
where the integration variable $\vy$ belongs to the domain 
$\Omega_\Delta \backslash \Omega_p$ which is the complement to the support 
$\Omega_\Delta$ of the filter of the region $\Omega_p$ occupied by the particle.
By assuming, for the sake of definiteness, a top-hat kernel, 
\begin{equation}
\label{eqn:top_hat_filter}
K(\vy-\vx)= \left\{
\begin{array}{l}
\displaystyle \frac{1}{\Delta^3} \qquad \lvert \vy-\vx \rvert < \Delta \\ \\
\displaystyle 0 \qquad \lvert \vy-\vx \rvert > \Delta \, ,
\end{array}
\right.
\end{equation}
the convolution integral (\ref{eqn:advective_filtered}) can be transformed by
incompressibility of the field $\vv_S$ into a surface integral
\begin{eqnarray}
\label{eqn:advective_filtered_gauss}
\vh_{SS}(\vx,t) = \frac{1}{\Delta^3}
\int_{\partial(\Omega_\Delta \backslash \Omega_p)} \vv_S \, 
\left( \vv_S \cdot \vn \right)  \, dS_{\vy} \, .
\end{eqnarray}
In expression (\ref{eqn:advective_filtered_gauss}) the integration point $\vy$
runs on $\partial \Omega_\Delta \cup \partial \Omega_p$ when the 
particle lays entirely within the filter width, right panel of figure 
\ref{fig:sketch_palle}. In the other case when the particle partially intercepts the
filter boundary, see the right sketch of the figure, the point $\vy$ runs on $\left(\partial \Omega_p \cap \Omega_\Delta \right) \cup
\partial \Omega_\Delta \backslash \left(\partial \Omega_\Delta \cap \Omega_p \right)$.
For detailed calculations it is convenient to define the vectors
$\vr=\vy-\vx$, $\vR=\vy-\vx_p$ and $\vd=\vx-\vx_p$. In fact, when 
$\vy \in \partial \Omega_\Delta$ the integration in 
(\ref{eqn:advective_filtered_gauss}) is better evaluated in terms of the $\vr$ 
variable while for $\vy \in \partial \Omega_p$ the use of the integration 
variable $\vR$ ease the calculations. In particular for 
$\vy \in \partial \Omega_\Delta$ the outward positive normal is $\vn=\hat{\vr}$ 
where the hat denote $\hat{\vr}=\vr/r$ while for $\vy \in \partial \Omega_p$
the positive normal is $\vn = -\hat{\vR}$. Note that in the new variables the 
singular velocity field depends on $\vR$, $\vv_S(\vR,t)$ with $\vR=\vd+\vr$. 
By exploiting the expression (\ref{eqn:velo_singular_close}) for $\vv_S$, 
the integrand function in equation (\ref{eqn:advective_filtered_gauss})
for $\vy \in \partial \Omega_\Delta$ is
\begin{eqnarray}
\label{eqn:int_y_filter}
\vv_S \left(\vv_S \cdot \vn \right) &=& \frac{1}{\left(8 \pi \mu R\right)^2}
\left\{\left[
(\vD_p^* \cdot \hat{\vr})+(\vD_p^* \cdot \hat{\vR})(\hat{\vR}\cdot\hat{\vr}) 
\right] \vD_p^* \right. \nonumber  \\
&+& \left. (\vD_p^* \cdot \hat{\vR}) \left[ 
(\vD_p^* \cdot \hat{\vr}) +(\vD_p^* \cdot \hat{\vR}) (\hat{\vR}\cdot\hat{\vr}) 
\right] \hat{\vR} \right\}
\end{eqnarray}
or 
\begin{eqnarray}
\label{eqn:int_y_particle}
\vv_S \left(\vv_S \cdot \vn \right) = \frac{2}{\left(8 \pi \mu R\right)^2}
(\vD_p^* \cdot \hat{\vR}) \left[ \vD_p^* + (\vD_p^* \cdot \hat{\vR}) \hat{\vR} \right]
\end{eqnarray}
when $\vy \in \partial \Omega_p$. 

Let us discuss the case when the particle is entirely within the filter kernel, 
see left panel of figure \ref{fig:sketch_palle}. The contributions to the surface 
integral coming from
$\vy \in \partial \Omega_p$ identically vanishes while the contribution from
$\vy \in \partial \Omega_\Delta$ in equation (\ref{eqn:advective_filtered_gauss}) 
can be explicitly calculated by using a system of spherical coordinates centered in $\vx$, 
i.e. by integrating with respect to $\vr$ the expression 
(\ref{eqn:int_y_filter}). After tedious but straightforward calculations it can be proved that each
term arising from (\ref{eqn:int_y_filter}) gives a finite contribution to the
integral thus providing the following estimate for $\vh_{SS}$, namely
\begin{eqnarray}
\label{eqn:h_SS_estimate}
\vh_{SS} \sim \frac{\lvert \vD_p^* \rvert^2}{\mu^2 \Delta^3} 
f\left(\frac{d}{\Delta}\right)
\end{eqnarray}
where $f\left(d/\Delta\right)$ is a regular function of the ratio 
$d/\Delta$, with $0\le d/\Delta<1-a/\Delta$.

 The same conclusion holds when the calculations are 
repeated for the case when the particle intercepts the filter boundary, 
left panel of figure \ref{fig:sketch_palle}. The mixed advective terms
$\vv_S\cdot\nabla \vu_R$ and $\vu_R\cdot\nabla\vv_S$ can be calculated 
by means of the same procedure assuming that the regular contribution
$\vu_R$ and its gradients are constant on the filter length-scale $\Delta$.
In such conditions, the corresponding coarse-grained contributions scale as
\begin{eqnarray}
\label{eqn:h_mixed_estimate}
\vh_{RS}\sim \frac{\lvert \vD_p^* \rvert^2}{\mu^2 \sigma_R \Delta^2} \, \quad
\vh_{SR}\sim \frac{\lvert \vD_p^* \rvert^2}{\mu^2 \sigma^2_R \Delta} \,.
\end{eqnarray}
The estimates (\ref{eqn:h_SS_estimate}) and (\ref{eqn:h_mixed_estimate}) can be
used to compare the order of magnitude of the advective terms against the order of
magnitude of the feedback term in equations
(\ref{eqn:ns_regularized}) which scales as
\begin{eqnarray}
\label{eqn:forcing_estimate}
\frac{\vD_p(t-\epsilon_R)}{\rho_f} g\left[\vx-\vx_p(t-\epsilon_R),\epsilon_R\right] 
\sim \frac{\lvert \vD_p \rvert}{\rho_f \sigma_R^3}\, .
\end{eqnarray}
From the above estimates, it follows
\begin{eqnarray}
\label{eqn:h_estimate}
\frac{\mathcal{O}\left(\vh_{SS}\right)}
{\mathcal{O}\left(\frac{\vD_p}{\rho_f}g\right)} \sim 
Re_p \left(\frac{\sigma_R}{\Delta}\right)^3 
\qquad
\frac{\mathcal{O}\left(\vh_{RS}; \vh_{SR} \right)}
{\mathcal{O}\left(\frac{\vD_p}{\rho_f}g\right)} \sim 
Re_p \left(\frac{\sigma_R}{\Delta}\right)^2 \, ,
\end{eqnarray}
where $Re_p$ is the particle Reynolds number calculated by using the 
particle radius $a$, the particle-to-fluid slip velocity $w_{rel}$ and the
kinematic viscosity $\nu$. It follows that all these terms can be neglected in 
the limit of small particle Reynolds number.

Let us give an example of the detailed calculations of the integral 
(\ref{eqn:advective_filtered_gauss}) in the two cases reported in figure
\ref{fig:sketch_palle}. We first address the case reported in the left 
panel of the figure where the particle is entirely inside the filter kernel
hence the integral is splitted on $\partial \Omega_\Delta$ where the integrand
function is given by equation (\ref{eqn:int_y_filter}) and $\partial \Omega_p$
where the expression (\ref{eqn:int_y_particle}) must be adopted. Let us first 
discuss
the integration of equation (\ref{eqn:int_y_filter}) on $\partial \Omega_\Delta$.
For convenience we fix the polar axis along the direction $\ve_3$ such that
$r_1=r \sin{\phi}\, \cos{\theta},\, r_2=r \sin{\phi}\, \sin{\theta},\, r_3=r \cos{\phi}$.
Due to the symmetries of the problem only the terms involving the contributions
from $r_3, R_3, R_1^2 R_3, R_2^2 R_3, R_3^3$ give a contribution to the integral.
In the following we address the term $(\vD_p^*\cdot \hat{\vr})\vD_p^*$ wich reduces
to
\begin{eqnarray}
\label{eqn:int_1}
\frac{1}{\Delta^3} \frac{\vD_p^* {D_p^3}^*}{(8 \pi \mu)^2}
\int_{\partial \Omega_\Delta} \frac{\hat{r_3}}{R^2} \Delta^2 \sin{\phi} d\phi d\theta
\end{eqnarray}
where $R=\sqrt{\Delta^2+d^2+2d\Delta \cos{\phi}}$. By defining $b=d/\Delta$ we get
\begin{eqnarray}
\label{eqn:int_2}
\frac{2 \pi}{\Delta^3} \frac{\vD_p^* {D_p^3}^*}{(8 \pi \mu)^2}
\int_{-1}^1 \frac{\xi}{1+b^2+2b\xi} d\xi
\end{eqnarray}
and
\begin{eqnarray}
\label{eqn:int_3}
\frac{2 \pi}{\Delta^3} \frac{\vD_p^* {D_p^3}^*}{(8 \pi \mu)^2}
\frac{1}{4b^2}\left[4b+2(b^2+1) \ln{\frac{|b-1|}{|b+1|}} \right].
\end{eqnarray}
The above expression holds for  $0\le b<1-a/\Delta$ and is apparently singular
for $b\to0$. However for small values of $b$ we have 
$\ln{|b-1|}=\ln{(1-b)}\simeq -b$ and $\ln{(1+b)}\simeq b$. It follows that the
term in square brackets in expression (\ref{eqn:int_3}) goes like $-4b^3$ and
equation (\ref{eqn:int_3}) vanishes for $d/\Delta\to0$. A term which
gives a finite contribution the expression (\ref{eqn:advective_filtered_gauss})
is indeed given by 
$(\vD_p^*\cdot \hat{\vR})(\hat{\vR}\cdot \hat{\vr})\vD_p^*$ wich reduces
to ${D_p^3}^* \hat{R_3} (\hat{\vR}\cdot \hat{\vr})$. Hence the expression is 
transformed into
\begin{eqnarray}
\label{eqn:int_4}
\frac{1}{\Delta^3} \frac{\vD_p^* {D_p^3}^*}{(8 \pi \mu)^2}
\int_{\partial \Omega_\Delta} \frac{\hat{R_3} (\hat{\vR}\cdot \hat{\vr})}{R^2} 
\Delta^2 \sin{\phi} d\phi d\theta
\end{eqnarray}
where $(\hat{\vR}\cdot \hat{\vr})=(1+b\cos{\phi})/\sqrt{1+b^2+2b\cos{\phi}}$.
After substitution we get the integral
\begin{eqnarray}
\label{eqn:int_5}
\frac{2 \pi}{\Delta^3} \frac{\vD_p^* {D_p^3}^*}{(8 \pi \mu)^2}
\int_{-1}^1 \frac{(1+b\xi)^2}{(1+b^2+2b\xi)^2} d\xi\, ,
\end{eqnarray}
which can be integrated providing the following expression
\begin{eqnarray}
\label{eqn:int_6}
\frac{2 \pi}{\Delta^3} \frac{\vD_p^* {D_p^3}^*}{(8 \pi \mu)^2}
\frac{1}{4b} \left[4b+2(1-b^2)\ln\frac{|b+1|}{|b-1|} \right]\,.
\end{eqnarray}
For small values of $b$ we have $\ln{|b+1|}-\ln{|b-1|} \simeq 2b$ hence the term
in square bracket goes like $8b-4b^2$ resulting in a finite limit of expression 
(\ref{eqn:int_6}). Let us know discuss the integration of the field given by 
(\ref{eqn:int_y_particle}) on $\partial \Omega_p$. The calculation is 
strightforward when the integral is computed with respect the variables $\vR$. 
In fact, it can be prooved that each contribution arising from expression 
(\ref{eqn:int_y_particle}) vanishes in agreement with the fact that the 
field $\vv_S$ is spherically simmetric with respect to the natural variable $\vR$.
We complete the discussion by discussing the integral
(\ref{eqn:advective_filtered_gauss}) when the particle partially intersect the filter 
boundary, see the right panel of figure \ref{fig:sketch_palle}.
In this case the field (\ref{eqn:int_y_filter}) must be used on 
$\partial \Omega_\Delta \backslash \left(\partial \Omega_\Delta \cap \Omega_p \right)$
and the expression (\ref{eqn:int_y_particle}) on 
$\left(\partial \Omega_p \cap \Omega_\Delta \right)$. The same calculations reported above
can be easily repeated by taking into account that the angle $\phi$ or $\alpha$
assume values in $[0:\phi_{max}]$ and $[0:\alpha_{max}]$ respectively and that the ratio
$b=d/\Delta>1-a/\Delta$. Such limitations exclude any singular behaviors of the integrals
both on $\partial \Omega_\Delta \backslash \left(\partial \Omega_\Delta \cap \Omega_p \right)$ and $\left(\partial \Omega_p \cap \Omega_\Delta \right)$.
\subsection{Evaluation of the self-induced disturbance flow} 
\label{app:self_velocity}
The self-disturbance flow produced by the $p$th particle in a generic time step
$t_n\rightarrow t_{n+1}$ can be evaluated by integrating the complete
equation for the disturbance field, namely equation (\ref{eqn:velo_reg_eq}), that
we report below in a slighly different notation where the subscript $_R$ is omitted
\begin{eqnarray}
\label{eqn:velo_reg_eq_tot}
\frac{\partial \vv}{\partial t} - \nu \nabla^2 \vv
+\frac{1}{\rho_f}\nabla {\rm q} = - \frac{1}{\rho_f}
\vD_p(t-\epsilon_R) \, g\left[ \vx-\vx_p(t-\epsilon_R),\epsilon_R \right]
\end{eqnarray}
with the initial condition $\vv(\vx,t_n) = 0$. For the sake of simplicity let us
consider an Euler-like time integration algorithm. In order to achieve the solution
$\vv(\vx,t_{n+1})$ the operator in (\ref{eqn:velo_reg_eq_tot}) is successively
splitted into three steps, namely the forcing step, the diffusion step and the 
projection step which enforces the condition $\nabla \cdot \vv=0$. Actually the 
forcing step gives
\begin{eqnarray}
\label{eqn:forcing_step}
\tilde{\vv}(\vx,t_{n+1})= - \frac{\Delta t}{\rho_f}
\vD_p(t_n-\epsilon_R) \, g\left[ \vx-\vx_p(t_n-\epsilon_R),\epsilon_R \right].
\end{eqnarray}
The diffusion step is readily achieved thanks to the semigroup property of 
solutions of the heat equation and the property (\ref{eqn:group_green}), 
namely
\begin{eqnarray}
\label{eqn:diffusion_step}
\vv_*(\vx,t_{n+1}) = \int \tilde{\vv}(\vxi,t_{n+1})
g\left( \vx-\vxi,\Delta t\right) d\vxi \, ,
\end{eqnarray}
which results in the pseudo-velocity
\begin{eqnarray}
\label{eqn:pseudo_velo_num}
\vv_*(\vx,t_{n+1}) = -\frac{\Delta t}{\rho_f}
\vD_p(t_n-\epsilon_R) \, g\left[ \vx-\vx_p(t_n-\epsilon_R),\epsilon_R+\Delta t \right].
\end{eqnarray}
 The divergence-free solution is achieved in terms of the decomposition 
$\vv(\vx,t_{n+1})=\vv_*(\vx,t_{n+1})+\nabla \Phi$ and the projection step
$\nabla^2 \phi = - \nabla \cdot \vv_*$. By using the expression 
(\ref{eqn:pseudo_velo_num}),
after some algebra, the solution $\vv(\vx,t_{n+1})$ can be evaluated in 
a closed form as
\begin{eqnarray}
\label{eqn:disturbance_velo}
\vv(\vx,t_{n+1}) = \frac{1}{\left(2 \pi \sigma^2\right)^{3/2}} 
\left\{\left[e^{-\eta^2} -\frac{f(\eta)}{2\eta^3}\right]  \vD^n 
- \left(\vD^n \cdot \hat{\vr}\right)
\left[e^{-\eta^2} -\frac{3f(\eta)}{2\eta^3}\right] \hat{\vr} \right\}.
\end{eqnarray}
In the above expression we have defined $\vD^n=\vD(t_n-\epsilon_R)$, 
$\vr=\vx-\vx_p(t_n-\epsilon_R)$, the hat denotes $\hat{\vr}=\vr/r$,
$\eta=r/\sqrt{2} \sigma$ is the dimensionless distance with 
$\sigma=\sqrt{2\nu(\epsilon_R + \Delta t)}$  and
$\displaystyle f(\eta)=\frac{\sqrt{\pi}}{2}\mbox{erf}(\eta)-\eta e^{-\eta^2}$.

\end{document}